\documentclass[10pt,aps,prb, twocolumn,superscriptaddress,floatfix, nofootinbib]{revtex4-2}
\usepackage{amsmath,amssymb,amstext,dsfont,tikz,graphicx,physics,mathtools,bm,
simpler-wick,ifthen}
\usepackage[normalem]{ulem}
\usepackage[colorlinks=true, allcolors=purple]{hyperref}
\usetikzlibrary{quantikz2}
\usepackage{circuitikz}
\usepackage{algpseudocode}
\usepackage{algorithm}
\usepackage{algpseudocode}
\usepackage{hyperref}
\usepackage[nameinlink,capitalise]{cleveref}
\usepackage{bbm,tikz}
\usepackage{multirow}   
\usetikzlibrary{arrows}

\def\bra#1{\mathinner{\langle{#1}|}}
\def\ket#1{\mathinner{|{#1}\rangle}}
\def\bs#1{\boldsymbol{#1}}

\makeatletter 
    
\renewcommand\onecolumngrid{% <<<<<<
\do@columngrid{one}{\@ne}%
\def\set@footnotewidth{\onecolumngrid}% <<<<<<<<<<<<<<<<
\def\footnoterule{\kern-6pt\hrule width 1.5in\kern6pt}%
}

\renewcommand\twocolumngrid{% <<<<<<
        \def\footnoterule{% restore rule
        \dimen@\skip\footins\divide\dimen@\thr@@
        \kern-\dimen@\hrule width.5in\kern\dimen@}
        \do@columngrid{mlt}{\tw@}
}%

\makeatother

\newcommand{\Z}{\mathbb{Z}}
\newcommand{\D}{\mathcal{D}}
\newcommand{\mcZ}{\mathcal{Z}}
\newcommand{\mcX}{\mathcal{X}}
\newcommand{\til}{\tilde}

\begin{document}

\title{Non-Clifford gates between stabilizer codes via non-Abelian topological order}

\author{Rohith Sajith}
\affiliation{Department of Physics, Harvard University, Cambridge, MA 02138, USA}

\author{Zijian Song}
\affiliation{Department of Physics and Astronomy, University of California, Davis, California 95656, USA}

\author{Brenden Roberts}
\affiliation{Department of Physics, Harvard University, Cambridge, MA 02138, USA}

\author{Varun Menon}
\affiliation{Department of Physics, Harvard University, Cambridge, MA 02138, USA}

\author{Yabo Li}
\affiliation{Center for Quantum Phenomena, Department of Physics,
New York University, 726 Broadway, New York, New York 10003, USA}

\date{\today}

\begin{abstract}
    We propose protocols to implement non-Clifford logical gates between stabilizer codes by entangling into a non-Abelian topological order as an intermediate step. Generalizing previous approaches, we provide a framework that generates a large class of non-Clifford and non-diagonal logical gates between qudit surface codes by gauging the topological symmetry of symmetry-enriched topological orders. As our main example, we concretely detail a protocol that utilizes the quantum double of $S_3$ to generate a controlled-charge conjugation ($C\mathcal{C}$) gate between a qubit and qutrit surface code. Both the preparation of non-Abelian states and logical state injection between the Abelian and non-Abelian codes are executed via finite-depth quantum circuits with measurement and feedforward. We discuss aspects of the fault-tolerance of our protocol, presenting insights on how to construct a heralded decoder for the quantum double of $S_3.$  We also outline how analogous protocols can be used to obtain logical gates between qudit surface codes by entangling into $\mathcal{D}(G),$ where $G$ is a semidirect product of Abelian groups. This work serves as a step towards classifying the computational power of non-Abelian quantum phases beyond the paradigm of anyon braiding on near-term quantum devices.
\end{abstract}

\makeatletter 
    
\renewcommand\onecolumngrid{% <<<<<<
\do@columngrid{one}{\@ne}%
\def\set@footnotewidth{\onecolumngrid}% <<<<<<<<<<<<<<<<
\def\footnoterule{\kern-6pt\hrule width 1.5in\kern6pt}%
}

\makeatother

\maketitle
{
  \hypersetup{linkcolor=black}
  \tableofcontents
}

\section{Introduction}

The discovery of topological order has profoundly transformed our understanding of quantum phases of matter and their potential applications in quantum computation~\cite{wen1990topological,wen1990ground,kitaev_fault_2023}. Over the past few decades, there has been substantial progress in the development of quantum error-correcting codes and their use in fault-tolerant quantum computation. A prominent class of such codes are the so-called topological codes (TCs). The paradigmatic example of a TC is the zero-temperature $\mathbb{Z}_2$ surface code, which realizes $\Z_2$ topological order and serves as a robust topological quantum memory~\cite{dennis2002topological}. Remarkably, such states have been realized in several near-term quantum platforms, including Rydberg atom-arrays, trapped ions, and superconducting qubits~\cite{bluvstein_logical_2023, evered_high_2023, moses_race_2023, acharya_quantum_2024}. Recently, Quantinuum's trapped ion platform has further realized both a $\mathbb{Z}_3$ toric code and the $D_4$ quantum double, both distinct topological phases from Kitaev's original qubit toric code \cite{iqbal_qutrit_2024, iqbal_non-Abelian_2024}. With these generalized topological orders at our disposal experimentally, a natural question arises: how can they be harnessed for quantum computation?

One promising approach to achieving universal topological quantum computation is to utilize the topological properties of non-Abelian anyons~\cite{kitaev_fault_2023}. In such schemes, computation is typically performed through the braiding and fusion measurements of non-Abelian anyons~\cite{kitaev_fault_2023,mochon2003anyons,mochon2004anyon,cui2015universal}. Several theoretical models are known to support universal quantum computation via these operations, including the Fibonacci anyon theory, the Majorana fermion model, and the $S_3$ quantum double~\cite{freedman2003topological,bravyi2002fermionic,mochon2004anyon,cui2015universal}. Despite significant recent experimental progress in preparing and manipulating non-Abelian anyons~\cite{iqbal_non-Abelian_2024}, a fully fault-tolerant method for their control remains an open challenge.

On the other hand, although manipulating non-Abelian anyons remains experimentally challenging, there exists a subclass of topological orders whose ground states can be prepared using measurement and feedforward techniques~\cite{bravyi2022adaptive,verresen_efficiently_2021,tantivasadakarn_hierarchy_2022,tantivasadakarn_shortest_2022,ren2024efficient, li_symmetry_2023}. These ground states can be generated with finite-depth quantum circuits, and decoders for such non-Abelian models have also been partially studied in recent work~\cite{chen_universal_2025}.

\subsection{Main results and ideas}

In this work, we utilize non-Abelian surface codes, which are generalizations of Abelian qudit surface codes, to perform quantum computation without relying on anyon fusion and braiding. Specifically, we present protocols that prepare non-Abelian topological order as an intermediate resource for implementing non-Clifford logical gates between Abelian qudit surface codes. In addition to performing lattice surgery between different qudit surface codes, our protocol offers a potential route to universal quantum computation. A key advantage of our approach is that the qudit surface code remains in its ground state throughout the entire process, thereby simplifying the decoding procedure. Similar ideas are also explored in recent literature~\cite{huang2025generating,davydova_universal_2025}.

As our main example, we demonstrate a protocol that yields a logical controlled-charge conjugation ($C\mathcal{C}$) gate between a qubit and a qutrit surface code via the entanglement of a non-Abelian $\mathcal{D}(S_3)$ code. Given a single qubit and a single qutrit, recall that 
\begin{equation}
    C\mathcal{C} = \ket{0}\bra{0} \otimes I + \ket{1}\bra{1} \otimes \mathcal{C},
\end{equation}
where $\mathcal{C}$ is the qutrit charge-conjugation operation that leaves $\ket{\tilde{0}}$ invariant but swaps $\ket{\tilde{1}}$ with $\ket{\tilde{2}}.$

Our protocol can be summarized as follows: Consider $\Z_2$ and $\Z_3$ surface codes both initialized in arbitrary logical states. We slide the $\mathbb{Z}_2$ surface code from left to right over the $\mathbb{Z}_3$ surface code by sequentially gauging and ungauging the charge-conjugation symmetry of the $\Z_3$ surface code. In the overlapping region of qubits and qutrits, we gauge the system to the $S_3$ quantum double. This results in a tripartite system consisting of the $\mathbb{Z}_2$ surface code, the $S_3$ quantum double, and the $\mathbb{Z}_3$ surface code. Gapped domain walls exist between the $S_3$ quantum double and the $\mathbb{Z}_2$ surface code, as well as between the $S_3$ quantum double and the $\mathbb{Z}_3$ surface code. As a result, logical operators of the stabilizer codes, or equivalently Abelian anyon lines, can be deformed across these domain walls into non-Abelian anyon lines of the $S_3$ quantum double. The anyon tunneling rules across these gapped boundaries are studied in Ref.~\cite{ren2023topological}. 

At a high level, the $\bs{\bar{X}}$ logical operator of the $\mathbb{Z}_2$ code—corresponding to the $m$ anyon line—maps to a charge-conjugation gauge flux in the non-Abelian $S_3$ code. As a result, sliding the logical information from the $\mathbb{Z}_2$ code through the $S_3$ quantum double from left to right implements a global $\mathcal{C}$ operation on the qutrit code precisely when the logical $\bs{\bar{Z}}$ operator is equal to $-1$. This realizes the action of a $C\mathcal{C}$ gate. Although a full proof of universality for the $\mathbb{Z}_2 \times \mathbb{Z}_3$ qudit surface code is beyond the scope of this work, we explicitly construct magic states for the underlying $\mathbb{Z}_2$ surface code using our $C\mathcal{C}$ gate in Appendix~\ref{magicstate}. This enables universal quantum computation using the qubit surface code, provided that the Clifford group can be implemented separately. A pictorial summary of our protocol is presented in Fig.~\ref{fig:overviewS3}.

We emphasize that our protocols are more general and can be used to extract logical operations from the quantum double $\mathcal{D}(G)$, where $G$ is any non-Abelian group realized as a split extension of two Abelian groups. Consider, for instance, the case $G = \mathbb{Z}_p \rtimes_{\psi} \mathbb{Z}_q$. Applying our protocol in this context yields a controlled-anyon automorphism gate:
\begin{equation}
C\mathrm{\psi}: |a b \rangle \mapsto | a^{t} b\rangle,
\end{equation}
where $a \in \mathbb{Z}_p$ labels the target qudit, $b \in \mathbb{Z}_q$ labels the control qudit, and $t$ is determined by the group action $\psi: \Z_q \rightarrow \text{Aut}(\Z_p).$

In addition to demonstrating the logical action of our protocol, we analyze aspects of its fault tolerance with respect to errors occurring after state preparation and those occurring before state preparation. To correct these errors, we propose a heralded decoding scheme for the $S_3$ quantum double. Unlike decoders designed for fault-tolerant quantum computation with non-Abelian anyons, our decoder is relatively simple, utilizing probabilistic syndrome measurements arising from the commuting projector model for the quantum double of $S_3.$ We present arguments for the fault-tolerance of our protocol to local stochastic Pauli noise, while a detailed analysis of measurement errors and circuit-level noise within the protocol is left for future work.  

This paper is organized as follows: in Section \ref{CodeConventions}, we discuss Abelian surface codes and the procedures of code extension and shrinkage. In Section \ref{CCGateMain}, we detail the procedure for a $C\mathcal{C}$ gate between a $\Z_2$ surface code and a $\Z_3$ surface code. In Section \ref{CAuto}, we discuss how the previous section can be generalized to obtain other non-Clifford logical gates from qudit surface codes. In Section \ref{SlidingErrorCorrection}, we detail how error correction can be performed during the course of our $C\mathcal{C}$ gate protocol.

\section{Surface Code Preliminaries}
\label{CodeConventions}

\subsection{Stabilizers and Logical Operators}

In this section, we detail out conventions for the surface codes used in the rest of the work. The bulk stabilizers of the standard $\Z_2$ surface code are
\begin{equation}
\raisebox{-.5\height} {
    \begin{tikzpicture}[scale=0.7]
            \draw (0.5, 1.5) -- (1.5, 1.5);
            \draw (0.5, 0.5) -- (1.5, 0.5);
            \draw (0.5, 0.5) -- (0.5, 1.5);
            \draw (1.5, 0.5) -- (1.5, 1.5);
            
            \node[magenta] at (0.5, 1) {$Z$};
            \node[magenta] at (1, 1.5) {$Z$};
            \node[magenta] at (1.5, 1) {$Z$};
            \node[magenta] at (1, 0.5) {$Z$};
            
            \draw (5, 1) -- (6, 1);
            \draw (6, 1) -- (6, 2);
            \draw (6, 1) -- (7, 1);
            \draw (6, 1) -- (6, 0);
            
            \node[blue] at (5.5, 1) {$X$};
            \node[blue] at (6, 1.5) {$X$};
            \node[blue] at (6.5, 1) {$X$};
            \node[blue] at (6, 0.5) {$X$};
    \end{tikzpicture} 
}
\label{ToricCodeStabs}
\end{equation} 
and the bulk stabilizers for the $\Z_3$ qutrit surface code are 
\begin{equation}
\raisebox{-0.5\height} {
    \begin{tikzpicture}[scale=0.7]
    \draw[brown] (0.5, 1.5) -- (1.5, 1.5);
    \draw[brown] (0.5, 0.5) -- (1.5, 0.5);
    \draw[brown] (0.5, 0.5) -- (0.5, 1.5);
    \draw[brown] (1.5, 0.5) -- (1.5, 1.5);
    
    \node[green] at (0.5, 1) {$\mcZ^{\dag}$};
    \node[green] at (1, 1.5) {$\mcZ$};
    \node[green] at (1.5, 1) {$\mcZ$};
    \node[green] at (1, 0.5) {$\mcZ^{\dag}$};

    \draw[brown] (5, 1) -- (6, 1);
    \draw[brown] (6, 1) -- (6, 2);
    \draw[brown] (6, 1) -- (7, 1);
    \draw[brown] (6, 1) -- (6, 0);
    
    \node[red] at (5.5, 1) {$\mcX$};
    \node[red] at (6, 1.5) {$\mcX$};
    \node[red] at (6.5, 1) {$\mcX^{\dag}$};
    \node[red] at (6, 0.5) {$\mcX^{\dag}$};
\end{tikzpicture}}
\end{equation}

Violations of qubit vertex stabilizers and qubit plaquette stabilizers will be referred to as $e$ and $m$ anyons, respectively. Similarly, violations of qutrit vertex and qutrit plaquette stabilizers will be referred to as $\til{e}$ and $\til{m}$ anyons. In order to encode logical information, we place both codes on lattices with smooth boundary conditions on the top and bottom boundaries and rough boundary conditions on the left and right boundaries. Specifically, this means that $e$ ($\til{e}$) anyon string operators may terminate on the rough boundaries without creating any anyons and $m$ ($\til{m}$) anyon string operators may terminate on the smooth boundaries of the qubit (qutrit) surface code without creating any anyons. Such string operators are the logical operators of the code. We remark that stabilizers at the rough and smooth boundaries are obtained by simply truncating the bulk plaquette and star operators, respectively. 

The $\Z_2$ surface code admits logical operators $\bs{\bar{Z}}$ and $\bs{\bar{X}}$ satisfying $\bs{\bar{Z}}\bs{\bar{X}} = -\bs{\bar{X}}\bs{\bar{Z}}$ and $\bs{\bar{Z}}^2 = \bs{\bar{X}}^2 = 1$, and thus stores one logical qubit. The configurations of the logical operators are displayed below.
\begin{equation}
\raisebox{-0.5\height} {
    \begin{tikzpicture}[scale=0.65]
    \draw (0,0) -- (10, 0);
    \draw (0,-1) -- (10,-1);
    \draw[line width = 0.5mm, magenta] (0,-2) -- (10,-2);
    \draw (0,-3) -- (10,-3);
    \draw (0,-4) -- (10,-4);

    \node at (10.75, -2) {$\bs{\bar{Z}}$};

    \node[magenta] at (0.5, -1.75) {$Z$};
    \node[magenta] at (1.5, -1.75) {$Z$};
    \node[magenta] at (2.5, -1.75) {$Z$};
    \node[magenta] at (3.5, -1.75) {$Z$};
    \node[magenta] at (4.5, -1.75) {$Z$};
    \node[magenta] at (5.5, -1.75) {$Z$};
    \node[magenta] at (6.5, -1.75) {$Z$};
    \node[magenta] at (7.5, -1.75) {$Z$};
    \node[magenta] at (8.5, -1.75) {$Z$};
    \node[magenta] at (9.5, -1.75) {$Z$};

    \node[blue] at (3.5, -0.25) {$X$};
    \node[blue] at (3.5, -1.25) {$X$};
    \node[blue] at (3.5, -2.25) {$X$};
    \node[blue] at (3.5, -3.25) {$X$};
    \node[blue] at (3.5, -4.25) {$X$};
    \node at (3.5, 0.5) {$\bs{\bar{X}}$};

    \draw (1, 0) -- (1, -4);
    \draw (2, 0) -- (2, -4);
    \draw (3, 0) -- (3, -4);
    \draw (4, 0) -- (4, -4);
    \draw (5, 0) -- (5, -4);
    \draw (6, 0) -- (6, -4);
    \draw (7, 0) -- (7, -4);
    \draw (8, 0) -- (8, -4);
    \draw (9, 0) -- (9, -4);

    \draw[line width = 0.5mm, blue] (3, 0) -- (4, 0);
    \draw[line width = 0.5mm, blue] (3, -1) -- (4, -1);
    \draw[line width = 0.5mm, blue] (3, -2) -- (4, -2);
    \draw[line width = 0.5mm, blue] (3, -3) -- (4, -3);
    \draw[line width = 0.5mm, blue] (3, -4) -- (4, -4);
\end{tikzpicture}}
\end{equation}

Similarly, the $\Z_{3}$ code admits logical operators $\bs{\bar{\mcZ}}$ and $\bs{\bar{\mcX}}$ satisfying the algebra $\bs{\bar{\mcZ}} \bs{\bar{\mcX}} = \omega \bs{\bar{\mcX}} \bs{\bar{\mcZ}}$ and $\bs{\bar{\mcZ}}^3 = \bs{\bar{\mcX}}^3 = 1$ for $\omega = e^{2\pi i / 3}$ and thus stores one logical qutrit. The configurations of the logical operators are displayed below.
\begin{equation}
\raisebox{-0.5\height}{
    \begin{tikzpicture}[scale=0.65]
    \draw[brown] (0,0) -- (10, 0);
    \draw[line width = 0.5mm, green] (0,-1) -- (10,-1);
    \draw[brown] (0,-2) -- (10,-2);
    \draw[line width = 0.5mm, green] (0,-3) -- (10,-3);
    \draw[brown] (0,-4) -- (10,-4);

    \node at (10.75, -1) {$\bs{\bar{\mcZ}}$};
    \node at (10.75, -3) {$\bs{\bar{\mcZ}^\dag}$};

    \node[green] at (0.5, -0.75) {$\mcZ$};
    \node[green] at (1.5, -0.75) {$\mcZ$};
    \node[green] at (2.5, -0.75) {$\mcZ$};
    \node[green] at (3.5, -0.75) {$\mcZ$};
    \node[green] at (4.5, -0.75) {$\mcZ$};
    \node[green] at (5.5, -0.75) {$\mcZ$};
    \node[green] at (6.5, -0.75) {$\mcZ$};
    \node[green] at (7.5, -0.75) {$\mcZ$};
    \node[green] at (8.5, -0.75) {$\mcZ$};
    \node[green] at (9.5, -0.75) {$\mcZ$};

    \node[green] at (0.5, -2.75) {$\mcZ^{\dag}$};
    \node[green] at (1.5, -2.75) {$\mcZ^{\dag}$};
    \node[green] at (2.5, -2.75) {$\mcZ^{\dag}$};
    \node[green] at (3.5, -2.75) {$\mcZ^{\dag}$};
    \node[green] at (4.5, -2.75) {$\mcZ^{\dag}$};
    \node[green] at (5.5, -2.75) {$\mcZ^{\dag}$};
    \node[green] at (6.5, -2.75) {$\mcZ^{\dag}$};
    \node[green] at (7.5, -2.75) {$\mcZ^{\dag}$};
    \node[green] at (8.5, -2.75) {$\mcZ^{\dag}$};
    \node[green] at (9.5, -2.75) {$\mcZ^{\dag}$};

    \node[red] at (3.5, -0.25) {$\mcX$};
    \node[red] at (3.5, -1.25) {$\mcX$};
    \node[red] at (3.5, -2.25) {$\mcX$};
    \node[red] at (3.5, -3.25) {$\mcX$};
    \node[red] at (3.5, -4.25) {$\mcX$};
    \node at (3.5, 0.5) {$\bs{\bar{\mcX}}$};

    \node[red] at (7.5, -0.25) {$\mcX^{\dag}$};
    \node[red] at (7.5, -1.25) {$\mcX^{\dag}$};
    \node[red] at (7.5, -2.25) {$\mcX^{\dag}$};
    \node[red] at (7.5, -3.25) {$\mcX^{\dag}$};
    \node[red] at (7.5, -4.25) {$\mcX^{\dag}$};
    \node at (7.5, 0.5) {$\bs{\bar{\mcX}^\dag}$};

    \draw[brown] (1, 0) -- (1, -4);
    \draw[brown] (2, 0) -- (2, -4);
    \draw[brown] (3, 0) -- (3, -4);
    \draw[brown] (4, 0) -- (4, -4);
    \draw[brown] (5, 0) -- (5, -4);
    \draw[brown] (6, 0) -- (6, -4);
    \draw[brown] (7, 0) -- (7, -4);
    \draw[brown] (8, 0) -- (8, -4);
    \draw[brown] (9, 0) -- (9, -4);

    \draw[line width = 0.5mm, red] (3, 0) -- (4, 0);
    \draw[line width = 0.5mm, red] (3, -1) -- (4, -1);
    \draw[line width = 0.5mm, red] (3, -2) -- (4, -2);
    \draw[line width = 0.5mm, red] (3, -3) -- (4, -3);
    \draw[line width = 0.5mm, red] (3, -4) -- (4, -4);

    \draw[line width = 0.5mm, red] (7, 0) -- (8, 0);
    \draw[line width = 0.5mm, red] (7, -1) -- (8, -1);
    \draw[line width = 0.5mm, red] (7, -2) -- (8, -2);
    \draw[line width = 0.5mm, red] (7, -3) -- (8, -3);
    \draw[line width = 0.5mm, red] (7, -4) -- (8, -4);
\end{tikzpicture}}
\end{equation}

Both here and in future sections we draw the lattice for the $\Z_3$ code in brown to distinguish it from the $\Z_2$ code, which is drawn in black.

The $\Z_3$ code has a $\Z_2$ symmetry given by a global charge-conjugation operation $U_{\mathcal{C}} = \prod_{e} \mathcal{C}_{e}$ that leaves its stabilizers invariant but applies a logical qutrit charge-conjugation operation $\bs{\mathcal{C}}.$ It is precisely this symmetry that will be gauged (as defined in the following section) to obtain the $S_3$ quantum double.

\subsection{$\Z_2$ Gauging Map}

In this section, we review the $\Z_2$ gauging map, which will be the main tool to prepare topologically ordered states throughout this work~\cite{yoshida_topological_2016,kubica2018ungauging}.

The $\Z_2$ gauging map is a map between states with $\Z_2$ global symmetry to states with $\Z_2$ gauge symmetry. We define the $\Z_2$ symmetric state $|+\rangle^{\otimes N_v}$ on the Hilbert space of qubits on the vertices $v$ of a square lattice with a global symmetry of $U_{v} = \prod_v X_{v}.$ The output state of the map lies in the Hilbert space of qubits on the edges $e$ of the square lattice. The gauging map sends the $\Z_2$ symmetric operators $X_{v}$ and $Z_{v}Z_{v'}$ to $\prod_{\langle e, v \rangle} X_{e}$ and $Z_{e}$, respectively, where $v$ and $v'$ share edge $e.$

This map can be implemented via a finite-depth circuit with measurement and feedforward~\cite{tantivasadakarn_long_2021,bravyi2022adaptive}. In particular, we start with a $\Z_2$ symmetric state $\ket{+}^{\otimes N_v}$ on the vertices and the state $\ket{0}^{\otimes N_e}$ on the edges. The surface code ground state $\ket{\Z_2}$ can be prepared on the edges via the following circuit:
\begin{equation}
    \ket{\Z_2} = \bra{+}^{\otimes N_v} \prod_{\langle e, v \rangle} CX_{v \rightarrow e} \ket{0}^{\otimes N_{e}} \ket{+}^{\otimes N_{v}}.
\end{equation}
One can check that the stabilizers for the new ground state $\ket{\Z_2}$ are exactly those in Eq.~\eqref{ToricCodeStabs}. In practice, the overlap is instead performed using projective measurement---we discuss how to deal with incorrect measurement outcomes later in this section. 

The conceptual idea is that the unitary entangling the edge qubits and vertex qubits, $\prod_{\langle e, v \rangle} CX_{v \rightarrow e}$, creates a 2D cluster state, which is a symmetry-protected topological phase protected by the $0$-form symmetry $U_v = \prod_{v} X_{v}$ and the $1$-form symmetry $U_e = \prod_{e \in \gamma} Z_{e},$ where $\gamma$ is any closed loop of edge qubits on the square lattice. It is a general property of such SPTs that measuring out the vertex qubits in the symmetric basis will cause the symmetry of the edge qubits to be spontaneously broken. This is due to the mixed anomaly between the zero-form and one-form symmetries~\cite{li_measuring_2023, wen_classifying_2013}. In this particular case, the state that spontaneously breaks the $1$-form symmetries is the $\Z_2$ toric code.

\subsection{Code Extension and Shrinkage}

Now we discuss how to use the gauging map to perform code extension. The setup is that we already have a surface code initialized in an arbitrary logical state, and we would like to extend the code in a particular direction (for example, to the right), or equivalently increase its horizontal code distance. 

First, we introduce ancilla qubits initialized in the $\ket{+}$ state on the vertices and in the $\ket{0}$ state on the edges of the extended lattice in the manner shown in Eq.~\eqref{CodeExtension}. We then apply the 2D cluster state entangler $U_{CX} = \prod_{\langle v, e \rangle} CX_{v \rightarrow e}$ between these ancilla qubits and the qubits on the rightmost column of the surface code. We have drawn the ancilla vertex qubits in red and ancilla qubits on edges in turquoise.

\begin{equation}
\raisebox{-0.5\height} {
    \begin{tikzpicture}[scale=0.7]
    \draw (0,0) -- (6, 0);
    \draw (0,-1) -- (6,-1);
    \draw (0,-2) -- (6,-2);
    \draw (0,-3) -- (6,-3);
    \draw (0,-4) -- (6,-4);
    \draw[dashed] (6, 0) -- (7, 0);
    \draw[dashed] (6, -1) -- (7, -1);
    \draw[dashed] (6, -2) -- (7, -2);
    \draw[dashed] (6, -3) -- (7, -3);
    \draw[dashed] (6, -4) -- (7, -4);

    % \node[blue] at (3.5, -0.25) {$X$};
    % \node[blue] at (3.5, -1.25) {$X$};
    % \node[blue] at (3.5, -2.25) {$X$};
    % \node[blue] at (3.5, -3.25) {$X$};
    % \node[blue] at (3.5, -4.25) {$X$};
    % \node at (3.5, 0.5) {$\bs{\bar{X}}$};

    \draw (1, 0) -- (1, -4);
    \draw (2, 0) -- (2, -4);
    \draw (3, 0) -- (3, -4);
    \draw (4, 0) -- (4, -4);
    \draw (5, 0) -- (5, -4);
    \draw[dashed] (6, 0) -- (6, -4);

    % \draw[line width = 0.5mm, blue] (3, 0) -- (4, 0);
    % \draw[line width = 0.5mm, blue] (3, -1) -- (4, -1);
    % \draw[line width = 0.5mm, blue] (3, -2) -- (4, -2);
    % \draw[line width = 0.5mm, blue] (3, -3) -- (4, -3);
    % \draw[line width = 0.5mm, blue] (3, -4) -- (4, -4);

    \foreach \i in {6,...,6} {
        \foreach \j in {0,...,-4} {
            \draw[purple] (\i, \j) circle[radius=1.5pt];
            \fill[purple] (\i, \j) circle[radius=1.5pt];
        }
    }

    \foreach \i in {6,...,6} {
        \foreach \j in {0, ..., -3} {
            \draw[teal] (\i+0.5, \j) circle[radius=1.5pt];
            \fill[teal] (\i+0.5, \j) circle[radius=1.5pt];
            \draw[teal] (\i, \j-0.5) circle[radius=1.5pt];
            \fill[teal] (\i, \j-0.5) circle[radius=1.5pt];
        }
    }

    \foreach \i in {6,..., 6} {
        \draw[teal] (\i+0.5, -4) circle[radius=1.5pt];
        \fill[teal] (\i+0.5, -4) circle[radius=1.5pt];
    }

    % \foreach \j in {0, ..., -4} {
    %     \draw[teal] (5.5, \j) circle[radius=1.5pt];
    %     \fill[teal] (5.5, \j) circle[radius=1.5pt];
    % }

    \foreach \i in {6, ..., 6} {
        \foreach \j in {0, ..., -4} {
            if \i < 9 then
                \draw[line width = 0.5mm, cyan] (\i+0.075, \j) -- (\i+0.5-0.075, \j);
            \draw[line width = 0.5mm, cyan] (\i-0.075, \j) -- (\i-0.5+0.075, \j);
            
        }
    }

    \foreach \i in {6,...,6} {
        \foreach \j in {0, ..., -3} {
            \draw[line width = 0.5mm, cyan] (\i, \j-0.075) -- (\i, \j-0.5+0.075);

            \draw[line width = 0.5mm, cyan] (\i, \j-0.075-0.5) -- (\i, \j-1+0.075);
        
        }
    }

    \node[cyan] at (6.25,0.25) {$CX$};
\end{tikzpicture}}
\label{CodeExtension}
\end{equation}

As a second step, we measure out the red vertex qubits in the $X$ basis. This has the effect of creating the desired $\Z_2$ surface code state up to some number of $e$ anyons on the rightmost column of the code, whose positions come from the $X = -1$ measurement outcomes. However, these anyons can be readily removed by acting with $Z$ operators directly to the right of a violated stabilizer, as shown:
\begin{equation}
\raisebox{-0.5\height} {
    \begin{tikzpicture}[scale=0.7]
    \draw (0,0) -- (6, 0);
    \draw (0,-1) -- (6,-1);
    \draw (0,-2) -- (6,-2);
    \draw (0,-3) -- (6,-3);
    \draw (0,-4) -- (6,-4);
    \draw[dashed] (6, 0) -- (7, 0);
    \draw[dashed] (6, -1) -- (7, -1);
    \draw[dashed] (6, -2) -- (7, -2);
    \draw[dashed] (6, -3) -- (7, -3);
    \draw[dashed] (6, -4) -- (7, -4);

    \node[blue] at (3.5, -0.25) {$X$};
    \node[blue] at (3.5, -1.25) {$X$};
    \node[blue] at (3.5, -2.25) {$X$};
    \node[blue] at (3.5, -3.25) {$X$};
    \node[blue] at (3.5, -4.25) {$X$};
    \node at (3.5, 0.5) {$\bs{\bar{X}}$};

    \draw (1, 0) -- (1, -4);
    \draw (2, 0) -- (2, -4);
    \draw (3, 0) -- (3, -4);
    \draw (4, 0) -- (4, -4);
    \draw (5, 0) -- (5, -4);
    \draw[dashed] (6, 0) -- (6, -4);

    \draw[line width = 0.5mm, blue] (3, 0) -- (4, 0);
    \draw[line width = 0.5mm, blue] (3, -1) -- (4, -1);
    \draw[line width = 0.5mm, blue] (3, -2) -- (4, -2);
    \draw[line width = 0.5mm, blue] (3, -3) -- (4, -3);
    \draw[line width = 0.5mm, blue] (3, -4) -- (4, -4);

    \foreach \i in {6,...,6} {
        \foreach \j in {0,-2,-4} {
            \node[purple] at (\i, \j) {$\bs{e}$};
            \node[magenta] at (\i+0.5, \j) {$\bs{Z}$};
        }
    }
\end{tikzpicture}}
\label{EAnyonCorrection}
\end{equation}

Since the application of these $Z$ operators commutes with both $\bs{\bar{X}}$ (as shown above) and $\bs{\bar{Z}}$ (trivially), such a correction creates logical coherence between the surface code bulk and newly created code.

We now show that we can perform shrinkage of the left boundary by measuring out qubits in the $Z$ basis. However, this has the byproduct of producing $m$ anyons at the left boundary of the code corresponding to $Z=-1$ measurement outcomes. We draw a sample configuration below with measured links dashed and $Z=-1$ outcomes in red:
\begin{equation}
\label{MAnyons}
\raisebox{-0.5\height} {
    \begin{tikzpicture}[scale=0.7]
    \draw (1,0) -- (7, 0);
    \draw (1,-1) -- (7,-1);
    \draw (1,-2) -- (7,-2);
    \draw (1,-3) -- (7,-3);
    \draw (1,-4) -- (7,-4);
    \draw[dashed] (0, 0) -- (1, 0);
    \draw[dashed] (0, -1) -- (1, -1);
    \draw[red, dashed,thick] (0, -2) -- (1, -2);
    \draw[red, dashed,thick] (0, -3) -- (1, -3);
    \draw[dashed] (0, -4) -- (1, -4);

    % \node[blue] at (3.5, -0.25) {$X$};
    % \node[blue] at (3.5, -1.25) {$X$};
    % \node[blue] at (3.5, -2.25) {$X$};
    % \node[blue] at (3.5, -3.25) {$X$};
    % \node[blue] at (3.5, -4.25) {$X$};
    % \node at (3.5, 0.5) {$\bs{\bar{X}}$};

    \draw (2, 0) -- (2, -4);
    \draw (3, 0) -- (3, -4);
    \draw (4, 0) -- (4, -4);
    \draw (5, 0) -- (5, -4);
    \draw (6, 0) -- (6, -4);
    
    \draw[dashed] (1, 0) -- (1, -1);
    \draw[red, dashed,thick] (1, -1) -- (1, -2);
    \draw[dashed] (1, -2) -- (1, -3);
    \draw[red, dashed,thick] (1, -3) -- (1, -4);

    % \draw[line width = 0.5mm, blue] (3, 0) -- (4, 0);
    % \draw[line width = 0.5mm, blue] (3, -1) -- (4, -1);
    % \draw[line width = 0.5mm, blue] (3, -2) -- (4, -2);
    % \draw[line width = 0.5mm, blue] (3, -3) -- (4, -3);
    % \draw[line width = 0.5mm, blue] (3, -4) -- (4, -4);

    \foreach \i in {0,...,0} {
        \foreach \j in {-1,-3} {
            \node[red] at (\i+0.5+1, \j-0.5) {$\bs{m}$};
        }
    }

\end{tikzpicture}}
\end{equation}

It is a property of the surface code state that the measurement outcomes in the $\ket{1}$ state must appear in closed loop configurations on the dual lattice, including non-contractible lines that go from the top boundary to the bottom boundary. While measurement of all the qubits in the $Z$ basis will collapse the logical state of the surface code, partial measurement need not. In fact, measurement of some number of columns on the left boundary fixes a partial loop configuration that terminates on some number of $m$ anyons; the remaining loop configuration fluctuates according to the logical state of the surface code. After such a partial measurement, we can always eliminate the $m$ anyons in a way that there are no non-contractible $Z=-1$ lines, or $m$ anyon lines aligning with the measurement record. As an example, we can eliminate the $m$ anyons in Eq.~\eqref{MAnyons} in the following way, forming a contractible loop:
\begin{equation}
\raisebox{-0.5\height} {
    \begin{tikzpicture}[scale=0.7]
    \draw (1,0) -- (7, 0);
    \draw (1,-1) -- (7,-1);
    \draw (1,-2) -- (7,-2);
    \draw (1,-3) -- (7,-3);
    \draw (1,-4) -- (7,-4);
    \draw[dashed] (0, 0) -- (1, 0);
    \draw[dashed] (0, -1) -- (1, -1);
    \draw[red, dashed,thick] (0, -2) -- (1, -2);
    \draw[red, dashed,thick] (0, -3) -- (1, -3);
    \draw[dashed] (0, -4) -- (1, -4);

    % \node[blue] at (3.5, -0.25) {$X$};
    % \node[blue] at (3.5, -1.25) {$X$};
    % \node[blue] at (3.5, -2.25) {$X$};
    % \node[blue] at (3.5, -3.25) {$X$};
    % \node[blue] at (3.5, -4.25) {$X$};
    % \node at (3.5, 0.5) {$\bs{\bar{X}}$};

    \draw (2, 0) -- (2, -4);
    \draw (3, 0) -- (3, -4);
    \draw (4, 0) -- (4, -4);
    \draw (5, 0) -- (5, -4);
    \draw (6, 0) -- (6, -4);
    
    \draw[dashed] (1, 0) -- (1, -1);
    \draw[red, dashed,thick] (1, -1) -- (1, -2);
    \draw[dashed] (1, -2) -- (1, -3);
    \draw[red, dashed,thick] (1, -3) -- (1, -4);

    % \draw[line width = 0.5mm, blue] (3, 0) -- (4, 0);
    % \draw[line width = 0.5mm, blue] (3, -1) -- (4, -1);
    % \draw[line width = 0.5mm, blue] (3, -2) -- (4, -2);
    % \draw[line width = 0.5mm, blue] (3, -3) -- (4, -3);
    % \draw[line width = 0.5mm, blue] (3, -4) -- (4, -4);

    % \foreach \i in {0,...,0} {
    %     \foreach \j in {-1,-3} {
    %         \node[red] at (\i+0.5+1, \j-0.5) {$\bs{m}$};
    %     }
    % }

    \node[blue] at (0+0.5+1, -2) {$X$};
    \node[blue] at (0+0.5+1, -3) {$X$};

\end{tikzpicture}}
\end{equation}

In the case that the measurement record indicates that we have measured $Z = -1$ in a non-contractible line, we apply $\bs{\bar{X}}$ correction operator to the remaining surface code state. This is because the $\bs{\bar{Z}}$ logical operator counts the parity of the number of non-contractible lines in the state. If we measure out such a line, the parity of non-contractible lines in the remaining state has flipped, thus swapping the logical $\bs{\bar{\ket{0}}}$ state with the logical $\bs{\bar{\ket{1}}}$. Thus, this correction operation ensures that the logical state of the shrunken surface code is that same as that of the original surface code. Such a correction operation is shown below:

\begin{equation}
\raisebox{-0.5\height} {
    \begin{tikzpicture}[scale=0.7]
    \draw (1,0) -- (7, 0);
    \draw (1,-1) -- (7,-1);
    \draw (1,-2) -- (7,-2);
    \draw (1,-3) -- (7,-3);
    \draw (1,-4) -- (7,-4);
    \draw[red, dashed, thick] (0, 0) -- (1, 0);
    \draw[red, dashed, thick] (0, -1) -- (1, -1);
    \draw[red, dashed,thick] (0, -2) -- (1, -2);
    \draw[red, dashed,thick] (0, -3) -- (1, -3);
    \draw[red, dashed, thick] (0, -4) -- (1, -4);

    \node[blue] at (3.5, -0.25) {$X$};
    \node[blue] at (3.5, -1.25) {$X$};
    \node[blue] at (3.5, -2.25) {$X$};
    \node[blue] at (3.5, -3.25) {$X$};
    \node[blue] at (3.5, -4.25) {$X$};
    \node at (3.5, 0.5) {$\bs{\bar{X}}$};

    \draw (2, 0) -- (2, -4);
    \draw (3, 0) -- (3, -4);
    \draw (4, 0) -- (4, -4);
    \draw (5, 0) -- (5, -4);
    \draw (6, 0) -- (6, -4);
    
    \draw[dashed] (1, 0) -- (1, -1);
    \draw[dashed] (1, -1) -- (1, -2);
    \draw[dashed] (1, -2) -- (1, -3);
    \draw[dashed] (1, -3) -- (1, -4);

    \draw[line width = 0.5mm, blue] (3, 0) -- (4, 0);
    \draw[line width = 0.5mm, blue] (3, -1) -- (4, -1);
    \draw[line width = 0.5mm, blue] (3, -2) -- (4, -2);
    \draw[line width = 0.5mm, blue] (3, -3) -- (4, -3);
    \draw[line width = 0.5mm, blue] (3, -4) -- (4, -4);

    % \foreach \i in {0,...,0} {
    %     \foreach \j in {-1,-3} {
    %         \node[red] at (\i+0.5+1, \j-0.5) {$\bs{m}$};
    %     }
    % }

\end{tikzpicture}}
\end{equation}

\section{Controlled Charge-Conjugation Gate ($C\mathcal{C}$) from the $S_3$ quantum double}
\label{CCGateMain}

\begin{figure*}[t!]
    \centering
    \begin{tikzpicture}

    \node at (-2, 1.25) {$(a)$};
    \node at (0,0) {
        \begin{minipage}{0.18\textwidth} % First figure
            \begin{tikzpicture}
                \draw (0, 1) -- (1, 2);
                \draw (1, 2) -- (4, 2);
                \draw (4, 2) -- (3, 1);
                \draw (3, 1) -- (0, 1);
                \node at (2, 1.5) {$\mathcal{D}(\mathbb{Z}_2)$};
                \draw[blue] (3.5, 2) -- (2.5, 1);
                \node[blue] at (3.5, 2.25) {$\overline{X}$};
                \draw[magenta] (0.75,1.75) -- (3.75,1.75);
                \node[magenta] at (0.4, 1.75) {$\overline{Z}$};
                \draw (0+3, 1-1) -- (1+3, 2-1);
                \draw (1+3, 2-1) -- (4+3, 2-1);
                \draw (4+3, 2-1) -- (3+3, 1-1);
                \draw (3+3, 1-1) -- (0+3, 1-1);
                \node at (2+3, 1.5-1) {$\mathcal{D}(\mathbb{Z}_3)$};
                \draw[red] (3.5+3, 2-1) -- (2.5+3, 1-1);
                \node[red] at (3.5+3, 2.25-1) {$\overline{\mathcal{X}}$};
                \draw[green] (0.25+3, 1.25-1) -- (3.25+3, 1.25-1);
                \node[green] at (3.25+3+0.35, 1.25-1) {$\overline{\mathcal{Z}}$};
            \end{tikzpicture}
        \end{minipage}
    };

    \node at (6, 1.25) {$(b)$};

    \node at (8,0) {
        \begin{minipage}{0.18\textwidth} % First figure
            \begin{tikzpicture}
                \draw (0, 1) -- (1, 2);
                \draw (1, 2) -- (2.5, 2);
                \draw[dashed] (4, 2) -- (3, 1);
                \draw (1.5, 1) -- (0, 1);
                \node at (1.25, 1.5) {$\mathcal{D}(\mathbb{Z}_2)$};
                \draw (2.5, 2) -- (1.5, 1);
                \draw[dashed] (2.5, 2) -- (4, 2);
                \draw[dashed] (1.5, 1) -- (3, 1);

                \draw (0+1+0.25, 1-1.25) -- (1+1+0.25, 2-1.25);
                \draw (1+1+0.25, 2-1.25) -- (4+1+0.25, 2-1.25);
                \draw (4+1+0.25, 2-1.25) -- (3+1+0.25, 1-1.25);
                \draw (3+1+0.25, 1-1.25) -- (0+1+0.25, 1-1.25);
                % \draw[dashed] (2.5+1+0.25, 2-1.25) -- (1.5+1+0.25, 1-1.25);
                \node at (2+1+0.75-0.15, 1.5-1.25) {$\mathcal{D}(\mathbb{Z}_3)$};
                
                \draw[purple] (2.25+0.25, 1.25) circle[radius=1.5pt];
                \fill[purple] (2.25+0.25, 1.25) circle[radius=1.5pt];
                \draw[purple] (2.25+0.25, 1.25-1.125) -- (2.25+0.25, 1.25);
                \node[purple] at (2.25+0.25, 1.25-1.25) {$\mathcal{C}$};

                \node[scale=0.8] at (2.25+0.25-0.35, 1.25) {$\ket{+}$};
                \node[scale=0.8] at (2.25+0.5+0.25-0.35, 1.25+0.5) {$\ket{+}$};
                
                \draw[purple] (2.25+0.5+0.25, 1.25+0.5) circle[radius=1.5pt];
                \fill[purple] (2.25+0.5+0.25, 1.25+0.5) circle[radius=1.5pt];
                \draw[purple] (2.25+0.5+0.25, 1.25-1.125+0.5) -- (2.25+0.5+0.25, 1.25+0.5);
                \node[purple] at (2.25+0.5+0.25, 1.25-1.25+0.5) {$\mathcal{C}$};
            \end{tikzpicture}
        \end{minipage}
    };

    \node at (12.5, 1.25) {$(c)$};

    \node at (14,0) {
        \begin{minipage}{0.18\textwidth} % First figure
            \begin{tikzpicture}
                \draw (0, 1) -- (1, 2);
                \draw (1, 2) -- (4, 2);
                \draw (4, 2) -- (3, 1);
                \draw (3, 1) -- (0, 1);
                \node at (2, 1.5) {$\mathcal{D}(S_3)$};
                \draw[blue] (3.5, 2) -- (2.5, 1);
                \node[blue] at (3.5, 2.25) {$D$};
                \draw[red] (1.5, 2) -- (0.5, 1);
                \node[red] at (1.5, 2.25) {$F$};
                \draw[green] (0.25, 1.25) -- (3.25, 1.25);
                \node[green] at (3.25+0.35, 1.25) {$C$};
                \draw[magenta] (0.25+0.5, 1.25+0.5) -- (3.25+0.5, 1.25+0.5);
                \node[magenta] at (3.25+0.5+0.25, 1.25+0.5) {$B$};
            \end{tikzpicture}
        \end{minipage}
    };
    
    \node at (2-4, 1.25-3) {$(d)$};

    \node at (5.85+6.25-9-4,-0.25-3) {
        \begin{minipage}{0.18\textwidth}
            \begin{tikzpicture}
                \draw (0+2, 1) -- (1+2, 2);
                \draw (1+2, 2) -- (4+2, 2);
                \draw (4+2, 2) -- (3+2, 1);
                \draw (3+2, 1) -- (0+2, 1);
                \draw[dashed] (0+2, 1) -- (-1.5+2, 1);
                \draw[dashed] (1+2, 2) -- (1+2-1.5, 2);
                \draw[dashed] (1+2-1.5, 2) -- (-1.5+2, 1);
                \node at (1.25+2+1.5, 1.5) {$\mathcal{D}(\mathbb{Z}_2)$};
                \draw (2.5+2, 2) -- (1.5+2, 1);

                \draw (0+1+0.25-1.5, 1-1.25) -- (1+1+0.25-1.5, 2-1.25);
                \draw (1+1+0.25-1.5, 2-1.25) -- (4+1+0.25-1.5, 2-1.25);
                \draw (4+1+0.25-1.5, 2-1.25) -- (3+1+0.25-1.5, 1-1.25);
                \draw (3+1+0.25-1.5, 1-1.25) -- (0+1+0.25-1.5, 1-1.25);
                \draw (2.5+1+0.25-1.5, 2-1.25) -- (1.5+1+0.25-1.5, 1-1.25);
                \node at (2+1+0.75+0.25-3, 1.5-1.25) {$\mathcal{D}(\mathbb{Z}_3)$};

                \node at (2+1+0.75+0.25-3+1.5, 1.5-1.25) {$\mathcal{D}(S_3)$};
                
                \node[scale=0.7] at (1.5+0.2-0.2,1.5-0.2) {\begin{quantikz}
                    \meter{}
                \end{quantikz}};
                % \draw[densely dotted] (1.85, 1.4-0.2) -- (2.225, 1.4-0.2);

                \node[scale=0.7] at (1.5+0.5+0.2-0.15,1.5-0.2+0.7-0.2) {\begin{quantikz}
                    \meter{}
                \end{quantikz}};
                % \draw[densely dotted] (1.85+0.5, 1.4-0.2+0.7) -- (2.225+0.5+0.2, 1.4-0.2+0.7);

                \draw[blue] (2.2, 1.2) arc [
                    start angle = -90,
                    end angle = 26.5,
                    x radius = 0.8 cm,
                    y radius = 0.5 cm
                ];
                \node[blue, scale=0.7] at (2.2+0.5+0.5-0.25-0.1, 1.2+0.5-0.25-0.2) {$D$};
            \end{tikzpicture}
        \end{minipage}    
    };

    \node at (11-5, 1.25-3)  {$(e)$};

    \node at (5.85+6.25-5,-0.25-3) {
        \begin{minipage}{0.18\textwidth}
            \begin{tikzpicture}
                \draw (0+4.25, 1) -- (1+4.25, 2);
                \draw (1+4.25, 2) -- (4+4.25, 2);
                \draw (4+4.25, 2) -- (3+4.25, 1);
                \draw (3+4.25, 1) -- (0+4.25, 1);
                \node at (2+4.25, 1.5) {$\mathcal{D}(\mathbb{Z}_2)$};
                \draw[blue] (3.5+4.25, 2) -- (2.5+4.25, 1);
                \node[blue] at (3.5+4.25, 2.25) 
                {$\overline{X}\mathcal{C}$};
                \draw[magenta] (0.75+4.25,1.75) -- (3.75+4.25,1.75);
                \node[magenta] at (3.75+4.25+0.3,1.75) {$\overline{Z}$};

                \draw (0, 1-1) -- (1, 2-1);
                \draw (1, 2-1) -- (4, 2-1);
                \draw (4, 2-1) -- (3, 1-1);
                \draw (3, 1-1) -- (0, 1-1);
                \node at (2, 1.5-1) {$\mathcal{D}(\mathbb{Z}_3)$};
                \draw[red] (3.5, 2-1) -- (2.5, 1-1);
                \node[red] at (3.5, 2.25-1) {$\overline{\mathcal{X}}^{\textcolor{magenta}{\overline{Z}}}$};
                \draw[green] (0.25, 1.25-1) -- (3.25, 1.25-1);
                \node[green] at (3.25+0.35, 1.25-1) {$\overline{\mathcal{Z}}^{\textcolor{magenta}{\overline{Z}}}$};
    
            \end{tikzpicture}
        \end{minipage}    
    };

\end{tikzpicture}
    \caption{Implementation of logical $C\mathcal{C}$ gate between $\D(\Z_2)$ and $\D(\Z_3)$ surface codes. $(a)$ The qubit and qutrit surface codes are separately initialized in arbitrary logical states. Depicted are the $\overline{\mcZ}$ and $\overline{\mcX}$ logical operators of the qutrit code and the $\overline{Z}$ and $\overline{X}$ logical operators of the qubit code. $(b)$ Non-Clifford $C\mathcal{C}$ gates between ancilla qubits initialized in the $\ket{0}$ and qutrits of the $\Z_3$ code. This is the symmetry-enrichment step, where the $\Z_2$ charge-conjugation symmetry of the qutrit surface code is coupled to ancilla qubits. $(c)$ Applying the gauging map to the symmetry-enriched $\Z_3$ code yields $S_3$ topological order. After applying the gauging map to the entire $\Z_3$ code, a $S_3$ quantum double has been prepared with $A+B+2C$ boundary conditions on the left and right boundaries and $A+D+F$ boundary conditions on the top and bottom boundaries, which are the analogues of rough and smooth boundary conditions for the $S_3$ non-Abelian code. Logical information from both codes is now injected into the $S_3$ code. The $\overline{X}$ and $\overline{Z}$ logical operator from the qubit code transform into the $B$ and $D$ anyon operators, respectively, while the $\overline{\mcZ}$ and the $\overline{\mcX}$ map to the $C$ and $F$ anyon operators respectively. $(d)$ The ejection of the qubit and qutrit code from the non-Abelian code is done by simply measuring out qubits from the left side of the $\Z_2$ code in the $Z$ basis. Measurement outcomes of $Z = -1$ correspond to the endpoints of ground state $D$ anyon loops terminating at the left boundary. Feedforward is performed to return the stabilizers of the $\Z_3$ code back to their original form.  $(e)$ Once the $\D(\Z_2)$ code is completely ejected from the $\D(\Z_3)$ code an effective $C\mathcal{C}$ has been performed taking $\overline{\mcZ} \rightarrow \overline{\mcZ}^{\overline{Z}},$ $\overline{\mcX} \rightarrow \overline{\mcX}^{\overline{Z}},$ $\overline{Z} \rightarrow \overline{Z},$ and $\overline{X} \rightarrow \overline{X}\mathcal{C}.$  More details on the transfer of logical operators are discussed in Section~\ref{CCGateMain}.}
    \label{fig:overviewS3}
\end{figure*}
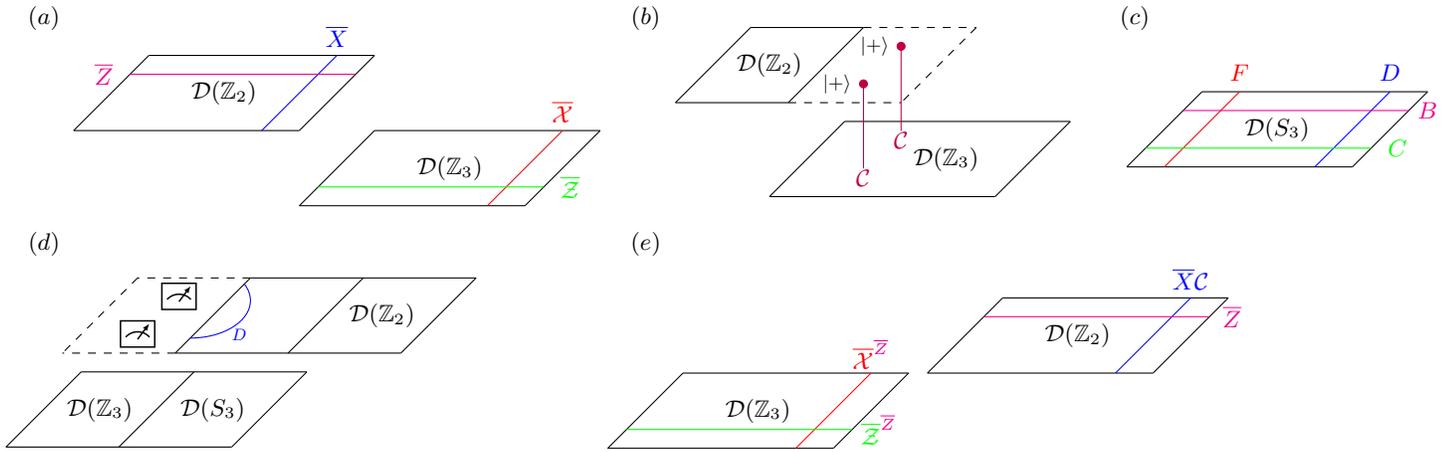

Here we detail the protocol that yields a $C\mathcal{C}$ between a qubit and a qutrit surface code utilizing the $S_3$ quantum double. The protocol is outlined in Fig.~\ref{fig:overviewS3}. It is worth noting that our protocol implements a $C\mathcal{C}$ gate between a qubit and a qutrit surface code, both of which can be initialized in arbitrary logical states. 

\subsection{Code Injection: Symmetry Enrichment of the Ground State}

As the first step, we show in this section that we can inject the logical state of both codes into the $S_3$ quantum double by performing the $\Z_2$ gauging map. This will construct a domain wall between the $\Z_2$ surface code and a newly created $S_3$ surface code. While we introduce the commuting projectors for the $S_3$ code in this section, a more detailed discussion regarding its anyon content and its boundaries can be found in Appendix~\ref{S3Review} and Appendix~\ref{S3Boundaries}, respectively.

Everywhere in the region in which we would like to create the $S_3$ surface code, we introduce ancilla qubits on the vertices initialized in the $\ket{+}$ state and on the edges initialized in the $\ket{0}$ state, on top of a $\mathbb{Z}_3$ surface code. As a first step, we apply a finite-depth circuit of local $C\mathcal{C}$ gates from the $\Z_2$ code to the $\Z_3$ code as shown.
\begin{equation}
    U_{C\mathcal{C}} = \prod_{\langle v, e \rangle^{{\rightarrow, \downarrow}}} C\mathcal{C}_{v \rightarrow e}
\end{equation}
\begin{equation}
\raisebox{-0.5\height} {
    \begin{tikzpicture}[scale=0.65]
    \draw (0,0) -- (6, 0);
    \draw (0,-1) -- (6,-1);
    \draw (0,-2) -- (6,-2);
    \draw (0,-3) -- (6,-3);
    \draw (0,-4) -- (6,-4);
    \draw[dashed] (6, 0) -- (10, 0);
    \draw[dashed] (6, -1) -- (10, -1);
    \draw[dashed] (6, -2) -- (10, -2);
    \draw[dashed] (6, -3) -- (10, -3);
    \draw[dashed] (6, -4) -- (10, -4);

    \draw[brown] (6+0.2, 0-0.2) -- (10+0.2, 0-0.2);
    \draw[brown] (6+0.2, -1-0.2) -- (10+0.2, -1-0.2);
    \draw[brown] (6+0.2, -2-0.2) -- (10+0.2, -2-0.2);
    \draw[brown] (6+0.2, -3-0.2) -- (10+0.2, -3-0.2);
    \draw[brown] (6+0.2, -4-0.2) -- (10+0.2, -4-0.2);

    % \node[blue] at (3.5, -0.25) {$X$};
    % \node[blue] at (3.5, -1.25) {$X$};
    % \node[blue] at (3.5, -2.25) {$X$};
    % \node[blue] at (3.5, -3.25) {$X$};
    % \node[blue] at (3.5, -4.25) {$X$};
    % \node at (3.5, 0.5) {$\bs{\bar{X}}$};

    \draw (1, 0) -- (1, -4);
    \draw (2, 0) -- (2, -4);
    \draw (3, 0) -- (3, -4);
    \draw (4, 0) -- (4, -4);
    \draw (5, 0) -- (5, -4);
    \draw[dashed] (6, 0) -- (6, -4);
    \draw[dashed] (7, 0) -- (7, -4);
    \draw[dashed] (8, 0) -- (8, -4);
    \draw[dashed] (9, 0) -- (9, -4);

    \draw[brown] (7+0.2, 0-0.2) -- (7+0.2, -4-0.2);
    \draw[brown] (8+0.2, 0-0.2) -- (8+0.2, -4-0.2);
    \draw[brown] (9+0.2, 0-0.2) -- (9+0.2, -4-0.2);
    % \draw[brown] (10+0.2, 0-0.2) -- (10+0.2, -4-0.2);

    % \draw[line width = 0.5mm, blue] (3, 0) -- (4, 0);
    % \draw[line width = 0.5mm, blue] (3, -1) -- (4, -1);
    % \draw[line width = 0.5mm, blue] (3, -2) -- (4, -2);
    % \draw[line width = 0.5mm, blue] (3, -3) -- (4, -3);
    % \draw[line width = 0.5mm, blue] (3, -4) -- (4, -4);

    \foreach \i in {6,...,9} {
        \foreach \j in {0,...,-4} {
            \draw[line width = 0.3mm, purple] (\i, \j) -- (\i+0.6, \j-0.15);
            \draw[purple] (\i, \j) circle[radius=1.5pt];
            \fill[purple] (\i, \j) circle[radius=1.5pt];
            \node[purple] at (\i+0.7, \j-0.2) {$\mathcal{C}$};
        }
    }

    \foreach \i in {7,...,9} {
        \foreach \j in {0,...,-3} {
            \draw[line width = 0.3mm, purple] (\i, \j) -- (\i+0.15, \j-0.6);
            \draw[purple] (\i, \j) circle[radius=1.5pt];
            \fill[purple] (\i, \j) circle[radius=1.5pt];
            \node[purple] at (\i+0.2, \j-0.7) {$\mathcal{C}$};
        }
    }

    \node[purple] at (9.5,0.25) {$C\mathcal{C}$};

\end{tikzpicture}}
\label{CCGate}
\end{equation}

We refer to this step of the $S_3$ code preparation as \emph{symmetry-enrichment} since we are coupling the $\Z_2$ symmetry $U_{v} = \prod_{v} X_{v}$ of the ancilla qubits to the $U_{\mathcal{C}} = \prod_{e} \mathcal{C}_{e}$ symmetry of the $\Z_3$ surface code with the $C\mathcal{C}$ gates. This is done by coupling each vertex qubit to the two edge qutrits directly to the left and directly below it.

Once this entangling unitary is performed between the qubits and qutrits, we can perform the $\Z_2$ gauging map exactly as described in Eq.~\eqref{CodeExtension} and Eq.~\eqref{EAnyonCorrection}. When the $\Z_2$ gauging map (or equivalently code extension process) is preceeded by the symmetry-enriching unitary, we refer to it as the charge-conjugation gauging map. This procedure creates a small slab of $S_3$ topological order in the region that the qubits and qutrits overlap. It further allows us to coherently inject the $\Z_2$ logical state into the non-Abelian topological order, as we will show later.

We now describe the commuting projectors of the bulk $S_3$ code as obtained from the symmetry-enrichment and gauging map circuits~\cite{verresen_efficiently_2021}. Before any entangling gates are applied, we have the following stabilizers in the region that the qubits and qutrits overlap:
\begin{equation}
\raisebox{-0.5\height} {
    \begin{tikzpicture}[scale=0.75]
    \draw[brown] (0.5, 1.5) -- (1.5, 1.5);
    \draw[brown] (0.5, 0.5) -- (1.5, 0.5);
    \draw[brown] (0.5, 0.5) -- (0.5, 1.5);
    \draw[brown] (1.5, 0.5) -- (1.5, 1.5);
    
    \node[green] at (0.5, 1) {$\mcZ^{\dag}$};
    \node[green] at (1, 1.5) {$\mcZ$};
    \node[green] at (1.5, 1) {$\mcZ$};
    \node[green] at (1, 0.5) {$\mcZ^{\dag}$};

    \draw[brown] (5-2.5, 1) -- (6-2.5, 1);
    \draw[brown] (6-2.5, 1) -- (6-2.5, 2);
    \draw[brown] (6-2.5, 1) -- (7-2.5, 1);
    \draw[brown] (6-2.5, 1) -- (6-2.5, 0);
    
    \node[red] at (5.5-2.5, 1) {$\mcX$};
    \node[red] at (6-2.5, 1.5) {$\mcX$};
    \node[red] at (6.5-2.5, 1) {$\mcX^{\dag}$};
    \node[red] at (6-2.5, 0.5) {$\mcX^{\dag}$};

    \draw (5.5, 1) -- (6.5, 1);
    \node[magenta] at (6, 1) {$Z$};

    \draw (5+2.5, 1) -- (6+2.5, 1);
    \draw (6+2.5, 1) -- (6+2.5, 2);
    \draw (6+2.5, 1) -- (7+2.5, 1);
    \draw (6+2.5, 1) -- (6+2.5, 0);

    \node[blue] at (6+2.5,1) {$X$};

\end{tikzpicture}}
\end{equation}

After applying $U_{C\mathcal{C}},$ the stabilizers now become: 
\begin{equation}
\raisebox{-0.5\height} {
    \begin{tikzpicture}
    \draw[brown] (0.5, 1.5) -- (1.5, 1.5);
    \draw[brown] (0.5, 0.5) -- (1.5, 0.5);
    \draw[brown] (0.5, 0.5) -- (0.5, 1.5);
    \draw[brown] (1.5, 0.5) -- (1.5, 1.5);

    \draw[gray] (0.5-0.3, 1.5+0.3) -- (1.5-0.3, 1.5+0.3);
    \draw[gray] (0.5-0.3, 0.5+0.3) -- (1.5-0.3, 0.5+0.3);
    \draw[gray] (0.5-0.3, 0.5+0.3) -- (0.5-0.3, 1.5+0.3);
    \draw[gray] (1.5-0.3, 0.5+0.3) -- (1.5-0.3, 1.5+0.3);

    \draw[magenta] (0.5-0.3, 1.5+0.3) circle[radius=1pt];
    \fill[magenta] (0.5-0.3, 1.5+0.3) circle[radius=1pt];
    \draw[magenta] (1.5-0.3, 1.5+0.3) circle[radius=1pt];
    \fill[magenta] (1.5-0.3, 1.5+0.3) circle[radius=1pt];
    \draw[magenta] (0.5-0.3, 0.5+0.3) circle[radius=1pt];
    \fill[magenta] (0.5-0.3, 0.5+0.3) circle[radius=1pt];
    \draw[magenta] (1.5-0.3, 0.5+0.3) circle[radius=1pt];
    \fill[magenta] (1.5-0.3, 0.5+0.3) circle[radius=1pt];

    \node[magenta, scale=0.6] at (0.5-0.3-0.2, 1.5+0.3) {$1$};
    \node[magenta, scale=0.6] at (0.5-0.3-0.2, 0.5+0.3) {$3$};
    \node[magenta, scale=0.6] at (1.5-0.3+0.2, 1.5+0.3) {$2$};

    \node[green, scale=0.7] at (0.5, 1) {$\mcZ^{-\textcolor{magenta}{Z_{1}}}$};
    \node[green, scale=0.7] at (1, 1.5) {$\mcZ^{\textcolor{magenta}{Z_{1}}}$};
    \node[green, scale=0.7] at (1.5, 1) {$\mcZ^{\textcolor{magenta}{Z_{2}}}$};
    \node[green, scale=0.7] at (1, 0.5) {$\mcZ^{\textcolor{magenta}{-Z_{3}}}$};

    \draw[brown] (5-2.5, 1) -- (6-2.5, 1);
    \draw[brown] (6-2.5, 1) -- (6-2.5, 2);
    \draw[brown] (6-2.5, 1) -- (7-2.5, 1);
    \draw[brown] (6-2.5, 1) -- (6-2.5, 0);

    \draw[gray] (5-2.5-0.3, 1+0.3) -- (6-2.5-0.3, 1+0.3);
    \draw[gray] (6-2.5-0.3, 1+0.3) -- (6-2.5-0.3, 2+0.3);
    \draw[gray] (6-2.5-0.3, 1+0.3) -- (7-2.5-0.3, 1+0.3);
    \draw[gray] (6-2.5-0.3, 1+0.3) -- (6-2.5-0.3, 0+0.3);

    \node[magenta, scale=0.6] at (5-2.5-0.3, 1+0.3+0.15) {$3$};
    \draw[magenta] (5-2.5-0.3, 1+0.3) circle[radius=1pt];
    \fill[magenta] (5-2.5-0.3, 1+0.3) circle[radius=1pt];

    \node[magenta, scale=0.6] at (6-2.5-0.3-0.15, 2+0.3) {$2$};
    \draw[magenta] (6-2.5-0.3, 2+0.3) circle[radius=1pt];
    \fill[magenta] (6-2.5-0.3, 2+0.3) circle[radius=1pt];

    \node[magenta, scale=0.6] at (6-2.5-0.3-0.15, 1+0.3+0.15) {$1$};
    \draw[magenta] (6-2.5-0.3, 1+0.3) circle[radius=1pt];
    \fill[magenta] (6-2.5-0.3, 1+0.3) circle[radius=1pt];
    
    \node[red, scale=0.7] at (5.5-2.5, 1) {$\mcX^{\textcolor{magenta}{Z_3}}$};
    \node[red, scale=0.7] at (6-2.5, 1.5) {$\mcX^{\textcolor{magenta}{Z_2}}$};
    \node[red, scale=0.7] at (6.5-2.5, 1) {$\mcX^{^{\textcolor{magenta}{-Z_1}}}$};
    \node[red, scale=0.7] at (6-2.5, 0.5) {$\mcX^{^{\textcolor{magenta}{-Z_1}}}$};

\end{tikzpicture}}
\end{equation}

\begin{equation}
\raisebox{-0.5\height} {
    \begin{tikzpicture}[scale=0.7]

    \draw (5.5, 1) -- (6.5, 1);
    \node[magenta] at (6, 1) {$Z$};

    \draw (5+2.5, 1) -- (6+2.5, 1);
    \draw (6+2.5, 1) -- (6+2.5, 2);
    \draw (6+2.5, 1) -- (7+2.5, 1);
    \draw (6+2.5, 1) -- (6+2.5, 0);

    \node[blue] at (6+2.5,1) {$X$};

    \draw[brown] (6+2.5+0.2, 1-0.2) -- (7+2.5+0.2, 1-0.2);
    \draw[brown] (6+2.5+0.2, 1-0.2) -- (6+2.5+0.2, 0-0.2);

    \node[purple] at (6+2.5+0.2+0.5, 1-0.2) {$\mathcal{C}$};
    \node[purple] at (6+2.5+0.2, 1-0.2-0.5) {$\mathcal{C}$};

    \end{tikzpicture}}
\end{equation}

The exponentiated operators indicate qutrit operators that are controlled on the eigenvalues of qubit operators. After applying $U_{CX}$, the qutrit stabilizers remain the same, but the qubit stabilizers get modified to
\begin{equation}
\raisebox{-0.5\height} {
    \begin{tikzpicture}[scale=0.9]

    \draw (5.5, 1) -- (6.5, 1);
    \node[magenta] at (6, 1) {$Z$};
    \node[magenta] at (5.5,1) {$Z$};
    \node[magenta] at (6.5,1) {$Z$};

    \draw (5+2.5, 1) -- (6+2.5, 1);
    \draw (6+2.5, 1) -- (6+2.5, 2);
    \draw (6+2.5, 1) -- (7+2.5, 1);
    \draw (6+2.5, 1) -- (6+2.5, 0);

    \node[blue] at (6+2.5,1) {$X$};
    \node[blue] at (6+2.5+0.5,1) {$X$};
    \node[blue] at (6+2.5,1+0.5) {$X$};
    \node[blue] at (6+2.5-0.5,1) {$X$};
    \node[blue] at (6+2.5,1-0.5) {$X$};

    \draw[brown] (6+2.5+0.2, 1-0.2) -- (7+2.5+0.2, 1-0.2);
    \draw[brown] (6+2.5+0.2, 1-0.2) -- (6+2.5+0.2, 0-0.2);

    \node[purple] at (6+2.5+0.2+0.5, 1-0.2) {$\mathcal{C}$};
    \node[purple] at (6+2.5+0.2, 1-0.2-0.5) {$\mathcal{C}$};
\end{tikzpicture}}
\end{equation}

Lastly, we measure all the vertex qubits in the $X$ basis. If we obtain $X_{v} = 1$ on all vertices, we are left with the following operators whose simultaneous $+1$ eigenspace describes the bulk of the ground state of $\mathcal{D}(S_{3})$:
\begin{equation}
\raisebox{-0.5\height} {
    \begin{tikzpicture}
    \draw[brown] (0.5-0.8, 1.5) -- (1.5-0.8, 1.5);
    \draw[brown] (0.5-0.8, 0.5) -- (1.5-0.8, 0.5);
    \draw[brown] (0.5-0.8, 0.5) -- (0.5-0.8, 1.5);
    \draw[brown] (1.5-0.8, 0.5) -- (1.5-0.8, 1.5);

    \draw[gray] (0.5-0.3-0.8, 1.5+0.3) -- (1.5-0.3-0.8, 1.5+0.3);
    \draw[gray] (0.5-0.3-0.8, 0.5+0.3) -- (1.5-0.3-0.8, 0.5+0.3);
    \draw[gray] (0.5-0.3-0.8, 0.5+0.3) -- (0.5-0.3-0.8, 1.5+0.3);
    \draw[gray] (1.5-0.3-0.8, 0.5+0.3) -- (1.5-0.3-0.8, 1.5+0.3);

    \node[magenta, scale=0.6] at (0.5-0.3+0.5-0.8, 1.5+0.3+0.15) {$1$};
    \node[magenta, scale=0.6] at (0.5-0.3-0.15-0.8, 1.5+0.3-0.5) {$2$};
    % \node[magenta, scale=0.6] at (0.5-0.3+0.5-0.8, 1.5+0.3+0.15-1) {$3$};
    % \node[magenta, scale=0.6] at (0.5-0.3-0.15-0.8+1, 1.5+0.3-0.5) {$4$};

    \draw[magenta] (0.5-0.3+0.5-0.8, 1.5+0.3) circle[radius=1pt];
    \fill[magenta] (0.5-0.3+0.5-0.8, 1.5+0.3) circle[radius=1pt];
    \draw[magenta] (0.5-0.3-0.8, 1.5+0.3-0.5) circle[radius=1pt];
    \fill[magenta] (0.5-0.3-0.8, 1.5+0.3-0.5) circle[radius=1pt];
    % \draw[magenta] (0.5-0.3+0.5-0.8, 1.5+0.3-1) circle[radius=1pt];
    % \fill[magenta] (0.5-0.3+0.5-0.8, 1.5+0.3-1) circle[radius=1pt];
    % \draw[magenta] (0.5-0.3-0.8+1, 1.5+0.3-0.5) circle[radius=1pt];
    % \fill[magenta] (0.5-0.3-0.8+1, 1.5+0.3-0.5) circle[radius=1pt];

    \node[green, scale=0.7] at (0.5-0.8, 1) {$\mcZ^{\dagger}$};
    \node[green, scale=0.7] at (1-0.8, 1.5) {$\mcZ$};
    \node[green, scale=0.7] at (1.5-0.8, 1) {$\mcZ^{\textcolor{magenta}{Z_{1}}}$};
    \node[green, scale=0.7] at (1-0.8, 0.5) {$\mcZ^{\textcolor{magenta}{-Z_{2}}}$};

    % \node[scale=0.7] at (1.7-0.3, 1.2) {$\text{+ h.c.}$};
    % \node[scale=0.7] at (1.7-0.2+3.5, 1.2) {$\text{+ h.c.}$};

    \draw[brown] (5-2.5, 1) -- (6-2.5, 1);
    \draw[brown] (6-2.5, 1) -- (6-2.5, 2);
    \draw[brown] (6-2.5, 1) -- (7-2.5, 1);
    \draw[brown] (6-2.5, 1) -- (6-2.5, 0);

    \draw[gray] (5-2.5-0.3, 1+0.3) -- (6-2.5-0.3, 1+0.3);
    \draw[gray] (6-2.5-0.3, 1+0.3) -- (6-2.5-0.3, 2+0.3);
    \draw[gray] (6-2.5-0.3, 1+0.3) -- (7-2.5-0.3, 1+0.3);
    \draw[gray] (6-2.5-0.3, 1+0.3) -- (6-2.5-0.3, 0+0.3);

    \node[magenta, scale=0.6] at (5-2.5-0.3+0.5, 1+0.3+0.15) {$1$};
    \draw[magenta] (5-2.5-0.3+0.5, 1+0.3) circle[radius=1pt];
    \fill[magenta] (5-2.5-0.3+0.5, 1+0.3) circle[radius=1pt];

    \node[magenta, scale=0.6] at (6-2.5-0.3-0.15, 2+0.3-0.5) {$2$};
    \draw[magenta] (6-2.5-0.3, 2+0.3-0.5) circle[radius=1pt];
    \fill[magenta] (6-2.5-0.3, 2+0.3-0.5) circle[radius=1pt];

    \node[red, scale=0.7] at (5.5-2.5, 1) {$\mcX^{\textcolor{magenta}{Z_1}}$};
    \node[red, scale=0.7] at (6-2.5, 1.5) {$\mcX^{\textcolor{magenta}{Z_2}}$};
    \node[red, scale=0.7] at (6.5-2.5, 1) {$\mcX^{\dag}$};
    \node[red, scale=0.7] at (6-2.5, 0.5) {$\mcX^{\dag}$};

    \node at (0, -0.5) {$\tilde{B}_p$};
    \node at (3.5, -0.5) {$\tilde{A}_v$};
\end{tikzpicture}}
\label{S3Stabs}
\end{equation}

\begin{equation}
\raisebox{-0.5\height} {
    \begin{tikzpicture}

    \draw (5.5+0.7, 1+0.5) -- (6.5+0.7, 1+0.5);
    \draw (5.5+0.7, 1-0.5) -- (6.5+0.7, 1-0.5);
    \draw (5.5+0.7, 1-0.5) -- (5.5+0.7, 1+0.5);
    \draw (6.5+0.7, 1-0.5) -- (6.5+0.7, 1+0.5);
    \node[magenta] at (5.5+0.7, 1-0.5+0.5) {$Z$};
    \node[magenta] at (5.5+0.7+1, 1-0.5+0.5) {$Z$};
    \node[magenta] at (5.5+0.7+1-0.5, 1-0.5+0.5-0.5) {$Z$};
    \node[magenta] at (5.5+0.7+1-0.5, 1-0.5+0.5-0.5+1) {$Z$};

    \draw (5+2.5+0.7, 1) -- (6+2.5+0.7, 1);
    \draw (6+2.5+0.7, 1) -- (6+2.5+0.7, 2);
    \draw (6+2.5+0.7, 1) -- (7+2.5+0.7, 1);
    \draw (6+2.5+0.7, 1) -- (6+2.5+0.7, 0);

    \node[blue] at (6+2.5+0.5+0.7,1) {$X$};
    \node[blue] at (6+2.5+0.7,1+0.5) {$X$};
    \node[blue] at (6+2.5-0.5+0.7,1) {$X$};
    \node[blue] at (6+2.5+0.7,1-0.5) {$X$};

    \draw[brown] (6+2.5+0.2+0.7, 1-0.2) -- (7+2.5+0.2+0.7, 1-0.2);
    \draw[brown] (6+2.5+0.2+0.7, 1-0.2) -- (6+2.5+0.2+0.7, 0-0.2);

    \node[purple] at (6+2.5+0.2+0.5+0.7, 1-0.2) {$\mathcal{C}$};
    \node[purple] at (6+2.5+0.2+0.7, 1-0.2-0.5) {$\mathcal{C}$};
    \node at (6.7, -0.5) {$B_p$};
    \node at (9.3, -0.5) {$A_v$};
    
    \end{tikzpicture}}
\end{equation}

We now discuss the algebra of these operators. The $B_p, A_v$ operators are inherited from the qubit surface code stabilizers and satisfy $B_p^2 = A_v^2 = 1,$ while the $\tilde{B}_p, \tilde{A}_v$ operators are inherited from the qutrit surface stabilizers and satisfy $\tilde{B}_p^3 = \tilde{A}_v^3 = 1.$ We further note that $[\tilde{B}_p, B_{p'}] = [\tilde{A}_v, B_{p'}] = 0$ for all possible vertices $v$ and plaquettes $p, p'.$ On the other hand, $\tilde{B}_{p'}$ commutes with $\tilde{A}_v$ for all $p' \ne p,$ where $p$ is the plaquette directly above and to the left of $v.$ As for $\tilde{B}_{p}$ and $\tilde{A}_v,$ we have the relation $\tilde{B}_{p} \tilde{A}_v = \omega^{-Z_1Z_4 + Z_2Z_3} \tilde{A}_v \tilde{B}_p,$ where the $Z_{i}$ are used to label the $Z$ operators on the plaquette $p$ as according to Eq.~\eqref{S3BpOperator}. The upshot of this is that $\tilde{B}_{p}$ and $\tilde{A}_v$ \textit{do} commute in the subspace where $B_p = 1,$ whereas they do not commute otherwise. 

$A_v$ commutes with $B_p$ for all plaquettes $p,$ but it only commutes with $\tilde{B}_{p'},$ and $\tilde{A}_{v'}$ for all $p' \ne p$ and for all $v' \ne v,$ where $p$ is the plaquette directly to the right and below $v.$ As for $\tilde{B}_p$ and $\tilde{A}_v$, we have the relations $A_v \tilde{A}_v = \tilde{A}^2_v A_v$ and $A_v \tilde{B}_p = \tilde{B}^2_p A_v.$

Despite the fact that the operators stabilizing the code space are not all mutually commuting, the projectors onto the $+1$ eigenspaces of all four operators commute. As a result, the ground state of quantum double of $S_3$ is referred to as a commuting projector model. In Appendix~\ref{QuantumDouble}, we discuss the relationship between this commuting projector model and Kitaev's $S_3$ quantum double. 

We refer to the violations of the modified qutrit stabilizers $\tilde{A}_v$ and $\tilde{B}_p$ as $C$ and $F$ anyons, respecively. Similarly, we refer to the violations of the modified qubit stabilizers $A_v$ and $B_p$ as $B$ and $D$ anyons, respectively. Only the $B$ anyon is Abelian, while the remaining anyons are non-Abelian. Of course, simply stating the presence of non-Abelian anyons does not completely specify the state of the stabilizers - we must also specify the internal state of the anyon. For a more detailed account of the anyons of the $S_3$ quantum double and how they relate to the commuting projectors in Eq.~\eqref{S3Stabs}, see Appendix~\ref{S3Review} and Appendix~\ref{QuantumDouble}.

Similar to $e$ anyons created in Eq.~\eqref{EAnyonCorrection}, Abelian $B$ anyons will be created within the $S_3$ quantum double if the measurements yield an outcome of $X_{v}=-1$~\cite{tantivasadakarn_long_2021}. These can be identically corrected by applying strings of $Z$ operators extending to the right boundary. Boundary projectors are obtained by truncating the bulk stabilizers appropriately, as discussed in Appendix \ref{S3Boundaries}.

After the $S_3$ code is extended by a single column via the gauging map, we remark that a round of error correction within the $S_3$ quantum double should be performed in order to ensure that the gauging map does not propagate non-local errors. See Section \ref{SlidingErrorCorrection} for more details.

\subsection{Code Injection: Symmetry Enrichment of the Logical Operators}

We now describe the transfer of logical operators of both codes into the $S_3$ quantum double. Beginning with the qubit code, we note that the $\bs{\bar{Z}}$ logical operator is simply extended into the $S_3$ code after the application of the charge-conjugation gauging map---it spans from the leftmost boundary of the $\Z_2$ code to the rightmost boundary of the $S_3$ code as shown:
\begin{equation}
\raisebox{-0.5\height} {
    \begin{tikzpicture}[scale=0.65]
    \draw (0,0) -- (6, 0);
    \draw (0,-1) -- (6,-1);
    \draw (0,-2) -- (6,-2);
    \draw (0,-3) -- (6,-3);
    \draw (0,-4) -- (6,-4);
    \draw (6, 0) -- (10, 0);
    \draw (6, -1) -- (10, -1);
    \draw (6, -2) -- (10, -2);
    \draw (6, -3) -- (10, -3);
    \draw (6, -4) -- (10, -4);

    \node at (3, -4.5) {$\bs{\mathcal{D}(\Z_2)}$};
    \node at (8, -4.5) {$\bs{\mathcal{D}(S_3)}$};

    \draw[brown] (6+0.2, 0-0.2) -- (10+0.2, 0-0.2);
    \draw[brown] (6+0.2, -1-0.2) -- (10+0.2, -1-0.2);
    \draw[brown] (6+0.2, -2-0.2) -- (10+0.2, -2-0.2);
    \draw[brown] (6+0.2, -3-0.2) -- (10+0.2, -3-0.2);
    \draw[brown] (6+0.2, -4-0.2) -- (10+0.2, -4-0.2);

    \draw (1, 0) -- (1, -4);
    \draw (2, 0) -- (2, -4);
    \draw (3, 0) -- (3, -4);
    \draw (4, 0) -- (4, -4);
    \draw (5, 0) -- (5, -4);
    \draw (6, 0) -- (6, -4);
    \draw (7, 0) -- (7, -4);
    \draw (8, 0) -- (8, -4);
    \draw (9, 0) -- (9, -4);

    \draw[line width = 0.5mm, magenta] (0,-2) -- (10,-2);

    \node at (10.75, -2) {$\bs{\bar{Z}}$};

    \node[magenta] at (0.5, -1.75) {$Z$};
    \node[magenta] at (1.5, -1.75) {$Z$};
    \node[magenta] at (2.5, -1.75) {$Z$};
    \node[magenta] at (3.5, -1.75) {$Z$};
    \node[magenta] at (4.5, -1.75) {$Z$};
    \node[magenta] at (5.5, -1.75) {$Z$};
    \node[magenta] at (6.5, -1.75) {$Z$};
    \node[magenta] at (7.5, -1.75) {$Z$};
    \node[magenta] at (8.5, -1.75) {$Z$};
    \node[magenta] at (9.5, -1.75) {$Z$};

    \draw[brown] (7+0.2, 0-0.2) -- (7+0.2, -4-0.2);
    \draw[brown] (8+0.2, 0-0.2) -- (8+0.2, -4-0.2);
    \draw[brown] (9+0.2, 0-0.2) -- (9+0.2, -4-0.2);
    % \draw[brown] (10+0.2, 0-0.2) -- (10+0.2, -4-0.2);

\end{tikzpicture}}
\end{equation}
This corresponds to the anyon map $e \leftrightarrow B$.

As for the logical $\bs{\bar{X}}$ operator, different representatives are obtained by multiplying by qubit star stabilizers. Within the qubit code, this results in a mere topological deformation of the operator. However, within the $S_3$ code, these stabilizers are dressed by extra $\mathcal{C}$ operators. As a result, representatives of the $\bs{\bar{X}}$ operator are given by membranes of $\mathcal{C}$ operators extending to the left, as illustrated below:
\begin{equation}
\raisebox{-0.5\height} {
    \begin{tikzpicture}[scale=0.8]
    \draw (2,0) -- (6, 0);
    \draw (2,-1) -- (6,-1);
    \draw (2,-2) -- (6,-2);
    \draw (2,-3) -- (6,-3);
    \draw (2,-4) -- (6,-4);
    \draw (6, 0) -- (10, 0);
    \draw (6, -1) -- (10, -1);
    \draw (6, -2) -- (10, -2);
    \draw (6, -3) -- (10, -3);
    \draw (6, -4) -- (10, -4);

    \draw[brown] (6+0.2, 0-0.2) -- (10+0.2, 0-0.2);
    \draw[brown] (6+0.2, -1-0.2) -- (10+0.2, -1-0.2);
    \draw[brown] (6+0.2, -2-0.2) -- (10+0.2, -2-0.2);
    \draw[brown] (6+0.2, -3-0.2) -- (10+0.2, -3-0.2);
    \draw[brown] (6+0.2, -4-0.2) -- (10+0.2, -4-0.2);

    % \draw (1, 0) -- (1, -4);
    % \draw (2, 0) -- (2, -4);
    \draw (3, 0) -- (3, -4);
    \draw (4, 0) -- (4, -4);
    \draw (5, 0) -- (5, -4);
    \draw (6, 0) -- (6, -4);
    \draw (7, 0) -- (7, -4);
    \draw (8, 0) -- (8, -4);
    \draw (9, 0) -- (9, -4);

    \draw[brown] (7+0.2, 0-0.2) -- (7+0.2, -4-0.2);
    \draw[brown] (8+0.2, 0-0.2) -- (8+0.2, -4-0.2);
    \draw[brown] (9+0.2, 0-0.2) -- (9+0.2, -4-0.2);
    % \draw[brown] (10+0.2, 0-0.2) -- (10+0.2, -4-0.2);

    \node[blue] at (8.5, 0) {$X$};
    \node[blue] at (8.5, -1) {$X$};
    \node[blue] at (8.5, -2) {$X$};
    \node[blue] at (8.5, -3) {$X$};
    \node[blue] at (8.5, -4) {$X$};
    \node at (8.5, 0.5) {$\bs{\bar{X}}$};

    \draw[line width = 0.3mm, blue] (8, 0) -- (9, 0);
    \draw[line width = 0.3mm, blue] (8, -1) -- (9, -1);
    \draw[line width = 0.3mm, blue] (8, -2) -- (9, -2);
    \draw[line width = 0.3mm, blue] (8, -3) -- (9, -3);
    \draw[line width = 0.3mm, blue] (8, -4) -- (9, -4);

    \foreach \i in {6, ..., 8} {
        \foreach \j in {0, ..., -4} {
            \node[purple] at (\i+0.2+0.5, \j-0.2) {$\mathcal{C}$};
        }
    }

    \foreach \i in {6, ..., 7} {
        \foreach \j in {0, ..., -3} {
            \node[purple] at (\i+0.2+1, \j-0.2-0.5) {$\mathcal{C}$};
        }
    } 

     \node at (4, -4.7) {$\bs{\mathcal{D}(\Z_2)}$};
    \node at (8, -4.7) {$\bs{\mathcal{D}(S_3)}$};
\label{DLogical}
\end{tikzpicture}}
\end{equation} 
The operator shown creates a $D$ anyon line extending between the top and bottom smooth boundaries of the $S_3$ code written as a finite depth two-dimensional circuit. This is in contrast to other works that write the $D$ anyon ribbon operator as a linear-depth one-dimensional circuit~\cite{lyons_protocols_2025, li_domain_2024}. The correspondence between these two representations is discussed in Ref.~\cite{li_symmetry_2023}.

Now we discuss the transfer of the $\Z_3$ logical operators. After applying $U_{C\mathcal{C}},$ $\bs{\bar{\mcX}}$ maps to
\begin{equation}
\raisebox{-0.5\height} {
    \begin{tikzpicture}[scale=0.7]
    \draw (0,0) -- (6, 0);
    \draw (0,-1) -- (6,-1);
    \draw (0,-2) -- (6,-2);
    \draw (0,-3) -- (6,-3);
    \draw (0,-4) -- (6,-4);
    \draw (6, 0) -- (10, 0);
    \draw (6, -1) -- (10, -1);
    \draw (6, -2) -- (10, -2);
    \draw (6, -3) -- (10, -3);
    \draw (6, -4) -- (10, -4);

    \draw[brown] (2+0.2, 0-0.2) -- (10+0.2, 0-0.2);
    \draw[brown] (2+0.2, -1-0.2) -- (10+0.2, -1-0.2);
    \draw[brown] (2+0.2, -2-0.2) -- (10+0.2, -2-0.2);
    \draw[brown] (2+0.2, -3-0.2) -- (10+0.2, -3-0.2);
    \draw[brown] (2+0.2, -4-0.2) -- (10+0.2, -4-0.2);

    % \node[blue] at (3.5, -0.25) {$Z$};
    % \node[blue] at (3.5, -1.25) {$Z$};
    % \node[blue] at (3.5, -2.25) {$Z$};
    % \node[blue] at (3.5, -3.25) {$Z$};
    % \node[blue] at (3.5, -4.25) {$Z$};
    % \node at (3.5, 0.5) {$\boldsymbol{W_{Z}}$};

    \draw (1, 0) -- (1, -4);
    \draw (2, 0) -- (2, -4);
    \draw (3, 0) -- (3, -4);
    \draw (4, 0) -- (4, -4);
    \draw (5, 0) -- (5, -4);
    \draw (6, 0) -- (6, -4);
    \draw (7, 0) -- (7, -4);
    \draw (8, 0) -- (8, -4);
    \draw (9, 0) -- (9, -4);

    % \draw[brown] (1+0.2, 0-0.2) -- (1+0.2, -4-0.2);
    % \draw[brown] (2+0.2, 0-0.2) -- (2+0.2, -4-0.2);
    \draw[brown] (3+0.2, 0-0.2) -- (3+0.2, -4-0.2);
    \draw[brown] (4+0.2, 0-0.2) -- (4+0.2, -4-0.2);
    \draw[brown] (5+0.2, 0-0.2) -- (5+0.2, -4-0.2);
    \draw[brown] (6+0.2, 0-0.2) -- (6+0.2, -4-0.2);
    \draw[brown] (7+0.2, 0-0.2) -- (7+0.2, -4-0.2);
    \draw[brown] (8+0.2, 0-0.2) -- (8+0.2, -4-0.2);
    \draw[brown] (9+0.2, 0-0.2) -- (9+0.2, -4-0.2);

    \node[red] at (5.5+0.2, -0.25-0.2)[scale=0.7] {$\mcX^{\textcolor{magenta}{Z_{0}}}$};
    \node[red] at (5.5+0.2, -1.25-0.2) [scale=0.7]{$\mcX^{\textcolor{magenta}{Z_{1}}}$};
    \node[red] at (5.5+0.2, -2.25-0.2) [scale=0.7] {$\mcX^{\textcolor{magenta}{Z_{2}}}$};
    \node[red] at (5.5+0.2, -3.25-0.2) [scale=0.7]{$\mcX^{\textcolor{magenta}{Z_{3}}}$};
    \node[red] at (5.5+0.2, -4.25-0.2) [scale=0.7]{$\mcX^{\textcolor{magenta}{Z_{4}}}$};
    \node at (5.5, 0.5) {$\bs{\bar{\mcX}}$};

    \draw[line width = 0.5mm, red] (5+0.2, 0-0.2) -- (6+0.2, 0-0.2);
    \draw[line width = 0.5mm, red] (5+0.2, -1-0.2) -- (6+0.2, -1-0.2);
    \draw[line width = 0.5mm, red] (5+0.2, -2-0.2) -- (6+0.2, -2-0.2);
    \draw[line width = 0.5mm, red] (5+0.2, -3-0.2) -- (6+0.2, -3-0.2);
    \draw[line width = 0.5mm, red] (5+0.2, -4-0.2) -- (6+0.2, -4-0.2);

    \foreach \j in {0, ..., -4} {
        \draw[purple] (5, \j) circle[radius=1.5pt];
        \fill[purple] (5, \j) circle[radius=1.5pt];
    }

    \node[magenta, scale=0.6] at (5-0.2, 0+0.2) {$0$}; 
    \node[magenta, scale=0.6] at (5-0.2, -1+0.2) {$1$};
    \node[magenta, scale=0.6] at (5-0.2, -2+0.2) {$2$};
    \node[magenta, scale=0.6] at (5-0.2, -3+0.2) {$3$};
    \node[magenta, scale=0.6] at (5-0.2, -4+0.2) {$4$};

    \node at (1, -4.7) {$\bs{\mathcal{D}(\Z_2)}$};
    \node at (8, -4.7) {$\bs{\mathcal{D}(S_3)}$};
\end{tikzpicture}}
\end{equation}

After applying $U_{CX},$ $\bs{\bar{\mcX}}$ does not change. However, after measuring out the vertex qubits, we must replace every vertex $Z_{i}$ operator with its result after the gauging map. Because they are charged under the $\Z_2$ symmetry $U_{v},$ each $Z_{i}$ turns into a nonlocal string after gauging as shown below:
\begin{equation}
\raisebox{-0.5\height} {
    \begin{tikzpicture}[scale=0.65]
    \draw (0,0) -- (6, 0);
    \draw (0,-1) -- (6,-1);
    \draw (0,-2) -- (6,-2);
    \draw (0,-3) -- (6,-3);
    \draw (0,-4) -- (6,-4);
    \draw (6, 0) -- (10, 0);
    \draw (6, -1) -- (10, -1);
    \draw (6, -2) -- (10, -2);
    \draw (6, -3) -- (10, -3);
    \draw (6, -4) -- (10, -4);

    \draw[brown] (2+0.2, 0-0.2) -- (10+0.2, 0-0.2);
    \draw[brown] (2+0.2, -1-0.2) -- (10+0.2, -1-0.2);
    \draw[brown] (2+0.2, -2-0.2) -- (10+0.2, -2-0.2);
    \draw[brown] (2+0.2, -3-0.2) -- (10+0.2, -3-0.2);
    \draw[brown] (2+0.2, -4-0.2) -- (10+0.2, -4-0.2);

    % \node[blue] at (3.5, -0.25) {$Z$};
    % \node[blue] at (3.5, -1.25) {$Z$};
    % \node[blue] at (3.5, -2.25) {$Z$};
    % \node[blue] at (3.5, -3.25) {$Z$};
    % \node[blue] at (3.5, -4.25) {$Z$};
    % \node at (3.5, 0.5) {$\boldsymbol{W_{Z}}$};

    \draw (1, 0) -- (1, -4);
    \draw (2, 0) -- (2, -4);
    \draw (3, 0) -- (3, -4);
    \draw (4, 0) -- (4, -4);
    \draw (5, 0) -- (5, -4);
    \draw (6, 0) -- (6, -4);
    \draw (7, 0) -- (7, -4);
    \draw (8, 0) -- (8, -4);
    \draw (9, 0) -- (9, -4);

    % \draw[brown] (1+0.2, 0-0.2) -- (1+0.2, -4-0.2);
    % \draw[brown] (2+0.2, 0-0.2) -- (2+0.2, -4-0.2);
    \draw[brown] (3+0.2, 0-0.2) -- (3+0.2, -4-0.2);
    \draw[brown] (4+0.2, 0-0.2) -- (4+0.2, -4-0.2);
    \draw[brown] (5+0.2, 0-0.2) -- (5+0.2, -4-0.2);
    \draw[brown] (6+0.2, 0-0.2) -- (6+0.2, -4-0.2);
    \draw[brown] (7+0.2, 0-0.2) -- (7+0.2, -4-0.2);
    \draw[brown] (8+0.2, 0-0.2) -- (8+0.2, -4-0.2);
    \draw[brown] (9+0.2, 0-0.2) -- (9+0.2, -4-0.2);

    \node[red, scale=0.7] at (5.5+0.2, -0.25-0.2) {$\mcX^{\textcolor{magenta}{\gamma_{0}}}$};
    \node[red, scale=0.7] at (5.5+0.2, -1.25-0.2) {$\mcX^{\textcolor{magenta}{\gamma_{1}}}$};
    \node[red, scale=0.7] at (5.5+0.2, -2.25-0.2) {$\mcX^{\textcolor{magenta}{\gamma_{2}}}$};
    \node[red, scale=0.7] at (5.5+0.2, -3.25-0.2) {$\mcX^{\textcolor{magenta}{\gamma_{3}}}$};
    \node[red, scale=0.7] at (5.5+0.2, -4.25-0.2) {$\mcX^{\textcolor{magenta}{\gamma_{4}}}$};
    \node at (5.5, 0.5) {$\bs{\bar{\mcX}}$};

    \draw[line width = 0.5mm, red] (5+0.2, 0-0.2) -- (6+0.2, 0-0.2);
    \draw[line width = 0.5mm, red] (5+0.2, -1-0.2) -- (6+0.2, -1-0.2);
    \draw[line width = 0.5mm, red] (5+0.2, -2-0.2) -- (6+0.2, -2-0.2);
    \draw[line width = 0.5mm, red] (5+0.2, -3-0.2) -- (6+0.2, -3-0.2);
    \draw[line width = 0.5mm, red] (5+0.2, -4-0.2) -- (6+0.2, -4-0.2);

    \foreach \j in {0, ..., -3} {
        \draw[purple] (5, \j-0.5) circle[radius=1.5pt];
        \fill[purple] (5, \j-0.5) circle[radius=1.5pt];
    }

    \foreach \i in {5, ..., 9} {
        \node[magenta] at (\i+0.5, 0)[scale=0.75] {$Z$};
    }

    \node[magenta] at (10.5, 0) {$\gamma_{Z}$};

    \node[magenta, scale=0.6] at (5-0.2, 0-0.5) {$0$}; 
    \node[magenta, scale=0.6] at (5-0.2, -1-0.5) {$1$};
    \node[magenta, scale=0.6] at (5-0.2, -2-0.5) {$2$};
    \node[magenta, scale=0.6] at (5-0.2, -3-0.5) {$3$};

    \node at (1, -4.7) {$\bs{\mathcal{D}(\Z_2)}$};
    \node at (8, -4.7) {$\bs{\mathcal{D}(S_3)}$};
\end{tikzpicture}}
\end{equation}
Here we define $\gamma_{i} = \gamma_{Z} \prod_{j \le i} Z_{j}$ to be a nonlocal string that each $\mcX$ operator in the logical operator is conditioned on. While we depicted one particular representative of these strings, any representative that terminates on the right boundary of the $S_3$ code is a valid one.

Based on the creation of $\mathcal{D}(S_3)$ to the right of $\mathcal{D}(\Z_2)$, we remark that the nonlocal string $\gamma_Z$ must end on the right boundary of the $S_3$ code. Physically, this comes from the fact that $\gamma_Z$ is the anyon line for both the $e$ anyon of the $\Z_2$ code and the $B$ anyon for the $S_3$ code, and $B$ anyons do not condense at the lefthand domain wall between the $\mathcal{D}(S_3)$ and $\mathcal{D}(\Z_2)$. At the level of the lattice stabilizers and the gauging duality circuit, an $X$ operator on a vertex qubit can be identified exactly with a product of $Z$ operators along any path extending to the right boundary, while the extra vertex qubits placed at the $\mathcal{D}(S_3)$ and $\mathcal{D}(\Z_2)$ interface prohibit such an identification at the left boundary.

Lastly, we consider the logical $\bs{\bar{\mcZ}}$ operator of the $\Z_3$ code. Similar to the analysis of $\bs{\bar{\mcX}},$ we simply have to exponentiate the operator by nonlocal $\til{\gamma_{i}}$ strings extending to the right boundary of the $S_3$ code:
\begin{equation}
\raisebox{-0.5\height} {
    \begin{tikzpicture}[scale=0.65]
    \draw (0,0) -- (6, 0);
    \draw (0,-1) -- (6,-1);
    \draw (0,-2) -- (6,-2);
    \draw (0,-3) -- (6,-3);
    \draw (0,-4) -- (6,-4);
    \draw (6, 0) -- (10, 0);
    \draw (6, -1) -- (10, -1);
    \draw (6, -2) -- (10, -2);
    \draw (6, -3) -- (10, -3);
    \draw (6, -4) -- (10, -4);

    \draw[brown] (2+0.2, 0-0.2) -- (10+0.2, 0-0.2);
    \draw[brown] (2+0.2, -1-0.2) -- (10+0.2, -1-0.2);
    \draw[brown] (2+0.2, -2-0.2) -- (10+0.2, -2-0.2);
    \draw[brown] (2+0.2, -3-0.2) -- (10+0.2, -3-0.2);
    \draw[brown] (2+0.2, -4-0.2) -- (10+0.2, -4-0.2);

    % \node[blue] at (3.5, -0.25) {$Z$};
    % \node[blue] at (3.5, -1.25) {$Z$};
    % \node[blue] at (3.5, -2.25) {$Z$};
    % \node[blue] at (3.5, -3.25) {$Z$};
    % \node[blue] at (3.5, -4.25) {$Z$};
    % \node at (3.5, 0.5) {$\boldsymbol{W_{Z}}$};

    \draw (1, 0) -- (1, -4);
    \draw (2, 0) -- (2, -4);
    \draw (3, 0) -- (3, -4);
    \draw (4, 0) -- (4, -4);
    \draw (5, 0) -- (5, -4);
    \draw (6, 0) -- (6, -4);
    \draw (7, 0) -- (7, -4);
    \draw (8, 0) -- (8, -4);
    \draw (9, 0) -- (9, -4);

    \draw[line width = 0.5mm, green] (2+0.2,-2-0.2) -- (10+0.2,-2-0.2);

    \foreach \i in {0,...,7} {
        \node[green] at (7-\i+2+0.5+0.2, -1.75-0.2-0.5) [scale=0.7]{$\mcZ^{\textcolor{magenta}{\til{\gamma_{\i}}}}$};
    }
    
    % \node[green] at (2.5+0.2, -1.75-0.2-0.5) {$\mcZ$};
    % \node[green] at (3.5+0.2, -1.75-0.2-0.5) {$\mcZ$};
    % \node[green] at (4.5+0.2, -1.75-0.2-0.5) {$\mcZ$};
    % \node[green] at (5.5+0.2, -1.75-0.2-0.5) {$\mcZ$};
    % \node[green] at (6.5+0.2, -1.75-0.2-0.5) {$\mcZ$};
    % \node[green] at (7.5+0.2, -1.75-0.2-0.5) {$\mcZ$};
    % \node[green] at (8.5+0.2, -1.75-0.2-0.5) {$\mcZ$};
    % \node[green] at (9.5+0.2, -1.75-0.2-0.5) {$\mcZ$};

    \draw[brown] (3+0.2, 0-0.2) -- (3+0.2, -4-0.2);
    \draw[brown] (4+0.2, 0-0.2) -- (4+0.2, -4-0.2);
    \draw[brown] (5+0.2, 0-0.2) -- (5+0.2, -4-0.2);
    \draw[brown] (6+0.2, 0-0.2) -- (6+0.2, -4-0.2);
    \draw[brown] (7+0.2, 0-0.2) -- (7+0.2, -4-0.2);
    \draw[brown] (8+0.2, 0-0.2) -- (8+0.2, -4-0.2);
    \draw[brown] (9+0.2, 0-0.2) -- (9+0.2, -4-0.2);
    % \draw[brown] (10+0.2, 0-0.2) -- (10+0.2, -4-0.2);

    \foreach \i in {7,..., 0} {
        \draw[purple] (\i+2+0.5, -2) circle[radius=1.5pt];
        \fill[purple] (\i+2+0.5, -2) circle[radius=1.5pt];
        \node[magenta, scale=0.6] at (7-\i+2+0.5, -1.8) {$\i$}; 
    }

    \node at (1, -4.7) {$\bs{\mathcal{D}(\Z_2)}$};
    \node at (8, -4.7) {$\bs{\mathcal{D}(S_3)}$};

    \node at (10.75, -2) {$\bs{\bar{\mcZ}}$};
\label{CAnyonLine}
\end{tikzpicture}}
\end{equation}
Here we use $\til{\gamma_{i}} = \prod_{j \le i} Z_{j}$ to denote a product of $Z$ operators extending to the right of a given vertex. We remark that this expression for the logical operator is exactly the form of the $C$ anyon ribbon operator presented in Ref. \cite{li_domain_2024}.

Within the $S_3$ code, the new representatives of the $\bs{\bar{X}}$ and $\bs{\bar{Z}}$ logical operators preserve the commutation relation $\bs{\bar{Z}} \bs{\bar{X}} = -\bs{\bar{X}} \bs{\bar{Z}}$. To verify the commutation relation between $\bs{\bar{\mcX}}$ and $\bs{\bar{\mcZ}}$, observe that they fail to commute only at the edge $e$ where they intersect. At this edge, both logical operators are conditioned on the same string $\gamma_{e}$. Since $\mcZ^{\gamma_{e}} \mcX^{\gamma_{e}} = \omega \mcX^{\gamma_{e}} \mcZ^{\gamma_{e}}$, it follows that the commutation relation $\bs{\bar{\mcZ}} \bs{\bar{\mcX}} = \omega \bs{\bar{\mcX}} \bs{\bar{\mcZ}}$ is preserved.

\begin{figure}
    \centering
    \includegraphics[width=\linewidth]{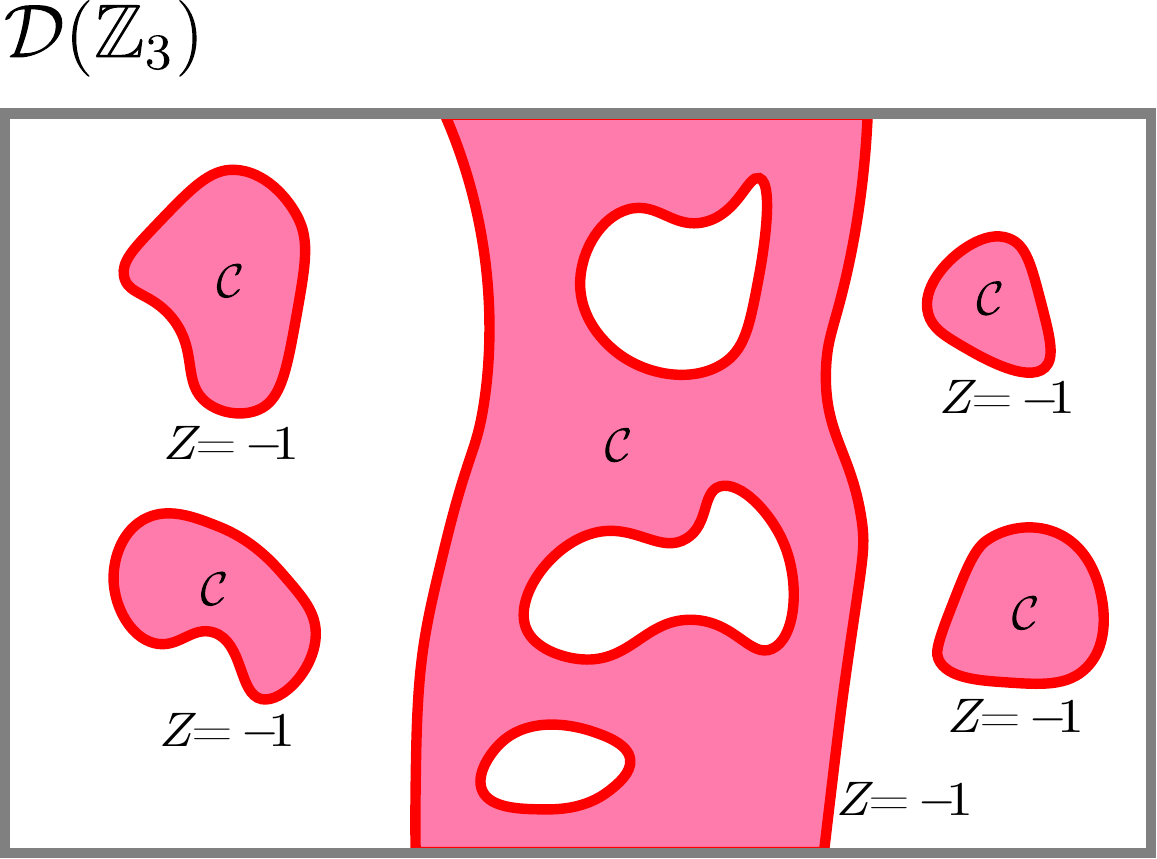}
    \caption{We schematically show the states in the superposition of the $\mathcal{D}(S_3)$ wavefunction in the $Z$ basis. Each state is a $\mathcal{D}(\Z_3)$ wavefunction with a charge-conjugation domain wall configuration applied to the state. After the qubits of the $S_3$ code are measured out in the $Z$ basis, one such state is obtained with the $Z = -1$ measurement outcomes bounding the domain walls.}
    \label{CorrectionFigure}
\end{figure}

\subsection{Code Ejection: $Z$ Measurement and $\mathcal{C}$ Feedforward}

To facilitate the code ejection procedure, let us first make some comments on the nature of the qubit part of the $S_3$ wavefunction. Viewing it in the $Z$ basis, we see that $Z = -1$ configurations must occur in closed loop configurations on the qubit dual lattice due to the qubit plaquette stabilizer $B_p$. Two loop configurations that are related by a contractible loop $\partial R$ can be related by multiplying by $A_v$ stabilizers within $R$; this has the effect of applying a membrane-like charge conjugation operation to region $R$ on the qutrits. We comment that if a contractible loop crosses over from the $\Z_2$ code to the $S_3$ code that charge conjugations act only in the part of the loop that is within $S_3.$ If there is a non-contractible line of $Z = -1$ qubits, it signifies a non-contractible charge conjugation domain wall, so to the right of this domain wall, charge conjugations are applied to the \emph{exterior} of the loops instead of the interior. In essence, within $\mathcal{D}(S_3),$ $Z = -1$ loops signify charge conjugation domain walls of the $\Z_3$ code, as shown in Figure~\ref{CorrectionFigure}. 

We now eliminate the qubits of the $\Z_2$ code on the left by performing measurements in the $Z$ basis in a column-by-column fashion. This process can be interpreted as ungauging the $\Z_2$ charge conjugation symmetry since we map the $\Z_2$ part of the $S_3$ code to a product state. As we measure from left to right, we will obtain a measurement record of $Z = -1$ outcomes that is a partial loop configuration. We remark that in the intermediate stages of the measurement, we produce some number of $m$ anyons at the left boundary of the $\Z_2$ code, aligning with the partial loop configuration of the measurement record. While we are within the $\Z_2$ code, these anyons can be actively eliminated in a manner that closes off the partial loop configuration in a contractible way (up to some logical correction if a non-contractible loop arises in the measurement record)---this was discussed initially in Sec.~\ref{CodeConventions}.

After the left $\Z_2$ code is completely eliminated, remaining $m$ anyons on the left boundary become $D$ anyons within the $S_3$ code, associated to violations of the $B_p$ operators. Notice that the qubit $Z$ operator commutes with $\tilde{A}_v$ and $\tilde{B}_p$, so no other anyons will be created. This anyon configuration is depicted below, where the dashed links denote qubits that have been measured out:
\begin{equation}
\raisebox{-0.5\height} {
    \begin{tikzpicture}[scale=0.85]
    \draw[dashed] (2,0) -- (6, 0);
    \draw[dashed] (2,-1) -- (6,-1);
    \draw[dashed] (2,-2) -- (6,-2);
    \draw[dashed] (2,-3) -- (6,-3);
    \draw[dashed] (2,-4) -- (6,-4);
    \draw (6, 0) -- (10, 0);
    \draw (6, -1) -- (10, -1);
    \draw (6, -2) -- (10, -2);
    \draw (6, -3) -- (10, -3);
    \draw (6, -4) -- (10, -4);

    \draw[brown] (6+0.2, 0-0.2) -- (10+0.2, 0-0.2);
    \draw[brown] (6+0.2, -1-0.2) -- (10+0.2, -1-0.2);
    \draw[brown] (6+0.2, -2-0.2) -- (10+0.2, -2-0.2);
    \draw[brown] (6+0.2, -3-0.2) -- (10+0.2, -3-0.2);
    \draw[brown] (6+0.2, -4-0.2) -- (10+0.2, -4-0.2);

    % \draw (1, 0) -- (1, -4);
    % \draw (2, 0) -- (2, -4);
    \draw[dashed] (3, 0) -- (3, -4);
    \draw[dashed] (4, 0) -- (4, -4);
    \draw[dashed] (5, 0) -- (5, -4);
    \draw[dashed] (6, 0) -- (6, -4);
    \draw (7, 0) -- (7, -4);
    \draw (8, 0) -- (8, -4);
    \draw (9, 0) -- (9, -4);

    \draw[brown] (7+0.2, 0-0.2) -- (7+0.2, -4-0.2);
    \draw[brown] (8+0.2, 0-0.2) -- (8+0.2, -4-0.2);
    \draw[brown] (9+0.2, 0-0.2) -- (9+0.2, -4-0.2);

    \draw[magenta, line width = 0.5mm] (6+1-2, 0) -- (7+1-2, 0);
    \draw[magenta, line width = 0.5mm] (6+1-2, -1) -- (7+1-2, -1);
    \draw[magenta, line width = 0.5mm] (7-2, -1) -- (7-2, -2);
    \draw[magenta, line width = 0.5mm] (6-2, -1) -- (6-2, -2);
    \draw[magenta, line width = 0.5mm] (5-2, -1) -- (5-2, -2);
    \draw[magenta, line width = 0.5mm] (5-2, -1) -- (4-2, -1);
    \draw[magenta, line width = 0.5mm] (5-2, 0) -- (4-2, 0);

    \draw[magenta, line width = 0.5mm] (4, -4) -- (5, -4);
    \draw[magenta, line width = 0.5mm] (5, -3) -- (5, -4);
    \draw[magenta, line width = 0.5mm] (6, -3) -- (6, -4);

    \node[blue] at (6.5, -3.5) {$D$};

    \node[magenta] at (4.5, 0.5) {$Z = -1$};

     \node at (4, -4.7) {$\bs{\mathcal{D}(\Z_2)}$};
    \node at (8, -4.7) {$\bs{\mathcal{D}(S_3)}$};
\end{tikzpicture}}
\end{equation}

While it naively seems necessary to apply active correction to the $D$ anyons within the $S_3$ code involving the $D$ anyon ribbon operators, this step is not necessary, given that we have access to the measurement record of qubits in the $Z$ basis. The main idea is that we postpone correction of the $D$ anyons until they turn into $m$ anyons on the representative of the $\Z_2$ code on the right. We create the representative using the same $\Z_2$ code extension procedure:
\begin{equation}
\raisebox{-0.5\height} {
    \begin{tikzpicture}[scale=0.7]
    \draw (0,0) -- (6, 0);
    \draw (0,-1) -- (6,-1);
    \draw (0,-2) -- (6,-2);
    \draw (0,-3) -- (6,-3);
    \draw (0,-4) -- (6,-4);
    \draw (6, 0) -- (10, 0);
    \draw (6, -1) -- (10, -1);
    \draw (6, -2) -- (10, -2);
    \draw (6, -3) -- (10, -3);
    \draw (6, -4) -- (10, -4);

    \draw[brown] (0+0.2, 0-0.2) -- (6+0.2, 0-0.2);
    \draw[brown] (0+0.2, -1-0.2) -- (6+0.2, -1-0.2);
    \draw[brown] (0+0.2, -2-0.2) -- (6+0.2, -2-0.2);
    \draw[brown] (0+0.2, -3-0.2) -- (6+0.2, -3-0.2);
    \draw[brown] (0+0.2, -4-0.2) -- (6+0.2, -4-0.2);

    \draw[dashed] (0, 0) -- (0, -4);
    \draw (1, 0) -- (1, -4);
    \draw (2, 0) -- (2, -4);
    \draw (3, 0) -- (3, -4);
    \draw (4, 0) -- (4, -4);
    \draw (5, 0) -- (5, -4);
    \draw (6, 0) -- (6, -4);
    \draw (7, 0) -- (7, -4);
    \draw (8, 0) -- (8, -4);
    \draw (9, 0) -- (9, -4);

    \draw[magenta, line width = 0.5mm] (0,0) -- (0,-1);
    \draw[magenta, line width = 0.5mm] (0,-2) -- (0,-3);

    \node[blue] at (0.5, -0.5) {$D$};
    \node[blue] at (0.5, -2.5) {$D$};

    \draw[brown] (1+0.2, 0-0.2) -- (1+0.2, -4-0.2);
    \draw[brown] (2+0.2, 0-0.2) -- (2+0.2, -4-0.2);
    \draw[brown] (3+0.2, 0-0.2) -- (3+0.2, -4-0.2);
    \draw[brown] (4+0.2, 0-0.2) -- (4+0.2, -4-0.2);
    \draw[brown] (5+0.2, 0-0.2) -- (5+0.2, -4-0.2);
    % \draw[brown] (6+0.2, 0-0.2) -- (6+0.2, -4-0.2);
    
    \node at (8, -4.7) {$\bs{\mathcal{D}(\Z_2)}$};
    \node at (3, -4.7) {$\bs{\mathcal{D}(S_3)}$};
\end{tikzpicture}}
\end{equation}

Measuring out qubits of the $S_3$ code in the $Z$ basis from left to right, we will obtain one particular loop configuration of $Z = -1$ and an insertion of charge conjugation domain walls on the $\Z_3$ code that align with it, as shown in Fig.~\ref{CorrectionFigure}. Such charge conjugation operations need to be corrected by applying the appropriate $\mathcal{C}$ feedfoward operation, which will return the $\Z_3$ stabilizers back to their original form.

 When the $Z = -1$ domain walls are contractible, we apply $\mathcal{C}$ to its interior (if the domain wall crosses the region of both $S_3$ and $\Z_2,$ then we only apply $\mathcal{C}$ to the part of the interior that is in $S_3$). If the measurement record indicates a non-contractible line, then it suffices to apply the logical operator $\bs{\bar{X}}$ to the state, along with a charge conjugation membrane on all the qutrit links to the right of the applied logical. Such a correction operation is shown below:
 
\begin{equation}
\raisebox{-0.5\height} {
    \begin{tikzpicture}[scale=0.7]
    \draw[dashed] (0,0) -- (2, 0);
    \draw[dashed] (0,-1) -- (2,-1);
    \draw[dashed] (0,-2) -- (2,-2);
    \draw[dashed] (0,-3) -- (2,-3);
    \draw[dashed] (0,-4) -- (2,-4);
    \draw (2, 0) -- (10, 0);
    \draw (2, -1) -- (10, -1);
    \draw (2, -2) -- (10, -2);
    \draw (2, -3) -- (10, -3);
    \draw (2, -4) -- (10, -4);

    \draw[brown] (0+0.2, 0-0.2) -- (6+0.2, 0-0.2);
    \draw[brown] (0+0.2, -1-0.2) -- (6+0.2, -1-0.2);
    \draw[brown] (0+0.2, -2-0.2) -- (6+0.2, -2-0.2);
    \draw[brown] (0+0.2, -3-0.2) -- (6+0.2, -3-0.2);
    \draw[brown] (0+0.2, -4-0.2) -- (6+0.2, -4-0.2);

    \draw[dashed] (0, 0) -- (0, -4);
    \draw[dashed] (1, 0) -- (1, -4);
    \draw[dashed] (2, 0) -- (2, -4);
    \draw (3, 0) -- (3, -4);
    \draw (4, 0) -- (4, -4);
    \draw (5, 0) -- (5, -4);
    \draw (6, 0) -- (6, -4);
    \draw (7, 0) -- (7, -4);
    \draw (8, 0) -- (8, -4);
    \draw (9, 0) -- (9, -4);

    \draw[magenta, line width = 0.5mm] (0,-4) -- (1,-4);
    \draw[magenta, line width = 0.5mm] (0,-3) -- (1,-3);
    \draw[magenta, line width = 0.5mm] (1,-3) -- (1,-2);
    \draw[magenta, line width = 0.5mm] (2,-2) -- (1,-2);
    \draw[magenta, line width = 0.5mm] (2,-1) -- (1,-1);
    \draw[magenta, line width = 0.5mm] (2,0) -- (1,0);

    \node[magenta] at (1.5,0.3) {$Z=-1$};

    \draw[brown] (1+0.2, 0-0.2) -- (1+0.2, -4-0.2);
    \draw[brown] (2+0.2, 0-0.2) -- (2+0.2, -4-0.2);
    \draw[brown] (3+0.2, 0-0.2) -- (3+0.2, -4-0.2);
    \draw[brown] (4+0.2, 0-0.2) -- (4+0.2, -4-0.2);
    \draw[brown] (5+0.2, 0-0.2) -- (5+0.2, -4-0.2);
    % \draw[brown] (6+0.2, 0-0.2) -- (6+0.2, -4-0.2);
    
    \node at (8, -4.7) {$\bs{\mathcal{D}(\Z_2)}$};
    \node at (3, -4.7) {$\bs{\mathcal{D}(S_3)}$};

    \foreach \i in {2, ..., 5} {
        \foreach \j in {0, ..., -4} {
            \node[purple] at (\i+0.2+0.5, \j-0.2) {$\mathcal{C}$};
        }
    }

    \foreach \i in {1, ..., 4} {
        \foreach \j in {0, ..., -3} {
            \node[purple] at (\i+0.2+1, \j-0.2-0.5) {$\mathcal{C}$};
        }
    } 

    \node[purple] at (1+0.2+0.5, -3-0.2) {$\mathcal{C}$};
    \node[purple] at (1+0.2+0.5, -4-0.2) {$\mathcal{C}$};
    \node[purple] at (1+0.2, -3-0.2-0.5) {$\mathcal{C}$};

    \node[blue] at (8.5, -0.25) {$X$};
    \node[blue] at (8.5, -1.25) {$X$};
    \node[blue] at (8.5, -2.25) {$X$};
    \node[blue] at (8.5, -3.25) {$X$};
    \node[blue] at (8.5, -4.25) {$X$};

    \draw[line width = 0.5mm, blue] (8, 0) -- (9, 0);
    \draw[line width = 0.5mm, blue] (8, -1) -- (9, -1);
    \draw[line width = 0.5mm, blue] (8, -2) -- (9, -2);
    \draw[line width = 0.5mm, blue] (8, -3) -- (9, -3);
    \draw[line width = 0.5mm, blue] (8, -4) -- (9, -4);
    
\end{tikzpicture}}
\end{equation}

This operation accounts for the modified $\Z_3$ stabilizers and the modification of the $\Z_2$ logical state due to the non-contractible line. It can be performed online while the $\Z_2$ qubits are being measured, or it can equivalently be performed at the end once all the qubits within $S_3$ are eliminated given access to the full measurement record.

Finally, once all the qubits of the $S_3$ code are eliminated, we find that remaining $D$ anyon lines that were not closed off become $m$ anyons in the right $\Z_2$ code, which can be corrected in a contractible manner as shown:
\begin{equation}
\raisebox{-0.5\height} {
    \begin{tikzpicture}[scale=0.7]
    \draw[dashed] (0,0) -- (6, 0);
    \draw[dashed] (0,-1) -- (6,-1);
    \draw[dashed] (0,-2) -- (6,-2);
    \draw[dashed] (0,-3) -- (6,-3);
    \draw[dashed] (0,-4) -- (6,-4);
    \draw (6, 0) -- (10, 0);
    \draw (6, -1) -- (10, -1);
    \draw (6, -2) -- (10, -2);
    \draw (6, -3) -- (10, -3);
    \draw (6, -4) -- (10, -4);

    \draw[brown] (0+0.2, 0-0.2) -- (6+0.2, 0-0.2);
    \draw[brown] (0+0.2, -1-0.2) -- (6+0.2, -1-0.2);
    \draw[brown] (0+0.2, -2-0.2) -- (6+0.2, -2-0.2);
    \draw[brown] (0+0.2, -3-0.2) -- (6+0.2, -3-0.2);
    \draw[brown] (0+0.2, -4-0.2) -- (6+0.2, -4-0.2);

    % \node[blue] at (3.5, -0.25) {$Z$};
    % \node[blue] at (3.5, -1.25) {$Z$};
    % \node[blue] at (3.5, -2.25) {$Z$};
    % \node[blue] at (3.5, -3.25) {$Z$};
    % \node[blue] at (3.5, -4.25) {$Z$};
    % \node at (3.5, 0.5) {$\boldsymbol{W_{Z}}$};

    \draw[dashed] (1, 0) -- (1, -4);
    \draw[dashed] (2, 0) -- (2, -4);
    \draw[dashed] (3, 0) -- (3, -4);
    \draw[dashed] (4, 0) -- (4, -4);
    \draw[dashed] (5, 0) -- (5, -4);
    \draw[dashed] (6, 0) -- (6, -4);
    \draw (7, 0) -- (7, -4);
    \draw (8, 0) -- (8, -4);
    \draw (9, 0) -- (9, -4);

    \draw[brown] (1+0.2, 0-0.2) -- (1+0.2, -4-0.2);
    \draw[brown] (2+0.2, 0-0.2) -- (2+0.2, -4-0.2);
    \draw[brown] (3+0.2, 0-0.2) -- (3+0.2, -4-0.2);
    \draw[brown] (4+0.2, 0-0.2) -- (4+0.2, -4-0.2);
    \draw[brown] (5+0.2, 0-0.2) -- (5+0.2, -4-0.2);
    % \draw[brown] (6+0.2, 0-0.2) -- (6+0.2, -4-0.2);

    \draw[magenta, line width = 0.5mm] (0+3,-4) -- (1+3,-4);
    \draw[magenta, line width = 0.5mm] (0+3,-3) -- (1+3,-3);
    \draw[magenta, line width = 0.5mm] (1+3,-3) -- (1+3,-2);
    \draw[magenta, line width = 0.5mm] (2+3,-2) -- (1+3,-2);
    \draw[magenta, line width = 0.5mm] (2+3,-2) -- (2+3,-1);
    \draw[magenta, line width = 0.5mm] (3+3,-2) -- (3+3,-1);

    \node[red] at (3+3+0.5,-1.5) {$m$};
    \node[blue] at (3+3+0.5,-2) {$X$};
    \node[blue] at (3+3+0.5,-3) {$X$};
    \node[blue] at (3+3+0.5,-4) {$X$};

    \node at (8, -4.7) {$\bs{\mathcal{D}(\Z_2)}$};
    \node at (3, -4.7) {$\bs{\mathcal{D}(S_3)}$};

\end{tikzpicture}}
\end{equation}
Throughout the course of our protocol, the $D$ anyons created from measurement will always be confined to the left boundary of the $S_3$ code. Any anyons present in the bulk of the $S_3$ code will be from outside errors - as a result, the error correction protocol discussed in Section ~\ref{SlidingErrorCorrection} will solely act in the bulk of the code and will not remove the anyons confined to the boundary. This error correction step should be performed after the ejection of each column.

\subsection{Code Ejection: Transformed Logical Operators}

Once code ejection is complete, we are left with the $\Z_2$ code to the right of the $\Z_3$ code, and a controlled-charge conjugation $C\mathcal{C}$ applied between them. We now show that the logical operators of both codes transform into their expected result under the $C\mathcal{C}$ gate.

To understand the transfer of the logical $\bs{\bar{Z}}$ operator, we note that the $Z = +1$ measurement outcomes in the code ejection procedure allow us to simply truncate the logical operator. As for contractible $Z = -1$ loops, we notice that they will always intersect the representative of $\bs{\bar{Z}}$ an even number of times, allowing us to similarly truncate the operator. As for non-contractible lines of $Z = -1$ observed in the measurement record, this results in a flip of $\bs{\bar{Z}}$ to $-\bs{\bar{Z}},$ or equivalently a swap of the logical $\bar{\ket{0}}$ and the logical $\bar{\ket{1}}$ state. However, in our code ejection procedure, everytime such an event occurs we apply the $\bs{\bar{X}}$ operator. Thus, in all cases, the $\bs{\bar{Z}}$ operator of the left $\mathcal{D}(\Z_2)$ will turn into the $\bs{\bar{Z}}$ of the right $\mathcal{D}(\Z_2)$ after the protocol is performed.

% \VM{Maybe this following discussion should be a separate sub-section? There should probably be elements of this discussion with less detail in the symmetry enrichment and code ejection sections to motivate those steps.}
As discussed above, the logical $\bs{\bar{X}}$ operator gets dressed by a membrane of $\mathcal{C}$ as shown in Eq.~\eqref{DLogical} as it passes through $\mathcal{D}(S_3)$. Thus, once the protocol is complete, we obtain the transformation $\bs{\bar{X}} \rightarrow \bs{\bar{X}}\bs{\bar{\mathcal{C}}},$ where $\bs{\bar{\mathcal{C}}}$ is a global charge conjugation operation on the $\Z_3$ code.

To understand the logical action on the $\bs{\bar{\mcZ}}$ and $\bs{\bar{\mcX}}$ operators, we must understand the nonlocal string that these operators are conditioned on and how it transfers into the representative of the $\Z_2$ code on the right. In particular, the nonlocal string will extend into the right $\Z_2$ code, as shown:
\begin{equation}
\raisebox{-0.5\height} {
    \begin{tikzpicture}[scale=0.65]
    \draw (0,0) -- (6, 0);
    \draw (0,-1) -- (6,-1);
    \draw (0,-2) -- (6,-2);
    \draw (0,-3) -- (6,-3);
    \draw (0,-4) -- (6,-4);
    \draw (6, 0) -- (10, 0);
    \draw (6, -1) -- (10, -1);
    \draw (6, -2) -- (10, -2);
    \draw (6, -3) -- (10, -3);
    \draw (6, -4) -- (10, -4);

    \draw[brown] (0+0.2, 0-0.2) -- (6+0.2, 0-0.2);
    \draw[brown] (0+0.2, -1-0.2) -- (6+0.2, -1-0.2);
    \draw[brown] (0+0.2, -2-0.2) -- (6+0.2, -2-0.2);
    \draw[brown] (0+0.2, -3-0.2) -- (6+0.2, -3-0.2);
    \draw[brown] (0+0.2, -4-0.2) -- (6+0.2, -4-0.2);

    % \node[blue] at (3.5, -0.25) {$Z$};
    % \node[blue] at (3.5, -1.25) {$Z$};
    % \node[blue] at (3.5, -2.25) {$Z$};
    % \node[blue] at (3.5, -3.25) {$Z$};
    % \node[blue] at (3.5, -4.25) {$Z$};
    % \node at (3.5, 0.5) {$\boldsymbol{W_{Z}}$};

    \draw (1, 0) -- (1, -4);
    \draw (2, 0) -- (2, -4);
    \draw (3, 0) -- (3, -4);
    \draw (4, 0) -- (4, -4);
    \draw (5, 0) -- (5, -4);
    \draw (6, 0) -- (6, -4);
    \draw (7, 0) -- (7, -4);
    \draw (8, 0) -- (8, -4);
    \draw (9, 0) -- (9, -4);

    \draw[brown] (1+0.2, 0-0.2) -- (1+0.2, -4-0.2);
    \draw[brown] (2+0.2, 0-0.2) -- (2+0.2, -4-0.2);
    \draw[brown] (3+0.2, 0-0.2) -- (3+0.2, -4-0.2);
    \draw[brown] (4+0.2, 0-0.2) -- (4+0.2, -4-0.2);
    \draw[brown] (5+0.2, 0-0.2) -- (5+0.2, -4-0.2);
    % \draw[brown] (6+0.2, 0-0.2) -- (6+0.2, -4-0.2);

    \node[red, scale=0.7] at (4.5+0.2, -0.25-0.2) {$\mcX^{\textcolor{magenta}{\gamma_{0}}}$};
    \node[red, scale=0.7] at (4.5+0.2, -1.25-0.2) {$\mcX^{\textcolor{magenta}{\gamma_{1}}}$};
    \node[red, scale=0.7] at (4.5+0.2, -2.25-0.2) {$\mcX^{\textcolor{magenta}{\gamma_{2}}}$};
    \node[red, scale=0.7] at (4.5+0.2, -3.25-0.2) {$\mcX^{\textcolor{magenta}{\gamma_{3}}}$};
    \node[red, scale=0.7] at (4.5+0.2, -4.25-0.2) {$\mcX^{\textcolor{magenta}{\gamma_{4}}}$};
    \node at (4.5, 0.5) {$\bs{\bar{\mcX}}$};

    \draw[line width = 0.5mm, red] (4+0.2, 0-0.2) -- (5+0.2, 0-0.2);
    \draw[line width = 0.5mm, red] (4+0.2, -1-0.2) -- (5+0.2, -1-0.2);
    \draw[line width = 0.5mm, red] (4+0.2, -2-0.2) -- (5+0.2, -2-0.2);
    \draw[line width = 0.5mm, red] (4+0.2, -3-0.2) -- (5+0.2, -3-0.2);
    \draw[line width = 0.5mm, red] (4+0.2, -4-0.2) -- (5+0.2, -4-0.2);

    \foreach \j in {0, ..., -3} {
        \draw[purple] (4, \j-0.5) circle[radius=1.5pt];
        \fill[purple] (4, \j-0.5) circle[radius=1.5pt];
    }

    \foreach \i in {4, ..., 9} {
        \node[magenta] at (\i+0.5, 0)[scale=0.8] {$Z$};
    }

    \node[magenta] at (10.5, 0) {$\gamma_{Z}$};

    \node[magenta, scale=0.6] at (4-0.2, 0-0.5) {${0}$}; 
    \node[magenta, scale=0.6] at (4-0.2, -1-0.5) {${1}$};
    \node[magenta, scale=0.6] at (4-0.2, -2-0.5) {${2}$};
    \node[magenta, scale=0.6] at (4-0.2, -3-0.5) {${3}$};

    \node at (8, -4.7) {$\bs{\mathcal{D}(\Z_2)}$};
    \node at (3, -4.7) {$\bs{\mathcal{D}(S_3)}$};

\label{NonlocalString}
\end{tikzpicture}}
\end{equation}

Recall that each $\mcX$ operator in the logical is conditioned on a string $\gamma_{i} = \gamma_{Z} \prod_{j < i} Z_{j}.$ Eq.~\eqref{NonlocalString} shows one possible representative $\gamma_Z.$ 

In the final step of the protocol, qubits of the $S_3$ code are measured out in the $Z$ basis. Subsequently, the $Z$ operators of the $\gamma_{Z}$ string that are in $\mathcal{D}(S_3)$ will get replaced by their results from measurement, allowing us to truncate each $\gamma_i$ to an operator living solely in the $\Z_2$ code.

Recall that that $Z = -1$ outcomes will come in some loop configuration once all the $m$ anyons remaining in the $\Z_2$ code are closed off into contractible loops according to the measurement record. All contractible loops of $Z = -1$ measurement outcomes intersect each $\gamma_{i}$ twice, so we are free to shorten the string without changing its overall sign. As for non-contractible lines, recall from our protocol above that whenever a non-contractible line is observed in the measurement record we apply a $\bs{\bar{X}}$ correction operation, so these also keep the overall sign of $\gamma_{i}$ invariant.

Thus, once the qubits of the $S_3$ code are fully measured out, we find that the $\bs{\bar{\mcX}}$ operator is now conditioned on a truncated nonlocal string that is precisely the logical $\bs{\bar{Z}}$ operator of the right $\Z_2$ code:

\begin{equation}
\raisebox{-0.5\height} {
    \begin{tikzpicture}[scale=0.7]
    % \draw (0,0) -- (6, 0);
    % \draw (0,-1) -- (6,-1);
    % \draw (0,-2) -- (6,-2);
    % \draw (0,-3) -- (6,-3);
    % \draw (0,-4) -- (6,-4);
    \draw (6+1+0.2, 0-0.2) -- (10+1+0.2, 0-0.2);
    \draw (6+1+0.2, -1-0.2) -- (10+1+0.2, -1-0.2);
    \draw (6+1+0.2, -2-0.2) -- (10+1+0.2, -2-0.2);
    \draw (6+1+0.2, -3-0.2) -- (10+1+0.2, -3-0.2);
    \draw (6+1+0.2, -4-0.2) -- (10+1+0.2, -4-0.2);

    \draw[brown] (0+0.2, 0-0.2) -- (6+0.2, 0-0.2);
    \draw[brown] (0+0.2, -1-0.2) -- (6+0.2, -1-0.2);
    \draw[brown] (0+0.2, -2-0.2) -- (6+0.2, -2-0.2);
    \draw[brown] (0+0.2, -3-0.2) -- (6+0.2, -3-0.2);
    \draw[brown] (0+0.2, -4-0.2) -- (6+0.2, -4-0.2);

    % \node[blue] at (3.5, -0.25) {$Z$};
    % \node[blue] at (3.5, -1.25) {$Z$};
    % \node[blue] at (3.5, -2.25) {$Z$};
    % \node[blue] at (3.5, -3.25) {$Z$};
    % \node[blue] at (3.5, -4.25) {$Z$};
    % \node at (3.5, 0.5) {$\boldsymbol{W_{Z}}$};

    % \draw (1, 0) -- (1, -4);
    % \draw (2, 0) -- (2, -4);
    % \draw (3, 0) -- (3, -4);
    % \draw (4, 0) -- (4, -4);
    % \draw (5, 0) -- (5, -4);
    % \draw (6, 0) -- (6, -4);
    \draw (7+1+0.2, 0-0.2) -- (7+1+0.2, -4-0.2);
    \draw (8+1+0.2, 0-0.2) -- (8+1+0.2, -4-0.2);
    \draw (9+1+0.2, 0-0.2) -- (9+1+0.2, -4-0.2);

    \draw[brown] (1+0.2, 0-0.2) -- (1+0.2, -4-0.2);
    \draw[brown] (2+0.2, 0-0.2) -- (2+0.2, -4-0.2);
    \draw[brown] (3+0.2, 0-0.2) -- (3+0.2, -4-0.2);
    \draw[brown] (4+0.2, 0-0.2) -- (4+0.2, -4-0.2);
    \draw[brown] (5+0.2, 0-0.2) -- (5+0.2, -4-0.2);
    % \draw[brown] (6+0.2, 0-0.2) -- (6+0.2, -4-0.2);

    \node[red, scale=0.7] at (3.5+0.2, -0.25-0.2) {$\mcX^{\textcolor{magenta}{\bs{\bar{Z}}}}$};
    \node[red, scale=0.7] at (3.5+0.2, -1.25-0.2) {$\mcX^{\textcolor{magenta}{\bs{\bar{Z}}}}$};
    \node[red, scale=0.7] at (3.5+0.2, -2.25-0.2) {$\mcX^{\textcolor{magenta}{\bs{\bar{Z}}}}$};
    \node[red, scale=0.7] at (3.5+0.2, -3.25-0.2) {$\mcX^{\textcolor{magenta}{\bs{\bar{Z}}}}$};
    \node[red, scale=0.7] at (3.5+0.2, -4.25-0.2) {$\mcX^{\textcolor{magenta}{\bs{\bar{Z}}}}$};
    
    \node at (3.5+0.2, 0.5-0.2) {$\bs{\bar{\mcX}^{\bs{\bar{Z}}}}$};

    \draw[line width = 0.5mm, red] (3+0.2, 0-0.2) -- (4+0.2, 0-0.2);
    \draw[line width = 0.5mm, red] (3+0.2, -1-0.2) -- (4+0.2, -1-0.2);
    \draw[line width = 0.5mm, red] (3+0.2, -2-0.2) -- (4+0.2, -2-0.2);
    \draw[line width = 0.5mm, red] (3+0.2, -3-0.2) -- (4+0.2, -3-0.2);
    \draw[line width = 0.5mm, red] (3+0.2, -4-0.2) -- (4+0.2, -4-0.2);

    \node at (3, -5) {$\bs{\mathcal{D}(\Z_3)}$};

    \foreach \i in {7, ..., 10} {
        \node[magenta] at (\i+0.5+0.2, -2-0.2) {$Z$};
    }

    \node[magenta] at (11.5+0.2, -2-0.2) {$\bs{\bar{Z}}$};
    \node at (9, -5) {$\bs{\mathcal{D}(\Z_2)}$};

\label{NonlocalString2}
\end{tikzpicture}}
\end{equation}

In other words, $\bs{\bar{\mcX}} \rightarrow \bs{\bar{\mcX}}^{\bs{\bar{Z}}}.$ Using an identical argument, we see that the nonlocal strings in Eq.~\eqref{CAnyonLine} can be truncated to the $\Z_2$ code, implying that $\bs{\bar{\mcZ}} \rightarrow \bs{\bar{\mcZ}}^{\bs{\bar{Z}}}$ as well.

Therefore, based on the above action on the logical operators, we have executed a $C\mathcal{C}$ from the qubit code to the qutrit code using this protocol.

\subsection{Summary of the protocol}

We summarize our protocol in the following algorithm:
\begin{enumerate}
    \item Initialize codes as in Fig. \ref{fig:overviewS3}(a) with the $\Z_2$ code to the left of the $\Z_3$ code. Initialize a square lattice of ancilla vertex qubits in the $\ket{+}$ state and ancilla edge qubits in the $\ket{0}$ state.
    \item We next perform code extension by a single column $c$ at the right edge of the current qubit code state.
    \begin{enumerate}
        \item If and only if $c$ contains both qubits and qutrits, apply $U_{C\mathcal{C}}$ on $c.$
        \item Apply $U_{CX}$ on $c.$
        \item Measure $X_v$ for all vertices $v$ in $c$ obtaining measurement outcomes $x_v.$
        \item If $x_v = -1,$ then apply $Z_{e}$ on the edge $e$ directly to the right of $v.$
    \end{enumerate}
    \item We next perform code shrinkage by a single column $c$ at the left edge of the current qubit code state.
    \begin{enumerate}
        \item Measure $Z_e$ for all edges $e$ in $c$ obtaining measurement outcomes $z_e$ and append $\{ z_e \}$ to the measurement record.
        \item If the measurement record now contains a new contractible loop $\partial R$, apply $\prod_{e} C_{e}$ to the portion of $R$ that contains qutrits.
        \item If the measurement record contains a new non-contractible line $\gamma,$ apply $\bs{\bar{X}}$ on the $\Z_2$ code and apply $\prod_{e} C_{e}$ to all qutrits to the right of $\gamma.$
        \item If and only if $c$ contains only qubits, identify the $m$ anyons using the measurement record. Apply finite depth $\prod X$ string operators to eliminate the $m$ anyons in a contractible manner.
    \end{enumerate}
    \item We now describe the error correction step, which is performed after each execution of steps 2 and 3 for each column $c.$
    \begin{enumerate}
        \item In the bulk of Abelian codes, measure vertex and plaquette syndromes and apply standard decoders for the qubit and qutrit surface codes.
        \item In the bulk of $S_3$ code, apply the $S_3$ non-Abelian decoder (discussed in Section \ref{SlidingErrorCorrection}).
    \end{enumerate}
\end{enumerate}

\section{Controlled-Automorphism Gate ($C\psi$) from non-Abelian surface codes}
\label{CAuto}

In this section, we demonstrate that a large class of non-Abelian topological orders can be used to perform non-Clifford logical gates between stabilizer codes by generalizing the protocol presented in the main text for $\mathcal{D}(S_3).$

We consider a $\Z_m$ surface code and a $\Z_n$ surface code. We refer to the $\Z_n$ code as the \emph{base} code and the $\Z_m$ code as the \emph{symmetry} code, as we will elaborate on later.
\begin{equation}
    \raisebox{-0.5\height}{
        \begin{tikzpicture}
                \draw (0, 1) -- (1, 2);
                \draw (1, 2) -- (4, 2);
                \draw (4, 2) -- (3, 1);
                \draw (3, 1) -- (0, 1);
                \node at (2, 1.5) {$\mathcal{D}(\mathbb{Z}_m)$};
                \draw[blue] (3.5, 2) -- (2.5, 1);
                \node[blue] at (3.5, 2.25){$\overline{X}$};
                \draw[magenta] (0.75,1.75) -- (3.75,1.75);
                \node[magenta] at (0.4, 1.75) {$\overline{Z}$};
                \draw (0+3, 1-1) -- (1+3, 2-1);
                \draw (1+3, 2-1) -- (4+3, 2-1);
                \draw (4+3, 2-1) -- (3+3, 1-1);
                \draw (3+3, 1-1) -- (0+3, 1-1);
                \node at (2+3, 1.5-1) {$\mathcal{D}(\mathbb{Z}_n)$};
                \draw[red] (3.5+3, 2-1) -- (2.5+3, 1-1);
                \node[red] at (3.5+3, 2.25-1) {$\overline{\mathcal{X}}$};
                \draw[green] (0.25+3, 1.25-1) -- (3.25+3, 1.25-1);
                \node[green] at (3.25+3+0.35, 1.25-1) {$\overline{\mathcal{Z}}$};
        \end{tikzpicture}
    }
\end{equation}
Here we show the logical operators $\overline{X}, \overline{Z}$ of the $\Z_m$ surface code and logical $\overline{\mathcal{X}}, \overline{\mathcal{Z}}$ operators of the $\Z_n$ surface code. They satisfy the algebras $\overline{Z} \ \overline{X} = e^{2\pi i/m} \overline{X} \ \overline{Z}$ and $\overline{\mcZ} \ \overline{\mcX} = e^{2\pi i/n} \overline{\mcX} \ \overline{\mcZ}$ as well as  $\bar{Z}^{m} = \bar{X}^{m} = 1$ and $\bar{\mcZ}^{n} = \bar{\mcX}^{n} = 1.$

We consider an automorphism $\psi: \Z_{m} \rightarrow \text{Aut}(\Z_{n}).$ This automorphism defines a split extension of Abelian groups:
\begin{equation}
    1 \rightarrow \Z_{m} \rightarrow G \rightarrow \Z_{n} \rightarrow 1,
\end{equation}
where $G$ is a non-Abelian group. 

Similar to the $\mathcal{D}(S_3)$ example, we then entangle the logical states of the base code and the symmetry code into $\mathcal{D}(G)$ by gauging the $\Z_{m}$ automorphism symmetry of the base code. The column-by-column gauging procedure generalizes almost identically from the $S_3$ example. We replace $C\mathcal{C}$ symmetry enrichment unitary with the $C\psi$ unitary, and we replace the $\Z_2$ gauging map with the $\Z_m$ gauging map~\cite{tantivasadakarn_hierarchy_2022}:
\begin{equation}
    \raisebox{-0.5\height}{
        \begin{tikzpicture}
                \draw (0, 1) -- (1, 2);
                \draw (1, 2) -- (4, 2);
                \draw (4, 2) -- (3, 1);
                \draw (3, 1) -- (0, 1);
                \node at (2, 1.5) {$\mathcal{D}(G)$};
                \draw[blue] (3.5, 2) -- (2.5, 1);
                \node[blue] at (3.5, 2.25) {$\sigma$};
                \draw[red] (1.5, 2) -- (0.5, 1);
                \node[red] at (1.5, 2.25) {$[\tilde{m}]_{\psi}$};
                \draw[green] (0.25, 1.25) -- (3.25, 1.25);
                \node[green] at (3.5+0.35, 1.25) {$[\tilde{e}]_{\psi}$};
                \draw[magenta] (0.25+0.5, 1.25+0.5) -- (3.25+0.5, 1.25+0.5);
                \node[magenta] at (3.25+0.5+0.25, 1.25+0.5) {$\phi$};
        \end{tikzpicture}
    }
\end{equation}

We depict the logical operators of the Abelian codes and their injection into the non-Abelian code $\mathcal{D}(G).$ Here $[\tilde{e}]_{\psi}$ and $[\tilde{m}]_{\psi}$ denote non-Abelian orbit anyons that come from the $\Z_{n}$ base code, and $\sigma$ and $\phi$ are the gauge flux and gauge charge of the $\Z_{m}$ anyon automorphism symmetry, respectively. We see that within the non-Abelian code $\mathcal{D}(G),$ the logical operators of the symmetry code have now been transformed into the automorphism symmetry charge and flux lines of the base code.

Lastly, when the codes are ejected using an analogous $Z$ measurement and $\psi$ feedforward procedure discussed above, we obtain a \emph{controlled-automorphism} gate $C\psi$ whose logical action is shown below:

\begin{equation}
    \raisebox{-0.5\height}{
        \begin{tikzpicture}[scale=0.85]
                \draw (0+4.25, 1) -- (1+4.25, 2);
                \draw (1+4.25, 2) -- (4+4.25, 2);
                \draw (4+4.25, 2) -- (3+4.25, 1);
                \draw (3+4.25, 1) -- (0+4.25, 1);
                \node at (2+4.25, 1.5) {$\mathcal{D}(\mathbb{Z}_m)$};
                \draw[blue] (3.5+4.25, 2) -- (2.5+4.25, 1);
                \node[blue] at (3.5+4.25, 2.25) 
                {$\overline{X}\bs{\psi}$};
                \draw[magenta] (0.75+4.25,1.75) -- (3.75+4.25,1.75);
                \node[magenta] at (3.75+4.25+0.3+0.1,1.75) {$\overline{Z}$};

                \draw (0, 1-1) -- (1, 2-1);
                \draw (1, 2-1) -- (4, 2-1);
                \draw (4, 2-1) -- (3, 1-1);
                \draw (3, 1-1) -- (0, 1-1);
                \node at (2, 1.5-1) {$\mathcal{D}(\mathbb{Z}_n)$};
                \draw[red] (3.5, 2-1) -- (2.5, 1-1);
                \node[red] at (3.5, 2.25-1+0.1) {$\psi^{\overline{Z}}(\overline{\mcX})$};
                \draw[green] (0.25, 1.25-1) -- (3.25, 1.25-1);
                \node[green] at (3.5+0.35+0.1, 1.25-1) {$\psi^{\overline{Z}}(\overline{\mcZ})$};
    
        \end{tikzpicture}
    }
\end{equation}
Here $\bs{\psi} = \prod_{e} \psi_{e}$ is the circuit that generates the automorphism symmetry of the $\Z_{n}$ code, which we assume satisfies $\bs{\psi}^{m} = 1.$ We further define
\begin{equation}
    \psi^{\overline{Z}}(\overline{\mcZ}) = \sum_{i = 0}^{m} \bar{\ket{i}}\bar{\bra{i}} \otimes \psi^{i}\overline{\mcZ} (\psi^{i})^{\dagger}
\end{equation}
as an operator that acts the automorphism $\bs{\psi}$ on the $\Z_n$ base code conditional on the logical state of of the $\Z_m$ symmetry code. As an example, for $G = S_3,$ $\psi$ is simply $\mathcal{C}$ and $\mathcal{C}^{\overline{Z}}(\overline{\mathcal{Z}}) \equiv \overline{\mathcal{Z}}^{\overline{Z}}.$

In particular, by using the group $\Z_2^2$ for the base code and the group $\Z_2$ for the symmetry code and a automorphism that swaps the two $\Z_2$ factors, we can obtain a non-Clifford Controlled-SWAP or Fredkin gate between 3 qubit surface codes by entangling into the $D_4 = \Z_2^2 \rtimes \Z_2$ quantum double. This non-Clifford gate can be used to perform universal quantum computation with qubit surface codes, provided that the entire Clifford group is also executable. All the circuits in Section \ref{CCGateMain} can be utilized in the $D_4$ case by simply replacing the $\mathcal{C}$ automorphism of the $S_3$ code with the $\text{SWAP}$ automorphism of two surface codes.

We remark that our procedure yields a different non-Clifford gate than those discussed in Ref. \cite{davydova_universal_2025} due to the difference between our $D_4$ model---which is obtained by gauging the SWAP symmetry of $\Z_2^2$ surface codes ---and the $\mathbb{Z}_2^3$ type-III twisted quantum double model introduced in the latter.

\section{Error Correction and Fault-tolerance of the $C\mathcal{C}$ Gate}
\label{SlidingErrorCorrection}

In this section, we comment on a potential route to performing error correction during the execution of the $C\mathcal{C}$ gate protocol and the extent to which it is fault-tolerant. We focus mainly on the $\mathcal{D}(S_3)$ example, as it is most relevant for near-term experiments, but similar arguments hold for any of the general protocols presented.

When only Abelian codes are present at the beginning and the end of the protocol, ordinary syndrome measurements of vertex and plaquette stabilizers can be performed. Once syndromes are identified, an appropriate decoder, such as a maximum-likelihood (ML) or minimum-weight perfect matching, can be used to correct errors. For qudit surface codes, standard decoders used for qubit surface codes do not apply, so generalized decoders must be used instead~\cite{watson_fast_2015}. Thus, as long as we operate below the threshold of both codes, the sliding protocol will not fail at these steps.

In the remainder of this section, we split the discussion of fault-tolerance within the non-Abelian code into three sections. First we discuss the importance of extending the codes column by column due to errors that occur before the gauging map. In Appendix \ref{S3Syndromes}, we discuss how single qubit and single qutrit $X, Z$ and $\mcX, \mcZ$ errors create violations of operators in Eq.~\eqref{S3Stabs}. Finally, we propose a procedure in which general errors can be corrected. We leave a numerical analysis of the threshold of our decoder against single qubit and qutrit errors to future work.

\subsection{Errors before the $\Z_2$ gauging map}

In Section \ref{CodeConventions}, we discussed how the gauging map can be performed column-by-column in order to extend surface codes to the right. A crucial reason for this is due to the non-locality of the gauging map; after its execution, operators charged under the gauged symmetry can become nonlocal, rendering an analysis of fault-tolerence significantly more challenging. Thus, it is important to understand how errors that occur before the code extension is performed propagate through the gauging map. As an example, if a $\mcZ_{e}$ error occurs within the $\Z_3$ code immediately before a single column $\bs{c}$ of gauging map is applied, we get the following error within $\mathcal{D}(S_3)$:

\begin{equation}
\raisebox{-0.5\height} {
    \begin{tikzpicture}
    \draw (4,0) -- (6, 0);
    \draw (4,-1) -- (6,-1);
    \draw (4,-2) -- (6,-2);
    \draw (4,-3) -- (6,-3);
    \draw (4,-4) -- (6,-4);
    \draw (6, 0) -- (8, 0);
    \draw (6, -1) -- (8, -1);
    \draw (6, -2) -- (8, -2);
    \draw (6, -3) -- (8, -3);
    \draw (6, -4) -- (8, -4);

    \draw[brown] (6+0.2, 0-0.2) -- (8+0.2, 0-0.2);
    \draw[brown] (6+0.2, -1-0.2) -- (8+0.2, -1-0.2);
    \draw[brown] (6+0.2, -2-0.2) -- (8+0.2, -2-0.2);
    \draw[brown] (6+0.2, -3-0.2) -- (8+0.2, -3-0.2);
    \draw[brown] (6+0.2, -4-0.2) -- (8+0.2, -4-0.2);

    % \draw (1, 0) -- (1, -4);
    % \draw (2, 0) -- (2, -4);
    % \draw (3, 0) -- (3, -4);
    % \draw (4, 0) -- (4, -4);
    \draw (5, 0) -- (5, -4);
    \draw (6, 0) -- (6, -4);
    \draw (7, 0) -- (7, -4);
    % \draw (8, 0) -- (8, -4);

    \draw[brown] (7+0.2, 0-0.2) -- (7+0.2, -4-0.2);
    \draw[brown] (8+0.2, 0-0.2) -- (8+0.2, -4-0.2);
    % \draw[brown] (10+0.2, 0-0.2) -- (10+0.2, -4-0.2);

    % \draw[line width = 0.3mm, blue] (8, 0) -- (9, 0);
    % \draw[line width = 0.3mm, blue] (8, -1) -- (9, -1);
    % \draw[line width = 0.3mm, blue] (8, -2) -- (9, -2);
    % \draw[line width = 0.3mm, blue] (8, -3) -- (9, -3);
    % \draw[line width = 0.3mm, blue] (8, -4) -- (9, -4);

    \draw[line width = 0.3mm, red] (7+0.2, -2-0.2) -- (7+0.2, -3-0.2);
    \node[red] at (7.6, -2.5-0.2)[scale=0.8] {$\mcZ_{e}^{\textcolor{magenta}{Z_1}}$};

    \draw[purple] (7+0.5, -2) circle[radius=1.5pt];
    \fill[purple] (7+0.5, -2) circle[radius=1.5pt];
    \node[magenta, scale=0.6] at (7+0.5, -1.8) {${1}$};

    \node at (7, 0.25) {$\bs{c}$};

    \node at (5, -4.7) {$\bs{\mathcal{D}(\Z_2)}$};
    \node at (7, -4.7) {$\bs{\mathcal{D}(S_3)}$};
\end{tikzpicture}}
\end{equation}

We remark that if the $S_3$ code is not prepared in a column-by-column fashion but instead prepared all at once, the $\mcZ$ error in the $\Z_3$ code would become an operator conditioned on a nonlocal string extending to the right in the $S_3$ code, turning it into a nonlocal error. When done column-by-column, the error only violates $A_v$ and $\tilde{A}_v$ stabilizers near the right boundary of the code, making its correction significantly easier. Analogously, we can show that $\mcX$ errors before the gauging map will only violate $\tilde{B}_p$ and $A_v$ stabilizers at the right boundary of the code.

To summarize, by performing error correction directly after just a single column of gauging map is applied, we are able to pass a state with purely local errors into the $S_3$ decoding procedure, as we detail next.

\subsection{Insights on a Heralded $S_3$ Decoding Strategy}
\label{S3ErrorCorrection}

In this section, we discuss a procedure in which errors can be corrected as the $S_3$ quantum double is prepared. As described in the previous section, this includes single qubit and single qutrit errors that occur before and after the $\Z_2$ charge-conjugation gauging map, as long as the $S_3$ code is prepared in a column-by-column fashion. It can also be used for any local errors applied after the $S_3$ code is prepared. In contrast to previous approaches, we directly correct the syndromes of the qubit-qutrit model instead of attempting to correcting anyon errors~\cite{wootton_error_2014}. As an aside, we comment that the approach of correcting the syndromes of the commuting projector model allows for a general, more efficient approach for decoding non-Abelian topological orders with large numbers of anyon types.

The first step of the decoding procedure is to correct the $B_p = -1$ syndromes, which are created, for example, by qubit $X$ errors. They are associated to both $D$ and $E$ anyon excitations as described in Appendix~\ref{QuantumDouble}. In the non-Abelian code $\mathcal{D}(S_3),$ such error strings probabilistically create violations of the $\tilde{A}_v$ and $\tilde{B}_p$ operators along the length of the string, as discussed in Appendix~\ref{S3Syndromes}. One simple approach to correct these errors is to measure $B_p$ operators and correct $B_{p} = -1$ errors analogously to the correction of $m$ anyons in the toric code, by pairing up with $X$ string operators in the bulk or connecting to the top or bottom boundary. However, such a correction probabilistically leaves behind $C$ and $F$ anyons that need to be corrected in future steps.

Here we also present a heralded decoding strategy, which takes into account the fact that all of the $B_p$ and the $\tilde{A}_v$ operators on different sites form a commuting set. Thus, we can measure both of these operators on all sites at once at the beginning of our decoding process to inform which $X$ correction string to apply. In particular, for a string of $X$ errors of length $\ell,$ the probability that there are no $\tilde{A}_v$ violations along the length of the string is $\frac{1}{3^\ell}.$ In the limit where the qubit and qutrit error rate is small, we employ a strategy of preferentially pairing up $B_p = -1$ syndromes along paths that contain $\tilde{A}_v$ syndromes. Doing so minimizes the number of new $C$ and $F$ anyons created. We leave a precise algorithm for determining such paths to future work. One can equivalently perform the aformentioned strategy by measuring $B_p$ and the $\tilde{B}_p$ operators as well.

The next step is to correct $\tilde{A}_v$ and $\tilde{B}_p$ syndromes, which originate from either qutrit errors or are created from the $X$ strings used to eliminate $B_p$ syndromes. Once all the $B_p$ operators are set to $1,$ all pairs of $\tilde{A}_v$ and $\tilde{B}_p$ operators will commute with each other as discussed in Section~\ref{CCGateMain}, so they can be measured simultaneously. As a result, the correction of $\tilde{A}_v$ syndromes can be performed independently of $\tilde{B}_p$ syndromes.

Without loss of generality, we consider the correction of $\tilde{A}_v$ syndromes with the case of $\tilde{B}_p$ syndromes following almost identically. We first discuss the choice of paths to eliminate these syndromes, and then we discuss the sequential, adaptive circuit used to do so on a given path.

Based on our choice of correction strings $\{\gamma_i\}$ used to eliminate $B_p = -1$ syndromes, we first correct the $C$ anyons that were created along the length of each $\gamma_i.$ Once this is finished, the remaining $C$ anyons are created by qutrit errors. In the limit where the qutrit error rate is small, we preferentially pair up $\tilde{A}_v = \omega, \omega^2$ syndromes on paths that have other syndromes along the path versus paths that have no syndromes. In particular, for two syndromes separated by a path $\gamma$ of length $\ell,$ the probability of having no other $\tilde{A}_v$ stabilizers along the path is $\frac{1}{2^\ell}.$ We leave a precise algorithm for determining $C$ anyon pairings to future work.

We now present the measurement and feedforward circuit used to remove a pair of $C$ anyons along a chosen path $\gamma$. The main idea is to sequentially measure $\tilde{A}_v$ syndromes along $\gamma$ and adaptively apply $\mcZ$ and $\mcZ^{\dagger}$ correction operations dependent on all previously obtained measurement outcomes. While the measurement of $\tilde{A}_v$ will create Abelian $B$ anyons along the length of $\gamma$, the advantage of this approach is that it replaces the correction of non-Abelian anyons with the correction of Abelian anyons, which is comparatively easier. For more details, see the adaptive string operators introduced in Appendix~\ref{ErrorCorrectionCircuits}. 

Once these non-Abelian syndromes are eliminated, there still are some number of Abelian $B$ anyons left to correct. These are readily eliminated by using an appropriate $\Z_2$ surface code decoder to pair up anyons using $Z$ string operators.

Thus, we have demonstrated that by first correcting $B_p$ syndromes, then correcting $\tilde{A}_v$ and $\tilde{B}_p$ syndromes, and finally $A_v$ syndromes, we can return the $S_3$ code to the ground state in the presence of qubit and qutrit errors. To place our procedure within a renormalization group (RG) decoder framework, we envision performing each of the aformentioned three steps at all scales before proceeding to the next step. We leave a study of an analytical and numerical calculation of a threshold of $\mathcal{D}(S_3)$ for our heralded decoding strategy against qubit and qutrit errors to future work.

\section{Discussion and Outlook}

In this paper, we have proposed a protocol that realizes non-Clifford logical operations between Abelian surface codes. The advantage of our approach is that the qudit surface codes remain in their ground state throughout the entire process, thereby simplifying the error correction process. 

We have explicitly demonstrated how the protocol applies to a qutrit surface code and a qubit surface code, realizing a controlled-charge conjugation ($C\mathcal{C}$) gate. Furthermore, we generalize the approach to surface codes based on arbitrary $\mathbb{Z}_p$ and $\mathbb{Z}_q$ gauge groups, where our protocol implements a nontrivial controlled-anyon automorphism. Given recent advances in preparing ground states of qubit and qutrit surface codes across multiple experimental platforms, our protocol can be directly implemented in these settings.

We have also discussed error correction and fault-tolerance procedures within our protocol. In particular, we have studied error correction in the presence of single-qubit and single-qutrit errors, both before and after the gauging process. We have proposed insights on how to construct a heralded decoder for the quantum double of $S_3$ that uses properties of the non-Abelian anyons to more effectively pair up syndromes. A theoretical or numerical analysis of the corresponding error thresholds of our decoding protocol remains an important direction for future work. Establishing such results would be the first step towards implementing our procedure on a quantum processor. One open question is how to perform error correction in the presence of measurement errors. 

We also note that in the recent work Ref.~\cite{davydova_universal_2025}, the authors establish a correspondence between circuit implementations and the path-integral formalism. Investigating the 3+1D spacetime picture, or a topological field theory (TFT) description, of our protocol could offer another promising direction. While the TFTs for the ground states of non-Abelian topological order have been explored, it remains an open question how to properly incorporate all anyons and symmetry defects into the TFT framework ~\cite{barkeshli_codimension_2023}. It is known that certain diagonal non-Clifford gates, such as the $\mathrm{CCZ}$ and $T$ gates, can be derived from path integrals of topological actions defined via cup products of $\Z_2$-valued cocycles. However, since such path integrals can only diagonal phase gates and the $C\mathcal{C}$ gate is non-diagonal, it would be interesting to explore how our protocol might fit into the TFT framework.

\vspace{5pt}

\begin{acknowledgments}
    We acknowledge Liyuan Chen, Yuanjie Ren, Anasuya Lyons, Chiu Fan Bowen Lo, Carolyn Zhang, Arpit Dua, Margarita Davydova, Andreas Bauer, Mikhail D. Lukin, Isaac Kim, and Ashvin Vishwanath for useful discussions. We thank Isaac Kim for reviewing the manuscript and providing valuable feedback. R.S. acknowledges support from the National Science Foundation Graduate Research Fellowship (NSF GRFP). B.~R.~acknowledges support from the Harvard Quantum Initiative Postdoctoral Fellowship in Science Engineering. Z.~S.~was supported by funds from the UC Multicampus Research Programs and Initiatives of the University of California, Grant Number M23PL5936. Y.~L. was supported by the U.S. National Science Foundation under Grant No. NSF DMR-2316598.
\end{acknowledgments}

\bibliography{refs.bib}

\appendix
\onecolumngrid

\newpage

\section{Review of the $S_{3}$ Quantum Double}
\label{S3Review}

In this section, we review the quantum double of $S_{3}$ and its categorical data. Since it is isomorphic to $D_3,$ the group $S_3$ can be generated by a rotation element $r$ and a reflection element $s$ satisfying the algebra $r^3 = s^2 = 1$ and $srs^{-1} = r^2.$ In particular, the group can be written as a split extension
\begin{equation}
    1 \rightarrow \Z_{3} \rightarrow S_{3} \rightarrow \Z_{2} \rightarrow 1,
\end{equation}
with an automorphism $\sigma: \Z_2 \rightarrow \text{Aut}(\Z_3)$ where the non-trivial element of $\Z_2$ swaps $r$ with $r^2.$

The group has three conjugacy classes: $[1] = \{1\}, [r] = \{r, r^2\}, $ and $[s] = \{s, rs, r^2s\}.$ It further has $3$ irreducible representations: the one-dimensional trivial irrep $\boldsymbol{1},$ the one-dimensional sign irrep $\boldsymbol{s}$ that assigns the value $-1$ to elements in $[s],$ and lastly a two-dimensional irrep $\boldsymbol{2}$, where
\begin{equation}
    r = \begin{pmatrix}
        0 & -1\\
        1 & -1
        \end{pmatrix}, \hspace{5mm}
    s = \begin{pmatrix}
        -1 & 1\\
        0 & 1
        \end{pmatrix}.
\end{equation}

Kitaev's quantum double is defined on an oriented square lattice with the Hilbert space $\mathbb{C}[S_3]$ on each edge \cite{kitaev_fault_2023}. We label this onsite space by an orthonormal basis $\{ \ket{g} \mid g \in S_3 \}.$ A natural set of operators that can be defined on this Hilbert space is a set of shift operators by the elements of $S_3$
\begin{equation}
    L^g_+\ket{h} = \ket{gh}, L^g_{-}\ket{h} = \ket{hg^{-1}},
\end{equation}
where $g, h \in S_3.$

We can further define projectors onto each individual basis state of $\mathbb{C}[S_3]$
\begin{equation}
    T^g_{+}\ket{h} = \delta_{g, h}\ket{h}, T^g_{-}\ket{h} = \delta_{g^{-1}, h}\ket{h}.
\end{equation}

The Hamiltonian for the quantum double of $S_3$ can be written as $H = -\sum_{v} A^{S_3}_v - \sum_{p} B^{S_3}_{p},$ where $A^{S_3}_v$ is a generalized vertex projector and $B^{S_3}_{p}$ is a generalized plaquette projector. We use the superscript to distinguish these operators from star and plaquette operators that will be defined later. We define $A^{S_3}_v = \frac{1}{6} \sum_{g \in S_3} A_v^g,$ where
\begin{equation}
\raisebox{-0.5\height} {
    \begin{tikzpicture}

    \node at (4.5, 1) {$A_v^g = $};

    \draw (5,1) to node[currarrow, rotate=180] {} (6,1);
    \draw (6, 1) to node[currarrow, rotate=90] {} (6, 2);
    \draw (6, 1) to node[currarrow, rotate=180] {} (7, 1);
    \draw (6, 1) to node[currarrow, rotate=90] {} (6, 0);
    
    \node[scale=0.9] at (5.5, 1+0.3) {$L_g^+$};
    \node[scale=0.9] at (6+0.3, 1.5) {$L_g^+$};
    \node[scale=0.9] at (6.5, 1-0.3) {$L_g^-$};
    \node[scale=0.9] at (6-0.3, 0.5) {$L_g^-$};

\end{tikzpicture}}
\end{equation}
is a generalized vertex shift operator by the element $g$.

The generalized plaquette operator $B_p^{h}$ is defined as as a projector onto the subspace the Hilbert space that has group-valued flux equal to $h$ on the plaquette $p$. We can write such an operator as
\begin{equation}
\raisebox{-0.5\height} {
    \begin{tikzpicture}

    \node at (4.5, 1) {$B_p^h = $};

    \node at (6.5, 1) {$\sum_{\{ g_i \}} \delta_{h, g_1g_2^{-1}g_3^{-1}g_4}$};

    \draw (8, 0.5) -- (8, 1.5);
    \draw (9.3, 1.5) -- (9.5, 1);
    \draw (9.5, 1) -- (9.3, 0.5);

    \draw (8.4, 1.3) to node[currarrow, rotate=180] {} (9, 1.3);
    \draw (8.4, 1.3) to node[currarrow, rotate=90] {} (8.4, 0.7);
    \draw (8.4, 0.7) to node[currarrow, rotate=180] {} (9, 0.7);
    \draw (9, 0.7) to node[currarrow, rotate=90] {} (9, 1.3);

    \node[scale=0.8] at (8.2, 1) {$g_1$};
    \node[scale=0.8] at (8.7, 1.5) {$g_2$};
    \node[scale=0.8] at (9.25, 1) {$g_3$};
    \node[scale=0.8] at (8.7, 0.5) {$g_4$};

    \draw (8.4+1.7, 1.3) to node[currarrow, rotate=180] {} (9+1.7, 1.3);
    \draw (8.4+1.7, 1.3) to node[currarrow, rotate=90] {} (8.4+1.7, 0.7);
    \draw (8.4+1.7, 0.7) to node[currarrow, rotate=180] {} (9+1.7, 0.7);
    \draw (9+1.7, 0.7) to node[currarrow, rotate=90] {} (9+1.7, 1.3);

    \node[scale=0.8] at (8.2+1.7, 1) {$g_1$};
    \node[scale=0.8] at (8.7+1.7, 1.5) {$g_2$};
    \node[scale=0.8] at (9.25+1.7, 1) {$g_3$};
    \node[scale=0.8] at (8.7+1.7, 0.5) {$g_4$};

    \draw (9.3+0.5, 1.5) -- (9.1+0.5, 1);
    \draw (9.1+0.5, 1) -- (9.3+0.5, 0.5);
    
    \draw (8+3.2, 0.5) -- (8+3.2, 1.5);

\end{tikzpicture}}
\label{S3BpOperator}
\end{equation}
The operator in the Hamiltonian $B^{S_3}_p$ is equal to $B_p^e,$ which projects onto states with no flux. 

We also define the notion of a site $s = (v, p)$ as a vertex $v$ along with a plaquette $p$ that is directly to the left and directly up from $v.$ It can be easily verified that all pairs of $A^{S_3}_v$ and $B^{S_3}_p$ commute on all pairs of sites, so the quantum double is a commuting projector model. The model has a unique ground state $\ket{S_3}$ when placed on a topologically trivial manifold.

We may now discuss the anyon content of the quantum double of $S_{3}.$ Each anyon is associated to an irredicuble representation of $\mathcal{D}(S_3),$ the Drinfel'd double or quantum double of $S_{3}$ \cite{kitaev_fault_2023}. It is well-known that for finite groups, irreps of $\mathcal{D}(S_3)$ are labelled by a pair $(C, R)$, in which $C$ is conjugacy classes and $R$ is the irreducible representations of the centralizer of a representative element of the conjugacy class. In this prescription, pure charges correspond to anyons with trivial conjugacy class $[1]$, and pure fluxes correspond to anyons with the trivial irrep $\boldsymbol{1}.$ All other anyons are dyons. We write a full list of anyons in Table \ref{tab:S3QuantumDouble}.
\begin{table}[]
    \centering
    \begin{tabular}{ |c|c|c|c|c|c|  }
 \hline
 \multicolumn{6}{|c|}{$S_3$ Quantum Double} \\
 \hline
 Conjugacy Class & Centralizer & Irrep & Dim & QD Label & SET Label \\
 \hline
 $[1]$  & $S_3$    & $\boldsymbol{1}$ & 1 & $A$ & 1 \\
 $[1]$   & $S_3$    & $\boldsymbol{s}$ & 1 & $B$ & $\phi$ \\
 $[1]$   & $S_3$    & $\boldsymbol{2}$ & 2 & $C$ & $[\tilde{e}]$ \\
 $[s]$   & $\Z_2$    & $\boldsymbol{1}$ & 3 & $D$ & $\sigma$ \\
 $[s]$   & $\Z_2$    & $\boldsymbol{s}$ & 3 & $E$ & $\phi\sigma$ \\
 $[r]$   & $\Z_3$    & $\boldsymbol{1}$ & 2 & $F$ & $[\tilde{m}]$ \\
 $[r]$   & $\Z_3$    & $\boldsymbol{\omega}$ & 2 & $G$ & $[\tilde{e}\tilde{m}]$ \\
 $[r]$   & $\Z_3$    & $\boldsymbol{\bar{\omega}}$ & 2 & $H$ & $[\tilde{e}\tilde{m}^2]$ \\
 
 \hline
\end{tabular}
    \caption{Correspondence between anyon data for the $\Z_3$ toric code with gauged charge conjugation symmetry and $\mathcal{D}(S_3).$ $\phi$ and $\sigma$ correspond to the gauge charge and gauge flux of the charge conjugation symmetry, respectively.}
    \label{tab:S3QuantumDouble}
\end{table}

It will also be useful for our purposes to view $\mathcal D(S_{3})$ as the result of gauging the $\Z_{2}$ charge-conjugation symmetry of $\mathcal D(\Z_{3}),$ or the qutrit toric code. The $\Z_{3}$ toric code has $9$ anyons generated by $\langle 1, \tilde{e}, \tilde{e}^{2} \rangle \times \langle 1, \tilde{m}, \tilde{m}^{2} \rangle,$ where $\tilde{e}$ is the $\Z_3$ gauge charge and $\tilde{m}$ is the $\Z_3$ gauge flux. The charge conjugation symmetry acts by swapping $\tilde{e} \leftrightarrow \tilde{e}^{2}$ and $\tilde{m} \leftrightarrow \tilde{m}^{2},$ a transformation which preserves the braiding statistics of the $\Z_3$ topological order. We can thus view $\mathcal{D}(\Z_{3})$ as a symmetry-enriched topological order enriched by a $\Z_{2}$ topological symmetry. 

We now discuss how to obtain the anyon content of $\mathcal{D}(S_3)$ via gauging. Gauging the charge conjugation symmetry of the $\Z_3$ topological order amounts to proliferating all closed charge-conjugation symmetry defects in the ground state of the theory. This will identify Abelian anyons that transform non-trivially under the anyon automorphism symmetry into a single \emph{orbit} non-Abelian anyon. For each anyon $a$ in $\mathcal{D}(\Z_3),$ we define a non-Abelian anyon $[a]$ to be an orbit of $a$ under charge conjugation. In addition, we must introduce a gauge charge and a gauge flux associated to the $\mathbb{Z}_{2}$ symmetry that is gauged. 

Upon doing this, we obtain a representation for anyons in $\mathcal{D}(S_{3})$ that is shown in Table \ref{tab:S3QuantumDouble}. A quick way to deduce the matching between the two representations is by utilizing the quantum dimensions of the anyons. In the quantum double prescription, the dimension of an anyon $([c], \boldsymbol{\rho})$ is equal to $|[c]| \cdot \dim(\boldsymbol{\rho}).$ Meanwhile, in the SET prescription, the quantum dimension of an orbit non-Abelian anyon will be the size of the orbit. Simply this information allows us to match all the anyons. We note that $\phi$ must be the $B$ anyon because it is the only other Abelian anyon other than the vacuum. $C$ is a pure charge, so it must correspond to $[\tilde{e}].$ The $F$ anyon is a pure flux, so it matches with $[\tilde{m}].$ Matching topological spins, we find that $[\tilde{e}\tilde{m}]$ and $[\tilde{e}\tilde{m}^2]$ correspond to anyons $G$ and $H.$ There is no orbit of $\Z_2$ charge conjugation of order $3,$ so $D$ and $E$ are equal to $\sigma$ and $\sigma \phi,$ respectively. 

For a more general prescription of how anyons in a topological order transform under the gauging of a topological symmetry, we refer the reader to Refs.~\cite{barkeshli_symmetry_2019,bulmash_gauging_2019}.

\section{Relation between Kitaev's $\mathcal{D}(S_3)$ model and the Qubit-Qutrit Model}
\label{QuantumDouble}

In this section, we briefly outline the connection between Kitaev's original model for $\mathcal{D}(S_3)$ and provide a mapping from it to the qubit-qutrit model presented in the main text.

First, it is useful to represent the algebra of left and right shift operators for the group $S_3$ in terms qubit and qutrit operators. We decompose the onsite Hilbert space $\mathbb{C}[S_3]$ into $\mathbb{C}^2 \otimes \mathbb{C}^3$ and use an orthonormal basis labelled by $\{ \ket{s^q} \otimes \ket{r^n} \mid  q \in \{0, 1\}, n \in \{0, 1, 2\} \},$ where we use $r$ to denote the reflection element and $s$ to denote the rotation element of $S_3.$ Then, the action of the left and right shift operators becomes
\begin{align*}
    L^r_+\ket{s^qr^n} &= \ket{s^qr^{n+(-1)^q}}. \\
    L^r_-\ket{s^qr^n} &= \ket{s^qr^{n-1}}. \\
    L^s_+\ket{s^qr^n} &= \ket{s^{q+1}r^{n}}. \\
    L^s_-\ket{s^qr^n} &= \ket{s^{q-1}r^{-n}}. \\
\end{align*}
    
In terms of qubit and qutrit operators, we can equate
\begin{align*}
    L^r_- &= I \otimes \mcX^{\dagger} \\
    L^r_+ &= \ket{0}\bra{0} \otimes \mcX + \ket{1}\bra{1} \otimes \mcX^{\dagger} = \mcX^{Z} \\
    L^s_+ &= X \otimes I \\
    L^s_- &= X \otimes \mathcal{C}. \\
\end{align*}

We now can rewrite the operators of the quantum double Hamiltonian in terms of of qubit and qutrits. We first remark that $A^{S_3}_v$ can be written as the product of two projectors:
\begin{equation}
    A^{S_3}_v = \left (\frac{1 + A_v^r + A_v^{r^2}}{3} \right) \left (\frac{1 + A_v^s}{2} \right).
\end{equation}
After using the operator dictionary, we find that $A_{v}^r = \tilde{A}_v$ and that $A_{v}^s = A_v$ as shown in Eq. \ref{S3Stabs}.

As for the plaquette projectors $B_p^h$, we can obtain them by substituting in $g_i = s^{q_i}r^{n_i}$ into the equation $g_1g_2^{-1}g_3^{-1}g_4 = h = s^{q_h}r^{n_h}$
\begin{align*}
    \label{PlaquetteProjectors}
    q_1+q_2+q_3+q_4 = q_h, \\
    (n_1 - n_2)(-1)^{q_1+q_h} - n_3(-1)^{q_1+q_2+q_h}+n_4 = n_h.
\end{align*}

The first condition on clearly $q_1, q_2, q_3, q_4$ yields the qubit plaquette projector upon setting $q_h = 0.$ Once this is enforced and we set $n_h = 0$ as well, the second condition can be simplified into 
\begin{equation}
    n_1 - n_2 - n_3(-1)^{q_2} + n_4(-1)^{q_1} = 0,
\end{equation}
which yields the modified qutrit plaquette projector $\tilde{B}_p.$ Thus, indeed Kitaev's $S_3$ model and the qubit-qutrit model obtained from the gauging map are equivalent.

We can also write down projectors onto the $\bs{A-H}$ anyons of the $S_3$ quantum double in the qubit-qutrit langauge for a given site $s = (v, p)$ ~\cite{laubscher_universal_2019}:
\begin{itemize}
    \item For the $\bs{A}$ anyon, $A_v = B_p = 1,$ and $\tilde{A}_v = \tilde{B}_p = 1.$
    \item For the $\bs{B}$ anyon, $A_v = -1, B_p = 1,$ and $\tilde{A}_v = \tilde{B}_p = 1.$
    \item For the $\bs{C}$ anyon, $A_v = B_p = 1,$ and $ \tilde{B}_p  = 1$ in all cases. Then, we have $2$ possibilities: $\tilde{A}_v = \omega$ and $\tilde{A}_v = \omega^2.$ The complete projector onto the $C$ anyon is a projector onto both of these internal states.
\end{itemize}

To describe the remaining flux and dyon anyons, we need to define operators that measure the flux on a plaquette. To do this, we split the plaquette projector condition into two cases. In the case where $q_h = 0,$ we define the projector $\tilde{B}_p^{+}$
\begin{equation}
    \raisebox{-0.5\height}{
        \begin{tikzpicture}
        \node[scale=0.8] at (0.5-0.8-1, 1)  {$\tilde{B}_p^{+} = $};
        
        \draw[brown] (0.5-0.8, 1.5) -- (1.5-0.8, 1.5);
        \draw[brown] (0.5-0.8, 0.5) -- (1.5-0.8, 0.5);
        \draw[brown] (0.5-0.8, 0.5) -- (0.5-0.8, 1.5);
        \draw[brown] (1.5-0.8, 0.5) -- (1.5-0.8, 1.5);
    
        \draw[dashed] (0.5-0.3-0.8, 1.5+0.3) -- (1.5-0.3-0.8, 1.5+0.3);
        \draw[dashed] (0.5-0.3-0.8, 0.5+0.3) -- (1.5-0.3-0.8, 0.5+0.3);
        \draw[dashed] (0.5-0.3-0.8, 0.5+0.3) -- (0.5-0.3-0.8, 1.5+0.3);
        \draw[dashed] (1.5-0.3-0.8, 0.5+0.3) -- (1.5-0.3-0.8, 1.5+0.3);

        \node[magenta, scale=0.5] at (0.5-0.3+0.5-0.8, 1.5+0.3+0.15) {${2}$};
        \node[magenta, scale=0.5] at (0.5-0.3-0.15-0.8, 1.5+0.3-0.5) {${1}$};
    
        \draw[magenta] (0.5-0.3+0.5-0.8, 1.5+0.3) circle[radius=1pt];
        \fill[magenta] (0.5-0.3+0.5-0.8, 1.5+0.3) circle[radius=1pt];
        \draw[magenta] (0.5-0.3-0.8, 1.5+0.3-0.5) circle[radius=1pt];
        \fill[magenta] (0.5-0.3-0.8, 1.5+0.3-0.5) circle[radius=1pt];

        \draw[magenta] (0.5-0.3+0.5-0.8, 1.5+0.3-1) circle[radius=1pt];
        \fill[magenta] (0.5-0.3+0.5-0.8, 1.5+0.3-1) circle[radius=1pt];
        \draw[magenta] (0.5-0.3-0.8+1, 1.5+0.3-0.5) circle[radius=1pt];
        \fill[magenta] (0.5-0.3-0.8+1, 1.5+0.3-0.5) circle[radius=1pt];

        \node[magenta, scale=0.5] at (0.5-0.3-0.8+1+0.15, 1.5+0.3-0.5) {$3$};
        \node[magenta, scale=0.5] at (0.5-0.3+0.5-0.8, 1.5+0.3-1-0.15) {$4$};

        \node[green, scale=0.7] at (0.5-0.8, 1) {$\mcZ^{\textcolor{magenta}{-Z_1}}$};
        \node[green, scale=0.7] at (1-0.8, 1.5) {$\mcZ^{\textcolor{magenta}{Z_1}}$};
        \node[green, scale=0.7] at (1.5-0.8, 1) {$\mcZ^{\textcolor{magenta}{Z_{1}Z_{2}}}$};
        \node[green, scale=0.6] at (1-0.8, 0.5) {$\mcZ^{\dagger}$};
    
        \end{tikzpicture}
    }
\end{equation}
Similarly, in the case where $q_h = 1,$ we define the projector $\tilde{B}_p^{-}$
\begin{equation}
    \raisebox{-0.5\height}{
        \begin{tikzpicture}
        \node[scale=0.8] at (0.5-0.8-1, 1)  {$\tilde{B}_p^{-} = $};
        
        \draw[brown] (0.5-0.8, 1.5) -- (1.5-0.8, 1.5);
        \draw[brown] (0.5-0.8, 0.5) -- (1.5-0.8, 0.5);
        \draw[brown] (0.5-0.8, 0.5) -- (0.5-0.8, 1.5);
        \draw[brown] (1.5-0.8, 0.5) -- (1.5-0.8, 1.5);
    
        \draw[dashed] (0.5-0.3-0.8, 1.5+0.3) -- (1.5-0.3-0.8, 1.5+0.3);
        \draw[dashed] (0.5-0.3-0.8, 0.5+0.3) -- (1.5-0.3-0.8, 0.5+0.3);
        \draw[dashed] (0.5-0.3-0.8, 0.5+0.3) -- (0.5-0.3-0.8, 1.5+0.3);
        \draw[dashed] (1.5-0.3-0.8, 0.5+0.3) -- (1.5-0.3-0.8, 1.5+0.3);

        \node[magenta, scale=0.5] at (0.5-0.3+0.5-0.8, 1.5+0.3+0.15) {${2}$};
        \node[magenta, scale=0.5] at (0.5-0.3-0.15-0.8, 1.5+0.3-0.5) {${1}$};
    
        \draw[magenta] (0.5-0.3+0.5-0.8, 1.5+0.3) circle[radius=1pt];
        \fill[magenta] (0.5-0.3+0.5-0.8, 1.5+0.3) circle[radius=1pt];
        \draw[magenta] (0.5-0.3-0.8, 1.5+0.3-0.5) circle[radius=1pt];
        \fill[magenta] (0.5-0.3-0.8, 1.5+0.3-0.5) circle[radius=1pt];

        \draw[magenta] (0.5-0.3+0.5-0.8, 1.5+0.3-1) circle[radius=1pt];
        \fill[magenta] (0.5-0.3+0.5-0.8, 1.5+0.3-1) circle[radius=1pt];
        \draw[magenta] (0.5-0.3-0.8+1, 1.5+0.3-0.5) circle[radius=1pt];
        \fill[magenta] (0.5-0.3-0.8+1, 1.5+0.3-0.5) circle[radius=1pt];

        \node[magenta, scale=0.5] at (0.5-0.3-0.8+1+0.15, 1.5+0.3-0.5) {$3$};
        \node[magenta, scale=0.5] at (0.5-0.3+0.5-0.8, 1.5+0.3-1-0.15) {$4$};

        \node[green, scale=0.7] at (0.5-0.8, 1) {$\mcZ^{\textcolor{magenta}{Z_1}}$};
        \node[green, scale=0.7] at (1-0.8, 1.5) {$\mcZ^{\textcolor{magenta}{-Z_1}}$};
        \node[green, scale=0.7] at (1.5-0.8+0.1, 1) {$\mcZ^{\textcolor{magenta}{-Z_{1}Z_{2}}}$};
        \node[green, scale=0.7] at (1-0.8, 0.5) {$\mcZ^{\dagger}$};
    
        \end{tikzpicture}
    }
\end{equation}
Using these modified projectors, we can enumerate the possibilities for the remaining anyons by using $\tilde{B_p^{+}}$ and $\tilde{B_p^{-}}$ to measure flux:
\begin{itemize}
    \item For the $\bs{D}$ anyon, $B_p= -1$ in all cases. We then have $3$ possibilities: (1) $\tilde{B}_p^{-} = 1, A_v = 1,$ (2) $\tilde{B}_p^- = \omega, \tilde{A}_vA_v = 1,$ (3) $\tilde{B}_p^- = \omega^2, \tilde{A}_v^2A_v = 1.$ The complete projector onto the $D$ anyon is a projector onto all $3$ of these internal states.
    \item For the $\bs{E}$ anyon, $B_p= -1$ in all cases. We then have $3$ possibilities: (1) $\tilde{B}_p^{-} = 1, A_v = -1,$ (2) $\tilde{B}_p^- = \omega, \tilde{A}_vA_v = -1,$ (3) $\tilde{B}_p^- = \omega^2, \tilde{A}_v^2A_v = -1.$ The complete projector onto the $E$ anyon is a projector onto all $3$ of these internal states.
    \item For the $\bs{F}$ anyon, $B_p = 1$ in all cases. We then have $2$ possibilities: (1) $\tilde{B}_p^+ = \omega, \tilde{A}_v = 1,$ (2) $\tilde{B}_p^+ = \omega^2, \tilde{A}_v = 1.$ The complete projector onto the $F$ anyon is a projector onto these $2$ internal states.
    \item For the $\bs{G}$ anyon, $B_p = 1$ in all cases. We then have $2$ possibilities: (1) $\tilde{B}_p^+ = \omega, \tilde{A}_v = \omega^2,$ (2) $\tilde{B}_p^+ = \omega^2, \tilde{A}_v = \omega.$ The complete projector onto the $G$ anyon is a projector onto these $2$ internal states.
    \item For the $\bs{H}$ anyon, $B_p = 1$ in all cases. We then have $2$ possibilities: (1) $\tilde{B}_p^+ = \omega, \tilde{A}_v = \omega,$ (2) $\tilde{B}_p^+ = \omega^2, \tilde{A}_v = \omega^2.$ The complete projector onto the $H$ anyon is a projector onto these $2$ internal states.
\end{itemize}

As a final comment, we remark on the quantum circuits needed to measure the operators $A_v, \tilde{A}_v, B_p, \tilde{B}_p.$ For the order-two operators $A_v$ and $B_p,$ we perform entangling gates between an ancilla qubit $a$ in the state $\ket{+}$ and the code qubits and qutrits and finally measure out the ancilla in the $X$ basis. Similarly, for the order-three operators $\tilde{A}_v$ and $\tilde{B}_p,$ we perform entangling gates between an ancilla qutrit $a$ in the state $\tilde{\ket{+}}$ and the code qubits and qutrits and finally measure out the ancilla in the $\mcX$ basis. Given the operators in Eq. \eqref{S3Stabs}, it is straightforward to construct these entangling gates; we refer the reader to Section IV of Ref. \cite{chen_universal_2025} for analogous circuits.

\section{Rough and Smooth Boundaries of $S_3$ quantum double}
\label{S3Boundaries}

In this section, we provide lattice realizations for the rough and smooth boundaries of the $S_3$ quantum double. In particular, the rough boundary is described by the Lagrangian algebra $A+B+2C,$ and the smooth boundary is described by the Lagrangian algebra $A+D+F.$ Recall from Appendix~\ref{S3Review} that the $B$ and $C$ anyons are the Abelian and non-Abelian charge of quantum dimension $2$, respectively, and the $D$ and $F$ anyons are non-Abelian fluxes of quantum dimension $3$ and $2,$ respectively. These boundaries are obtained by applying the $\Z_2$ charge conjugation gauging map to the left, right, top, and bottom boundaries of lattices shown in the main text.

For the rough boundary, we distinguish between the left and right sides of the system. The star projectors remain the same as the bulk. On the left side, the plaquette projectors are 
\begin{equation}
\raisebox{-0.5\height} {
    \begin{tikzpicture}
    \draw[brown] (0.5-0.8, 1.5) -- (1.5-0.8, 1.5);
    \draw[brown] (0.5-0.8, 0.5) -- (1.5-0.8, 0.5);
    % \draw[brown] (0.5-0.8, 0.5) -- (0.5-0.8, 1.5);
    \draw[brown] (1.5-0.8, 0.5) -- (1.5-0.8, 1.5);

    \draw[dashed] (0.5-0.3-0.8, 1.5+0.3) -- (1.5-0.3-0.8, 1.5+0.3);
    \draw[dashed] (0.5-0.3-0.8, 0.5+0.3) -- (1.5-0.3-0.8, 0.5+0.3);
    \draw[dashed] (0.5-0.3-0.8, 0.5+0.3) -- (0.5-0.3-0.8, 1.5+0.3);
    \draw[dashed] (1.5-0.3-0.8, 0.5+0.3) -- (1.5-0.3-0.8, 1.5+0.3);

    \node[magenta, scale=0.7] at (0.5-0.3+0.5-0.8, 1.5+0.3+0.15) {${1}$};
    \node[magenta, scale=0.7] at (0.5-0.3-0.15-0.8, 1.5+0.3-0.5) {${2}$};

    \draw[magenta] (0.5-0.3+0.5-0.8, 1.5+0.3) circle[radius=1pt];
    \fill[magenta] (0.5-0.3+0.5-0.8, 1.5+0.3) circle[radius=1pt];
    \draw[magenta] (0.5-0.3-0.8, 1.5+0.3-0.5) circle[radius=1pt];
    \fill[magenta] (0.5-0.3-0.8, 1.5+0.3-0.5) circle[radius=1pt];

    % \node[green, scale=0.6] at (0.5-0.8, 1) {$\mcZ$};
    \node[green, scale=0.7] at (1-0.8, 1.5) {$\mcZ$};
    \node[green, scale=0.7] at (1.5-0.8, 1) {$\mcZ^{\textcolor{magenta}{-Z_{1}}}$};
    \node[green, scale=0.7] at (1-0.8, 0.5) {$\mcZ^{\textcolor{magenta}{-Z_{2}}}$};

    \node[scale=0.7] at (1.7-0.3, 1.2) {$\text{+ h.c.}$};

    \draw (5.5+0.7, 1+0.5) -- (6.5+0.7, 1+0.5);
    \draw (5.5+0.7, 1-0.5) -- (6.5+0.7, 1-0.5);
    % \draw (5.5+0.7, 1-0.5) -- (5.5+0.7, 1+0.5);
    \draw (6.5+0.7, 1-0.5) -- (6.5+0.7, 1+0.5);
    % \node[magenta] at (5.5+0.7, 1-0.5+0.5) {$Z$};
    \node[magenta] at (5.5+0.7+1, 1-0.5+0.5) {$Z$};
    \node[magenta] at (5.5+0.7+1-0.5, 1-0.5+0.5-0.5) {$Z$};
    \node[magenta] at (5.5+0.7+1-0.5, 1-0.5+0.5-0.5+1) {$Z$};

\end{tikzpicture}}
\end{equation}
On the right side, the plaquette stabilizers are
\begin{equation}
\raisebox{-0.5\height} {
    \begin{tikzpicture}
    \draw[brown] (0.5-0.8, 1.5) -- (1.5-0.8, 1.5);
    \draw[brown] (0.5-0.8, 0.5) -- (1.5-0.8, 0.5);
    \draw[brown] (0.5-0.8, 0.5) -- (0.5-0.8, 1.5);
    % \draw[brown] (1.5-0.8, 0.5) -- (1.5-0.8, 1.5);

    \draw[dashed] (0.5-0.3-0.8, 1.5+0.3) -- (1.5-0.3-0.8, 1.5+0.3);
    \draw[dashed] (0.5-0.3-0.8, 0.5+0.3) -- (1.5-0.3-0.8, 0.5+0.3);
    \draw[dashed] (0.5-0.3-0.8, 0.5+0.3) -- (0.5-0.3-0.8, 1.5+0.3);
    \draw[dashed] (1.5-0.3-0.8, 0.5+0.3) -- (1.5-0.3-0.8, 1.5+0.3);

    \node[magenta, scale=0.7] at (0.5-0.3+0.5-0.8, 1.5+0.3+0.15) {${1}$};
    \node[magenta, scale=0.7] at (0.5-0.3-0.15-0.8, 1.5+0.3-0.5) {${2}$};

    \draw[magenta] (0.5-0.3+0.5-0.8, 1.5+0.3) circle[radius=1pt];
    \fill[magenta] (0.5-0.3+0.5-0.8, 1.5+0.3) circle[radius=1pt];
    \draw[magenta] (0.5-0.3-0.8, 1.5+0.3-0.5) circle[radius=1pt];
    \fill[magenta] (0.5-0.3-0.8, 1.5+0.3-0.5) circle[radius=1pt];

    \node[green, scale=0.7] at (0.5-0.8, 1) {$\mcZ$};
    \node[green, scale=0.7] at (1-0.8, 1.5) {$\mcZ$};
    % \node[green, scale=0.6] at (1.5-0.8, 1) {$\mcZ^{\textcolor{magenta}{-Z_{1}}}$};
    \node[green, scale=0.7] at (1-0.8, 0.5) {$\mcZ^{\textcolor{magenta}{-Z_{2}}}$};

    \node[scale=0.7] at (1.7-0.3, 1.2) {$\text{+ h.c.}$};

    \draw (5.5+0.7, 1+0.5) -- (6.5+0.7, 1+0.5);
    \draw (5.5+0.7, 1-0.5) -- (6.5+0.7, 1-0.5);
    \draw (5.5+0.7, 1-0.5) -- (5.5+0.7, 1+0.5);
    % \draw (6.5+0.7, 1-0.5) -- (6.5+0.7, 1+0.5);
    \node[magenta] at (5.5+0.7, 1-0.5+0.5) {$Z$};
    % \node[magenta] at (5.5+0.7+1, 1-0.5+0.5) {$Z$};
    \node[magenta] at (5.5+0.7+1-0.5, 1-0.5+0.5-0.5) {$Z$};
    \node[magenta] at (5.5+0.7+1-0.5, 1-0.5+0.5-0.5+1) {$Z$};

\end{tikzpicture}}
\end{equation}

On the top and bottom smooth boundaries, the plaquette stabilizers are the same as they are in bulk. On the top side, the star stabilizers are
\begin{equation}
    \raisebox{-0.5\height}{
        \begin{tikzpicture}
            \draw[brown] (5-2.5, 1) -- (6-2.5, 1);
    % \draw[brown] (6-2.5, 1) -- (6-2.5, 2);
    \draw[brown] (6-2.5, 1) -- (7-2.5, 1);
    \draw[brown] (6-2.5, 1) -- (6-2.5, 0);

    \draw[gray] (5-2.5-0.3, 1+0.3) -- (6-2.5-0.3, 1+0.3);
    % \draw[gray] (6-2.5-0.3, 1+0.3) -- (6-2.5-0.3, 2+0.3);
    \draw[gray] (6-2.5-0.3, 1+0.3) -- (7-2.5-0.3, 1+0.3);
    \draw[gray] (6-2.5-0.3, 1+0.3) -- (6-2.5-0.3, 0+0.3);

    \node[magenta, scale=0.7] at (5-2.5-0.3+0.5, 1+0.3+0.15) {$1$};
    \draw[magenta] (5-2.5-0.3+0.5, 1+0.3) circle[radius=1pt];
    \fill[magenta] (5-2.5-0.3+0.5, 1+0.3) circle[radius=1pt];

    % \node[magenta, scale=0.5] at (6-2.5-0.3-0.15, 2+0.3-0.5) {$2$};
    % \draw[magenta] (6-2.5-0.3, 2+0.3-0.5) circle[radius=1pt];
    % \fill[magenta] (6-2.5-0.3, 2+0.3-0.5) circle[radius=1pt];

    \node[red, scale=0.7] at (5.5-2.5, 1) {$\mcX^{\textcolor{magenta}{Z_1}}$};
    % \node[red, scale=0.6] at (6-2.5, 1.5) {$\mcX^{\textcolor{magenta}{Z_2}}$};
    \node[red, scale=0.7] at (6.5-2.5, 1) {$\mcX^{\dag}$};
    \node[red, scale=0.7] at (6-2.5, 0.5) {$\mcX^{\dag}$};

    \draw (5+2.5+0.7, 1) -- (6+2.5+0.7, 1);
    % \draw (6+2.5+0.7, 1) -- (6+2.5+0.7, 2);
    \draw (6+2.5+0.7, 1) -- (7+2.5+0.7, 1);
    \draw (6+2.5+0.7, 1) -- (6+2.5+0.7, 0);

    \node[blue] at (6+2.5+0.5+0.7,1) {$X$};
    % \node[blue] at (6+2.5+0.7,1+0.5) {$X$};
    \node[blue] at (6+2.5-0.5+0.7,1) {$X$};
    \node[blue] at (6+2.5+0.7,1-0.5) {$X$};

    \draw[brown] (6+2.5+0.2+0.7, 1-0.2) -- (7+2.5+0.2+0.7, 1-0.2);
    \draw[brown] (6+2.5+0.2+0.7, 1-0.2) -- (6+2.5+0.2+0.7, 0-0.2);

    \node[purple] at (6+2.5+0.2+0.5+0.7, 1-0.2) {$\mathcal{C}$};
    \node[purple] at (6+2.5+0.2+0.7, 1-0.2-0.5) {$\mathcal{C}$};

    \node[scale=0.7] at (1.7+3.3, 1) {$\text{+ h.c.}$};

    \end{tikzpicture}
    
    }
\end{equation}
On the bottom side, the star stabilizers are
\begin{equation}
    \raisebox{-0.5\height}{
        \begin{tikzpicture}
            \draw[brown] (5-2.5, 1) -- (6-2.5, 1);
    \draw[brown] (6-2.5, 1) -- (6-2.5, 2);
    \draw[brown] (6-2.5, 1) -- (7-2.5, 1);
    % \draw[brown] (6-2.5, 1) -- (6-2.5, 0);

    \draw[gray] (5-2.5-0.3, 1+0.3) -- (6-2.5-0.3, 1+0.3);
    \draw[gray] (6-2.5-0.3, 1+0.3) -- (6-2.5-0.3, 2+0.3);
    \draw[gray] (6-2.5-0.3, 1+0.3) -- (7-2.5-0.3, 1+0.3);
    % \draw[gray] (6-2.5-0.3, 1+0.3) -- (6-2.5-0.3, 0+0.3);

    \node[magenta, scale=0.5] at (5-2.5-0.3+0.5, 1+0.3+0.15) {$1$};
    \draw[magenta] (5-2.5-0.3+0.5, 1+0.3) circle[radius=1pt];
    \fill[magenta] (5-2.5-0.3+0.5, 1+0.3) circle[radius=1pt];

    \node[magenta, scale=0.5] at (6-2.5-0.3-0.15, 2+0.3-0.5) {$2$};
    \draw[magenta] (6-2.5-0.3, 2+0.3-0.5) circle[radius=1pt];
    \fill[magenta] (6-2.5-0.3, 2+0.3-0.5) circle[radius=1pt];

    \node[red, scale=0.7] at (5.5-2.5, 1) {$\mcX^{\textcolor{magenta}{Z_1}}$};
    \node[red, scale=0.7] at (6-2.5, 1.5) {$\mcX^{\textcolor{magenta}{Z_2}}$};
    \node[red, scale=0.7] at (6.5-2.5, 1) {$\mcX^{\dag}$};
    % \node[red, scale=0.6] at (6-2.5, 0.5) {$\mcX^{\dag}$};

    \draw (5+2.5+0.7, 1) -- (6+2.5+0.7, 1);
    \draw (6+2.5+0.7, 1) -- (6+2.5+0.7, 2);
    \draw (6+2.5+0.7, 1) -- (7+2.5+0.7, 1);
    % \draw (6+2.5+0.7, 1) -- (6+2.5+0.7, 0);

    \node[blue] at (6+2.5+0.5+0.7,1) {$X$};
    \node[blue] at (6+2.5+0.7,1+0.5) {$X$};
    \node[blue] at (6+2.5-0.5+0.7,1) {$X$};
    % \node[blue] at (6+2.5+0.7,1-0.5) {$X$};

    \draw[brown] (6+2.5+0.2+0.7, 1-0.2) -- (7+2.5+0.2+0.7, 1-0.2);
    % \draw[brown] (6+2.5+0.2+0.7, 1-0.2) -- (6+2.5+0.2+0.7, 0-0.2);

    \node[purple] at (6+2.5+0.2+0.5+0.7, 1-0.2) {$\mathcal{C}$};
    % \node[purple] at (6+2.5+0.2+0.7, 1-0.2-0.5) {$\mathcal{C}$};

    \node[scale=0.7] at (1.7+3.3, 1.3) {$\text{+ h.c.}$};

    \end{tikzpicture}
    
    }
\end{equation}

\section{Errors and Syndromes after $\mathcal{D}(S_3)$ state preparation}
\label{S3Syndromes}

Once the quantum double ground state $\ket{S_3}$ is prepared, we can derive the violated projectors due to the application of qubit $X$ and $Z$ errors as well as qutrit $\mcX$ and $\mcZ$ errors. Such errors will locally violate the commuting projectors introduced in Eq.~\eqref{S3Stabs}, which we refer to as $S_3$ syndromes. We compute the probabilities of various syndromes occurring in the presence of the aforementioned errors. Unlike the Abelian case, such errors are generally not ribbon operators for anyons within the non-Abelian phase. For our purposes, we simply care about eliminating syndromes and returning $\mathcal{D}(S_3)$ back to the ground state. In the case of the charge anyons $B$ and $C$, such violations correspond directly to anyons and their internal states. In contrast, for flux anyons $D$ and $F$, we choose to directly measure the flux operators that stabilize the ground state, which do not map directly onto the flux measurements that characterize the internal states of these anyons as discussed in Appendix \ref{QuantumDouble}.

On a single qutrit $\mcZ$ error, we consider the following pair of vertices $v, v'$ with an error $\mcZ_{e}$ acting in between them:
\begin{equation}
    \raisebox{-0.5\height} {
        \begin{tikzpicture}

            \draw[brown] (5-2.5, 1) -- (6-2.5, 1);
            \draw[brown] (6-2.5, 1) -- (6-2.5, 2);
            \draw[brown] (6-2.5, 1) -- (7-2.5, 1);
            \draw[brown] (6-2.5, 1) -- (6-2.5, 0);
            \draw[brown] (7-2.5, 0) -- (7-2.5, 2);
            \draw[brown] (7-2.5, 1) -- (8-2.5, 1);
            
            \draw[dashed] (5-2.5-0.3, 1+0.3) -- (6-2.5-0.3, 1+0.3);
            \draw[dashed] (6-2.5-0.3, 1+0.3) -- (6-2.5-0.3, 2+0.3);
            \draw[dashed] (6-2.5-0.3, 1+0.3) -- (7-2.5-0.3, 1+0.3);
            \draw[dashed] (6-2.5-0.3, 1+0.3) -- (6-2.5-0.3, 0+0.3);
            \draw[dashed] (7-2.5-0.3, 0+0.3) -- (7-2.5-0.3, 2+0.3);
            \draw[dashed] (7-2.5-0.3, 1+0.3) -- (8-2.5-0.3, 1+0.3);

            \node[scale=0.6] at (6-2.5-0.1, 1-0.1) {$v$};
            \node[scale=0.6] at (7-2.5-0.1+0.2, 1-0.1) {$v'$};
            \node[scale=0.6] at (6.6-2.5-0.1, 1-0.1+0.2) {$e$};

            % \node[magenta, scale=0.5] at (5-2.5-0.3+0.5, 1+0.3+0.15) {$Z_{1}$};
            % \draw[magenta] (5-2.5-0.3+0.5, 1+0.3) circle[radius=1pt];
            % \fill[magenta] (5-2.5-0.3+0.5, 1+0.3) circle[radius=1pt];
            
            % \node[magenta, scale=0.5] at (6-2.5-0.3-0.15, 2+0.3-0.5) {$Z_{2}$};
            % \draw[magenta] (6-2.5-0.3, 2+0.3-0.5) circle[radius=1pt];
            % \fill[magenta] (6-2.5-0.3, 2+0.3-0.5) circle[radius=1pt];

            % \node[red, scale=0.6] at (5.5-2.5, 1) {$\mcX^{\textcolor{magenta}{Z_1}}$};
            % \node[red, scale=0.6] at (6-2.5, 1.5) {$\mcX^{\textcolor{magenta}{Z_2}}$};
            % \node[red, scale=0.6] at (6.5-2.5, 1) {$\mcX^{\dag}$};
            % \node[red, scale=0.6] at (6-2.5, 0.5) {$\mcX^{\dag}$};

            % \draw (2, 0) -- (2, 2.5);
            % \draw (2, 0) -- (2.2, 0);
            % \draw (2,2.5) -- (2.2, 2.5);

            % \draw (5.8, 0) -- (5.8, 2.5);
            % \draw (5.8, 0) -- (5.6, 0);
            % \draw (5.8,2.5) -- (5.6, 2.5);

        \end{tikzpicture}
    }
\end{equation}
We would like to analyze the syndromes of the state $\mcZ_{e}\ket{S_3}$. We define two projectors onto the different eigenvalues of $\tilde{A}_v,$ or equivalently the internal states of the $C$ anyon:
\begin{equation}
    P^{\tilde{A}_v}_{\omega} = \frac{1 + \omega^2 \tilde{A}_v + \omega \tilde{A}^2_v}{3}, P^{\tilde{A}_v}_{\omega^2} = \frac{1 + \omega \tilde{A}_v + \omega^2 \tilde{A}_v^2}{3}.
\end{equation}

Using these projectors, notice that $P^{\tilde{A}_v}_{\omega} \mcZ_{e}\ket{S_3} = \mcZ_{e}\ket{S_3},$ meaning that there is a $C$ anyon on vertex $v$ with internal state $\tilde{A}_v = \omega.$ Similarly, we find that $\tilde{A}_{v'} \mcZ_{e}\ket{S_3} = \omega^{-Z_e} \mcZ_{e}\ket{S_3}$, meaning that there is also a $C$ anyon on vertex $v'$ but with an indefinite internal state that is conditioned on $Z_e$. Indeed, we can calculate the probability of being in each of the two internal states:
\begin{equation}
    \bra{S_3} \mcZ_{e}^{\dagger}P^{\tilde{A}_{v'}}_{\omega} \mcZ_{e} \ket{S_3} = \bra{S_3} \mcZ_{e}^{\dagger}P^{\tilde{A}_{v'}}_{\omega^2} \mcZ_{e} \ket{S_3} = \frac{1}{2}.
\end{equation}
In addition to the $C$ anyons created, we also violate $A_{v}$. In particular, we find that
\begin{equation}
    \bra{S_3} \mcZ_{e}^{\dagger} P^{A_v}_{-1} \mcZ_{e} \ket{S_3} = \frac{1}{2},
\end{equation}
where $P^{A_v}_{-1} = \frac{1-A_v}{2}$ is a projector onto states with a $B$ anyon on vertex $v$. Thus, the state $\mcZ_{e}\ket{S_3}$ is in an equal superposition of having $A_{v} = 1$ and $A_{v} = -1$. Despite there being no definite outcome for $P^{B}_{v}$, we say there is just a $C$ anyon on vertex $v$ due to the $S_3$ fusion rule $B \times C = C$. 

We can now analyze the case of qutrit $\mcX$ errors, which is similar to the $\mcZ$ case. Here we consider two adjacent plaquettes $p$ and $p'$ with an $\mcX_e$ error occurring on a connecting edge $e$:
\begin{equation}
\label{PlquettePair}
    \raisebox{-0.5\height}{
        \begin{tikzpicture}
            \draw[brown] (0.5-0.8, 1.5) -- (1.5-0.8, 1.5);
            \draw[brown] (0.5-0.8, 0.5) -- (1.5-0.8, 0.5);
            \draw[brown] (0.5-0.8, 0.5) -- (0.5-0.8, 1.5);
            \draw[brown] (1.5-0.8, 0.5) -- (1.5-0.8, 1.5);
            
            \draw[dashed] (0.5-0.3-0.8, 1.5+0.3) -- (1.5-0.3-0.8, 1.5+0.3);
            \draw[dashed] (0.5-0.3-0.8, 0.5+0.3) -- (1.5-0.3-0.8, 0.5+0.3);
            \draw[dashed] (0.5-0.3-0.8, 0.5+0.3) -- (0.5-0.3-0.8, 1.5+0.3);
            \draw[dashed] (1.5-0.3-0.8, 0.5+0.3) -- (1.5-0.3-0.8, 1.5+0.3);

            \node[scale=0.6] at (1-0.3-0.8, 0.5+0.3+0.1) {$e$};
            \node[scale=0.6] at (1-0.3-0.8-0.5-0.15, 0.5+0.3+0.1-0.5-0.1) {$e'$};
            \node[scale=0.6] at (1-0.3-0.8-0.5-0.15, 0.5+0.3+0.1-0.5-0.1+0.5) {$v$};
            \node[scale=0.6] at (1-0.3-0.8-0.5-0.15+1+0.3, 0.5+0.3+0.1-0.5-0.1+0.5) {$v'$};
            \node[scale=0.6] at (1-0.3-0.8, 0.5+0.3+0.1+0.4) {$p$};
            \node[scale=0.6] at (1-0.3-0.8, 0.5+0.3+0.1-0.6) {$p'$};

            \draw[brown] (0.5-0.8, 1.5-1) -- (1.5-0.8, 1.5-1);
            \draw[brown] (0.5-0.8, 0.5-1) -- (1.5-0.8, 0.5-1);
            \draw[brown] (0.5-0.8, 0.5-1) -- (0.5-0.8, 1.5-1);
            \draw[brown] (1.5-0.8, 0.5-1) -- (1.5-0.8, 1.5-1);
            
            \draw[dashed] (0.5-0.3-0.8, 1.5+0.3-1) -- (1.5-0.3-0.8, 1.5+0.3-1);
            \draw[dashed] (0.5-0.3-0.8, 0.5+0.3-1) -- (1.5-0.3-0.8, 0.5+0.3-1);
            \draw[dashed] (0.5-0.3-0.8, 0.5+0.3-1) -- (0.5-0.3-0.8, 1.5+0.3-1);
            \draw[dashed] (1.5-0.3-0.8, 0.5+0.3-1) -- (1.5-0.3-0.8, 1.5+0.3-1);
        \end{tikzpicture}
    }
\end{equation}

Since $B_p = B_{p'} = 1,$ this means that the appropriate operator to measure to determine the plaquette fluxes are $\tilde{B}_p^{+}$ and $\tilde{B}_{p'}^{+}.$ We see that $\tilde{B}_p^{+} \mcX_e \ket{S_3} = \omega^2 \mcX_e \ket{S_3},$ while $\tilde{B}_{p'}^{+} \mcX_e \ket{S_3} = \omega^{Z_{e'}} \mcX_e \ket{S_3},$ so plaquette $p$ has flux $r^2$ while plaquette $p'$ has indefinite flux that depends on $Z_{e'}$. Moreover, since all the $\tilde{A}_{v}$ operators are not violated, we have $F$ anyons on both $p$ and $p'.$ 

Writing down projectors onto internal states of the $F$ anyon requires measurement of $\tilde{B}_p^{+}.$ Similarly, later we will see that projectors onto the internal state of the $D$ anyon require the measurement of $\tilde{B}_p^{-}.$ In order to minimize the number of distinct operators we have to measure, we choose to do error correction by simply measuring $\tilde{B}_p.$ The appropriate projectors are
\begin{equation}
    \label{BpProjectors}
    P^{\tilde{B}_p}_{\omega} = \frac{1 + \omega^2 \tilde{B}_p + \omega \tilde{B^2_p}}{3}, P^{\tilde{B}_p}_{\omega^2} = \frac{1 + \omega \tilde{B}_p + \omega^2 \tilde{B^2_p}}{3}.
\end{equation}
Indeed, we can calculate the probability amplitudes of being in each eigenstate
\begin{equation}
    \bra{S_3} \mcX_{e}^{\dagger}P^{\tilde{B}_{p'}}_{\omega} \mcX_{e} \ket{S_3} = \bra{S_3} \mcX_{e}^{\dagger} P^{\tilde{B}_{p'}}_{\omega^2} \mcX_{e} \ket{S_3} = \frac{1}{2},
\end{equation}
so plaquette $p'$ is in an equal superposition of having $\tilde{B}_{p'} = \omega$ and $\tilde{B}_{p'} = \omega^2.$ 

Like before, we see that $\mcX_{e}$ does not commute with $A_v,$ and we have 
\begin{equation}
    \bra{S_3} \mcX_{e}^{\dagger} P^{A_v}_{-1} \mcX_{e} \ket{S_3} = \frac{1}{2},
\end{equation}
meaning we probabilistically have a violated $A_{v}$ syndrome.

When a single qubit $Z$ error is applied, it is clear that two adjacent qubit star stabilizer $A_v$ are flipped to $-1,$ and no other projectors are flipped. These correspond to an adjacent pair of Abelian $B$ anyons.

We lastly consider a qubit $X_e$ error that occurs in the setup in Eq.~\eqref{PlquettePair}. First, we immediately see that $B_p = B_{p'} = -1$, so the appropriate flux operators to measure are $\tilde{B}_p^-$ and $\tilde{B}_{p'}^-$. Since $A_{v'}\ket{S_3} = \tilde{B}_p^-\ket{S_3} = \ket{S_3}$, we see that the site $(v', p)$ has a $D$ anyon in the first internal state as described in Appendix \ref{QuantumDouble}. On the other hand, the $D$ anyon on plaquette $p'$ is in an indefinite internal state.

As discussed above, we perform error correction by measuring $\tilde{B}_p$ and $\tilde{B}_{p'},$ obtaining the following syndrome probabilities. Using the projectors defined in Eq.~\eqref{BpProjectors},
\begin{equation}
     \bra{S_3} X_{e} P_{\omega}^{\tilde{B}_{p'}} X_{e} \ket{S_3} = \bra{S_3} X_{e} P_{\omega^2}^{\tilde{B}_{p'}} X_{e}\ket{S_3} = \frac{1}{3}.
\end{equation}

Similarly, we see that $X_e$ violates the $\tilde{A}_{v'}$ operator on vertex $v'$ - we can compute the probability of the three internal states of the $C$ anyon on $v'$: 
\begin{equation}
    \bra{S_3} X_{e} P_{\omega}^{\tilde{A}_v} X_{e} \ket{S_3} = \bra{S_3} X_{e} P_{\omega^2}^{\tilde{A}_v} X_{e} \ket{S_3} = \frac{1}{3}.
\end{equation}
Thus, we see that the $X_{e}$ error creates a pair of $D$ anyons on plaquette $p$ and $p'$ as well as a $C$ anyon on vertex $v'$. In particular, this implies that a string of $X$ errors will create $\tilde{A}_v$ and $\tilde{B}_p$ violations along the length of the string.

The probabilities computed in this section (along with those associated for more general errors) can be fed into a simulation of the error correction of the $S_3$ ground state. In particular, after errors occur and syndromes are measured in the order prescribed in the main text, a simulation of error correction requires one to sample from the probability distribution for each syndrome. We emphasize that this is distinct from the Abelian case, where syndromes are instead deterministic.

\section{Correction Procedures for $S_3$ Error Correction}
\label{ErrorCorrectionCircuits}

In this appendix, we detail the sequential adaptive circuits needed to eliminate $\tilde{A}_v$ and $\tilde{B}_p$ syndromes. As mentioned in the main text, the first step of our error correction procedure is removing $B_p = -1$ syndromes using a heralded decoder that additionally utilizes the $\tilde{A}_v$ syndrome information. Once this is done, we can correct $\tilde{A}_v$ and $\tilde{B}_p$ independently. 

We split this appendix into two sections: First, we tackle the case of correcting $C$ anyons with no knowledge of what operators created the syndromes (i.e.~local or non-local operators) and demonstrate that they can be cleaned up to either vacuum or a $C$ anyon with a sequential adaptive circuit, up to some Abelian $B$ anyons. Second, we illustrate that this correction procedure can eliminate a path of local $\mcZ$ and $\mcZ^{\dagger}$ errors into Abelian $B$ anyons.

\subsection{Sequential Adaptive Circuit for $C$ Anyons}
To illustrate our procedure, we analyze the case of $\tilde{A}_v$ syndromes along a chain of $4$ sites:
\begin{equation}
\label{AvSyndrome}
    \raisebox{-0.5\height} {
        \begin{tikzpicture}[scale=1.5]

            \draw[brown] (5-2.5, 1) -- (6-2.5, 1);
            \draw[brown] (6-2.5, 1) -- (6-2.5, 2);
            \draw[brown] (6-2.5, 1) -- (7-2.5, 1);
            \draw[brown] (6-2.5, 1) -- (6-2.5, 0);
            \draw[brown] (7-2.5, 0) -- (7-2.5, 2);
            \draw[brown] (7-2.5, 1) -- (8-2.5, 1);
            \draw[brown] (8-2.5, 1) -- (10-2.5, 1);
            \draw[brown] (8-2.5, 0) -- (8-2.5, 2);
            \draw[brown] (9-2.5, 0) -- (9-2.5, 2);
            
            \draw[dashed] (5-2.5-0.3, 1+0.3) -- (6-2.5-0.3, 1+0.3);
            \draw[dashed] (6-2.5-0.3, 1+0.3) -- (6-2.5-0.3, 2+0.3);
            \draw[dashed] (6-2.5-0.3, 1+0.3) -- (7-2.5-0.3, 1+0.3);
            \draw[dashed] (6-2.5-0.3, 1+0.3) -- (6-2.5-0.3, 0+0.3);
            \draw[dashed] (7-2.5-0.3, 0+0.3) -- (7-2.5-0.3, 2+0.3);
            \draw[dashed] (7-2.5-0.3, 1+0.3) -- (8-2.5-0.3, 1+0.3);
            \draw[dashed] (8-2.5-0.3, 1+0.3) -- (10-2.5-0.3, 1+0.3);
            \draw[dashed] (8-2.5-0.3, 0+0.3) -- (8-2.5-0.3, 2+0.3);
            \draw[dashed] (9-2.5-0.3, 0+0.3) -- (9-2.5-0.3, 2+0.3);

            \node[scale=0.75] at (6-2.5-0.4, 1+0.4) {$\bs{\omega}$};
            \node[scale=0.75] at (7-2.5-0.4, 1+0.4) {$\bs{1}$};
            \node[scale=0.75] at (8-2.5-0.4, 1+0.4) {$\bs{1}$};
            \node[scale=0.75] at (9-2.5-0.45, 1+0.4) {$\bs{\omega^2}$};

            \node[scale=0.75] at (6-2.5-0.4+0.3, 1+0.1) {$v_1$};
            \node[scale=0.75] at (7-2.5-0.4+0.3, 1+0.1) {$v_2$};
            \node[scale=0.75] at (8-2.5-0.4+0.3, 1+0.1) {$v_3$};
            \node[scale=0.75] at (9-2.5-0.4+0.3, 1+0.1) {$v_4$};

            \node[scale=0.55] at (6.6-2.5-0.4+0.3, 1-0.1) {$e_1$};
            \node[scale=0.55] at (7.6-2.5-0.4+0.3, 1-0.1) {$e_2$};
            \node[scale=0.55] at (8.6-2.5-0.4+0.3, 1-0.1) {$e_3$};

        \end{tikzpicture}
    }
\end{equation}
As an example, we have shown a chain of $4$ sites with syndromes $\tilde{A}_{v_1} = \omega, \tilde{A}_{v_2} = 1, \tilde{A}_{v_3} = 1, \tilde{A}_{v_4} = \omega^2,$ meaning we have $C$ anyons on vertices $v_1$ and $v_4.$

We now demonstrate how to fuse such a syndrome configuration into either vacuum or a single $C$ anyon on vertex $v_4.$ The first step is to apply a $\mcZ_{e_1}^{\dagger}$ operation, changing the syndromes to 
\begin{equation}
    \raisebox{-0.5\height} {
        \begin{tikzpicture}[scale=1.5]

            \draw[brown] (5-2.5, 1) -- (6-2.5, 1);
            \draw[brown] (6-2.5, 1) -- (6-2.5, 2);
            \draw[brown] (6-2.5, 1) -- (7-2.5, 1);
            \draw[brown] (6-2.5, 1) -- (6-2.5, 0);
            \draw[brown] (7-2.5, 0) -- (7-2.5, 2);
            \draw[brown] (7-2.5, 1) -- (8-2.5, 1);
            \draw[brown] (8-2.5, 1) -- (10-2.5, 1);
            \draw[brown] (8-2.5, 0) -- (8-2.5, 2);
            \draw[brown] (9-2.5, 0) -- (9-2.5, 2);
            
            \draw[dashed] (5-2.5-0.3, 1+0.3) -- (6-2.5-0.3, 1+0.3);
            \draw[dashed] (6-2.5-0.3, 1+0.3) -- (6-2.5-0.3, 2+0.3);
            \draw[dashed] (6-2.5-0.3, 1+0.3) -- (7-2.5-0.3, 1+0.3);
            \draw[dashed] (6-2.5-0.3, 1+0.3) -- (6-2.5-0.3, 0+0.3);
            \draw[dashed] (7-2.5-0.3, 0+0.3) -- (7-2.5-0.3, 2+0.3);
            \draw[dashed] (7-2.5-0.3, 1+0.3) -- (8-2.5-0.3, 1+0.3);
            \draw[dashed] (8-2.5-0.3, 1+0.3) -- (10-2.5-0.3, 1+0.3);
            \draw[dashed] (8-2.5-0.3, 0+0.3) -- (8-2.5-0.3, 2+0.3);
            \draw[dashed] (9-2.5-0.3, 0+0.3) -- (9-2.5-0.3, 2+0.3);

            \node[scale=0.75] at (6-2.5-0.4, 1+0.4) {$\bs{1}$};
            \node[scale=0.75] at (7-2.5-0.55, 1+0.4) {$\bs{\omega^{-Z_{e_1}}}$};
            \node[scale=0.75] at (8-2.5-0.4, 1+0.4) {$\bs{1}$};
            \node[scale=0.75] at (9-2.5-0.45, 1+0.4) {$\bs{\omega^2}$};

            \node[scale=0.75] at (6-2.5-0.4+0.3, 1+0.1) {$v_1$};
            \node[scale=0.75] at (7-2.5-0.4+0.3, 1+0.1) {$v_2$};
            \node[scale=0.75] at (8-2.5-0.4+0.3, 1+0.1) {$v_3$};
            \node[scale=0.75] at (9-2.5-0.4+0.3, 1+0.1) {$v_4$};

            \node[scale=0.75] at (6.6-2.5-0.4+0.3, 1-0.1) {$e_1$};
            \node[scale=0.75] at (7.6-2.5-0.4+0.3, 1-0.1) {$e_2$};
            \node[scale=0.75] at (8.6-2.5-0.4+0.3, 1-0.1) {$e_3$};

        \end{tikzpicture}
    }
\end{equation}
Next we measure $\tilde{A}_{v_2},$ which effectively measures $Z_{e_1}.$ Since $A_{v_2}$ does not commute with $\tilde{A}_{v_2},$ such a measurement will put us into a superposition of having no anyon on $v_2$ and having a $B$ anyon on $v_2.$ We postpone the correction of $B$ anyons to the last step of the decoding protocol as discussed in the main text. Let us denote the measurement outcome as $z_{e_1}.$ We then apply an adaptive correction operation $\mcZ^{-z_{e_1}}_{e_2}.$ This modifies the syndromes to 
\begin{equation}
    \raisebox{-0.5\height} {
        \begin{tikzpicture}[scale=1.5]

            \draw[brown] (5-2.5, 1) -- (6-2.5, 1);
            \draw[brown] (6-2.5, 1) -- (6-2.5, 2);
            \draw[brown] (6-2.5, 1) -- (7-2.5, 1);
            \draw[brown] (6-2.5, 1) -- (6-2.5, 0);
            \draw[brown] (7-2.5, 0) -- (7-2.5, 2);
            \draw[brown] (7-2.5, 1) -- (8-2.5, 1);
            \draw[brown] (8-2.5, 1) -- (10-2.5, 1);
            \draw[brown] (8-2.5, 0) -- (8-2.5, 2);
            \draw[brown] (9-2.5, 0) -- (9-2.5, 2);
            
            \draw[dashed] (5-2.5-0.3, 1+0.3) -- (6-2.5-0.3, 1+0.3);
            \draw[dashed] (6-2.5-0.3, 1+0.3) -- (6-2.5-0.3, 2+0.3);
            \draw[dashed] (6-2.5-0.3, 1+0.3) -- (7-2.5-0.3, 1+0.3);
            \draw[dashed] (6-2.5-0.3, 1+0.3) -- (6-2.5-0.3, 0+0.3);
            \draw[dashed] (7-2.5-0.3, 0+0.3) -- (7-2.5-0.3, 2+0.3);
            \draw[dashed] (7-2.5-0.3, 1+0.3) -- (8-2.5-0.3, 1+0.3);
            \draw[dashed] (8-2.5-0.3, 1+0.3) -- (10-2.5-0.3, 1+0.3);
            \draw[dashed] (8-2.5-0.3, 0+0.3) -- (8-2.5-0.3, 2+0.3);
            \draw[dashed] (9-2.5-0.3, 0+0.3) -- (9-2.5-0.3, 2+0.3);

            \node[scale=0.75] at (6-2.5-0.4, 1+0.4) {$\bs{1}$};
            \node[scale=0.75] at (7-2.5-0.4, 1+0.4) {$\bs{1}$};
            \node[scale=0.75] at (8-2.5-0.6, 1+0.4) {$\bs{\omega}^{z_{e_1}Z_{e_2}}$};
            \node[scale=0.75] at (9-2.5-0.45, 1+0.4) {$\bs{\omega^2}$};

            \node[scale=0.75] at (6-2.5-0.4+0.3, 1+0.1) {$v_1$};
            \node[scale=0.75] at (7-2.5-0.4+0.3, 1+0.1) {$v_2$};
            \node[scale=0.75] at (8-2.5-0.4+0.3, 1+0.1) {$v_3$};
            \node[scale=0.75] at (9-2.5-0.4+0.3, 1+0.1) {$v_4$};

            \node[scale=0.75] at (6.6-2.5-0.4+0.3, 1-0.1) {$e_1$};
            \node[scale=0.75] at (7.6-2.5-0.4+0.3, 1-0.1) {$e_2$};
            \node[scale=0.75] at (8.6-2.5-0.4+0.3, 1-0.1) {$e_3$};

        \end{tikzpicture}
    }
\end{equation}
As before, we measure $\tilde{A}_{v_3},$ which effectively measures $Z_{e_2}.$ Let us denote the obtained eigenvalue of $Z_{e_2}$ as $z_{e_2}.$ Subsequently, we apply an adaptive correction operation $\mcZ^{-z_{e_1}z_{e_2}}_{e_3},$ which yields
\begin{equation}
    \raisebox{-0.5\height} {
        \begin{tikzpicture}[scale=1.5]

            \draw[brown] (5-2.5, 1) -- (6-2.5, 1);
            \draw[brown] (6-2.5, 1) -- (6-2.5, 2);
            \draw[brown] (6-2.5, 1) -- (7-2.5, 1);
            \draw[brown] (6-2.5, 1) -- (6-2.5, 0);
            \draw[brown] (7-2.5, 0) -- (7-2.5, 2);
            \draw[brown] (7-2.5, 1) -- (8-2.5, 1);
            \draw[brown] (8-2.5, 1) -- (10-2.5, 1);
            \draw[brown] (8-2.5, 0) -- (8-2.5, 2);
            \draw[brown] (9-2.5, 0) -- (9-2.5, 2);
            
            \draw[dashed] (5-2.5-0.3, 1+0.3) -- (6-2.5-0.3, 1+0.3);
            \draw[dashed] (6-2.5-0.3, 1+0.3) -- (6-2.5-0.3, 2+0.3);
            \draw[dashed] (6-2.5-0.3, 1+0.3) -- (7-2.5-0.3, 1+0.3);
            \draw[dashed] (6-2.5-0.3, 1+0.3) -- (6-2.5-0.3, 0+0.3);
            \draw[dashed] (7-2.5-0.3, 0+0.3) -- (7-2.5-0.3, 2+0.3);
            \draw[dashed] (7-2.5-0.3, 1+0.3) -- (8-2.5-0.3, 1+0.3);
            \draw[dashed] (8-2.5-0.3, 1+0.3) -- (10-2.5-0.3, 1+0.3);
            \draw[dashed] (8-2.5-0.3, 0+0.3) -- (8-2.5-0.3, 2+0.3);
            \draw[dashed] (9-2.5-0.3, 0+0.3) -- (9-2.5-0.3, 2+0.3);

            \node[scale=0.75] at (6-2.5-0.4, 1+0.4) {$\bs{1}$};
            \node[scale=0.75] at (7-2.5-0.4, 1+0.4) {$\bs{1}$};
            \node[scale=0.75] at (8-2.5-0.4, 1+0.4) {$\bs{1}$};
            \node[scale=0.5] at (9-2.5-0.6, 1+0.4) {$\bs{\omega^{-1 + z_{e_1}z_{e_2}Z_{e_3}}}$};

            \node[scale=0.75] at (6-2.5-0.4+0.3, 1+0.1) {$v_1$};
            \node[scale=0.75] at (7-2.5-0.4+0.3, 1+0.1) {$v_2$};
            \node[scale=0.75] at (8-2.5-0.4+0.3, 1+0.1) {$v_3$};
            \node[scale=0.75] at (9-2.5-0.4+0.3, 1+0.1) {$v_4$};

            \node[scale=0.75] at (6.6-2.5-0.4+0.3, 1-0.1) {$e_1$};
            \node[scale=0.75] at (7.6-2.5-0.4+0.3, 1-0.1) {$e_2$};
            \node[scale=0.75] at (8.6-2.5-0.4+0.3, 1-0.1) {$e_3$};

        \end{tikzpicture}
    }
\end{equation}
The final step is to measure $\tilde{A}_{v_4}.$ This allows us to obtain a measurement result $z_{e_3}$ for $Z_{e_3}.$ We now see that $\tilde{A}_{v_1} = \tilde{A}_{v_2} = \tilde{A}_{v_3} = 1,$ while $\tilde{A}_{v_4} = -1 + z_{e_1}z_{e_2}z_{e_3},$ meaning that whether the obtained syndrome fuse to vacuum or another syndrome depends on the measurement results. 

If the syndrome in Eq.~\eqref{AvSyndrome} was created by local operators along the length of the string, then we must have that $z_{e_1}z_{e_2}z_{e_3} = 1$ because the global topological charge of the $S_3$ code subject to purely local errors must be vacuum. On the other hand, if the syndromes on $v_1$ and $v_4$ were from different $C$ anyon vacuum pairs, then there is a non-zero probability that $z_{e_1}z_{e_2}z_{e_3} = -1,$ meaning that the two original $C$ anyons fuse to another $C$ anyon. We note that an analogous correction procedure exists for correcting $\tilde{B}_p$ syndromes utilizing adaptive $\mcX$ correction operations.

To understand this physically, recall from Section~\ref{CCGateMain} that in the $S_3$ wavefunction, $Z = -1$ measurement outcomes imply the presence of a change conjugation domain wall, or in the gauged picture, a $D$ anyon line. From the braiding properties of the $S_3$ quantum double, two $C$ anyons created from vacuum that re-fused along a path that crosses a $D$ anyon line will fuse to another $C$ anyon. Similarly, for the case of $\tilde{B}_p$ syndromes, two $F$ anyons created from vacuum that re-fused along a path that crosses a $D$ anyon line will fuse to another $F$ anyon. Thus, the value of $z_{e_1}z_{e_2}z_{e_3}$ tells us whether there are an even or odd number of charge conjugation domain walls separating the anyons at $v_1$ and $v_4$.

We remark that it is alternatively possible to correct the above syndromes purely unitarily with linear depth circuits. Such circuits will mirror expressions for the $C$ anyon found in Eq.~\eqref{CAnyonLine} and in other references~\cite{li_domain_2024, lyons_protocols_2025}.

\subsection{Correction of Local $\mcZ/\mcZ^{\dagger}$ errors}

We consider the correction a path $\mcZ$ and $\mcZ^{\dagger}$ errors acting on the $S_3$ ground state to demonstrate how the circuit from the previous section can fuse a chain of local errors to up to the presence of Abelian $B$ anyons. As described in the main text, an analogous circuit exists for correcting paths of $\mcX$ and $\mcX^{\dagger}$ errors by measuring $\tilde{B}_p$ syndromes.

In particular, let $\mcZ_{e_1}^{a_1}, \mcZ_{e_2}^{a_2},$ and $\mcZ_{e_3}^{a_3}$ errors act on edges $e_1, e_2, e_3,$ where $a_i \in \{1, -1\}$ are unknown to us. After measuring the $\tilde{A}_v$ operators, we obtain the following syndromes all along the length of the path:
\begin{equation}
\label{AvSyndrome_4}
    \raisebox{-0.5\height} {
        \begin{tikzpicture}[scale=1.5]

            \draw[brown] (5-2.5, 1) -- (6-2.5, 1);
            \draw[brown] (6-2.5, 1) -- (6-2.5, 2);
            \draw[brown] (6-2.5, 1) -- (7-2.5, 1);
            \draw[brown] (6-2.5, 1) -- (6-2.5, 0);
            \draw[brown] (7-2.5, 0) -- (7-2.5, 2);
            \draw[brown] (7-2.5, 1) -- (8-2.5, 1);
            \draw[brown] (8-2.5, 1) -- (10-2.5, 1);
            \draw[brown] (8-2.5, 0) -- (8-2.5, 2);
            \draw[brown] (9-2.5, 0) -- (9-2.5, 2);
            
            \draw[dashed] (5-2.5-0.3, 1+0.3) -- (6-2.5-0.3, 1+0.3);
            \draw[dashed] (6-2.5-0.3, 1+0.3) -- (6-2.5-0.3, 2+0.3);
            \draw[dashed] (6-2.5-0.3, 1+0.3) -- (7-2.5-0.3, 1+0.3);
            \draw[dashed] (6-2.5-0.3, 1+0.3) -- (6-2.5-0.3, 0+0.3);
            \draw[dashed] (7-2.5-0.3, 0+0.3) -- (7-2.5-0.3, 2+0.3);
            \draw[dashed] (7-2.5-0.3, 1+0.3) -- (8-2.5-0.3, 1+0.3);
            \draw[dashed] (8-2.5-0.3, 1+0.3) -- (10-2.5-0.3, 1+0.3);
            \draw[dashed] (8-2.5-0.3, 0+0.3) -- (8-2.5-0.3, 2+0.3);
            \draw[dashed] (9-2.5-0.3, 0+0.3) -- (9-2.5-0.3, 2+0.3);

            \node[scale=0.65] at (6-2.5-0.4, 1+0.4) {$\bs{\omega^{a_1}}$};
            \node[scale=0.65] at (7-2.5-0.53, 1+0.4) {$\bs{\omega^{a_2 - a_1z_{e_1}}}$};
            \node[scale=0.65] at (8-2.5-0.53, 1+0.4) {$\bs{\omega^{a_3 - a_2z_{e_2}}}$};
            \node[scale=0.65] at (9-2.5-0.48, 1+0.4) {$\bs{\omega^{-a_3z_{e_3}}}$};

            \node[scale=0.75] at (6-2.5-0.4+0.3, 1+0.1) {$v_1$};
            \node[scale=0.75] at (7-2.5-0.4+0.3, 1+0.1) {$v_2$};
            \node[scale=0.75] at (8-2.5-0.4+0.3, 1+0.1) {$v_3$};
            \node[scale=0.55] at (9-2.5-0.4+0.3, 1+0.1) {$v_4$};

            \node[scale=0.55] at (6.6-2.5-0.4+0.3, 1-0.1) {$e_1$};
            \node[scale=0.55] at (7.6-2.5-0.4+0.3, 1-0.1) {$e_2$};
            \node[scale=0.55] at (8.6-2.5-0.4+0.3, 1-0.1) {$e_3$};

        \end{tikzpicture}
    }
\end{equation}

Generalizing calculations from Appendix \ref{S3Syndromes}, one can show that each of the stabilizers in the bulk of the string is violated with probability $\frac{1}{2}.$ Here $z_{e_1}, z_{e_2}, z_{e_3}$ indicate that each of the vertex stabilizers have been collapsed to a definite eigenvalue. Note crucially that we do not have direct access to the $z_{e_i}$ values because the $a_{i}$ values are unknown. However, we can still execute our adaptive circuit moving from left to right.

Note that the measurement of $\tilde{A}_{v_1}$ allows us to learn $a_1.$ We now apply a $\mcZ^{-a_1}_{e_1}$ correction operation, changing the syndromes to 
\begin{equation}
\label{AvSyndrome_3}
    \raisebox{-0.5\height} {
        \begin{tikzpicture}[scale=1.5]

            \draw[brown] (5-2.5, 1) -- (6-2.5, 1);
            \draw[brown] (6-2.5, 1) -- (6-2.5, 2);
            \draw[brown] (6-2.5, 1) -- (7-2.5, 1);
            \draw[brown] (6-2.5, 1) -- (6-2.5, 0);
            \draw[brown] (7-2.5, 0) -- (7-2.5, 2);
            \draw[brown] (7-2.5, 1) -- (8-2.5, 1);
            \draw[brown] (8-2.5, 1) -- (10-2.5, 1);
            \draw[brown] (8-2.5, 0) -- (8-2.5, 2);
            \draw[brown] (9-2.5, 0) -- (9-2.5, 2);
            
            \draw[dashed] (5-2.5-0.3, 1+0.3) -- (6-2.5-0.3, 1+0.3);
            \draw[dashed] (6-2.5-0.3, 1+0.3) -- (6-2.5-0.3, 2+0.3);
            \draw[dashed] (6-2.5-0.3, 1+0.3) -- (7-2.5-0.3, 1+0.3);
            \draw[dashed] (6-2.5-0.3, 1+0.3) -- (6-2.5-0.3, 0+0.3);
            \draw[dashed] (7-2.5-0.3, 0+0.3) -- (7-2.5-0.3, 2+0.3);
            \draw[dashed] (7-2.5-0.3, 1+0.3) -- (8-2.5-0.3, 1+0.3);
            \draw[dashed] (8-2.5-0.3, 1+0.3) -- (10-2.5-0.3, 1+0.3);
            \draw[dashed] (8-2.5-0.3, 0+0.3) -- (8-2.5-0.3, 2+0.3);
            \draw[dashed] (9-2.5-0.3, 0+0.3) -- (9-2.5-0.3, 2+0.3);

            \node[scale=0.65] at (6-2.5-0.4, 1+0.4) {$\bs{1}$};
            \node[scale=0.65] at (7-2.5-0.53, 1+0.4) {$\bs{\omega^{a_2}}$};
            \node[scale=0.65] at (8-2.5-0.53, 1+0.4) {$\bs{\omega^{a_3 - a_2z_{e_2}}}$};
            \node[scale=0.65] at (9-2.5-0.48, 1+0.4) {$\bs{\omega^{-a_3z_{e_3}}}$};

            \node[scale=0.75] at (6-2.5-0.4+0.3, 1+0.1) {$v_1$};
            \node[scale=0.75] at (7-2.5-0.4+0.3, 1+0.1) {$v_2$};
            \node[scale=0.75] at (8-2.5-0.4+0.3, 1+0.1) {$v_3$};
            \node[scale=0.55] at (9-2.5-0.4+0.3, 1+0.1) {$v_4$};

            \node[scale=0.55] at (6.6-2.5-0.4+0.3, 1-0.1) {$e_1$};
            \node[scale=0.55] at (7.6-2.5-0.4+0.3, 1-0.1) {$e_2$};
            \node[scale=0.55] at (8.6-2.5-0.4+0.3, 1-0.1) {$e_3$};

        \end{tikzpicture}
    }
\end{equation}
Notice that this correction operation eliminates the $z_{e_1}$ dependence in the eigenvalue of $\tilde{A}_{v_2}.$ We subsequently measure $\tilde{A}_{v_2},$ which allows us to learn the value of $a_2.$ Repeating the procedure for $e_1,$ we apply a $\mcZ^{-a_2}_{e_2}$ correction operation, which gives us the syndromes
\begin{equation}
\label{AvSyndrome_2}
    \raisebox{-0.5\height} {
        \begin{tikzpicture}[scale=1.5]

            \draw[brown] (5-2.5, 1) -- (6-2.5, 1);
            \draw[brown] (6-2.5, 1) -- (6-2.5, 2);
            \draw[brown] (6-2.5, 1) -- (7-2.5, 1);
            \draw[brown] (6-2.5, 1) -- (6-2.5, 0);
            \draw[brown] (7-2.5, 0) -- (7-2.5, 2);
            \draw[brown] (7-2.5, 1) -- (8-2.5, 1);
            \draw[brown] (8-2.5, 1) -- (10-2.5, 1);
            \draw[brown] (8-2.5, 0) -- (8-2.5, 2);
            \draw[brown] (9-2.5, 0) -- (9-2.5, 2);
            
            \draw[dashed] (5-2.5-0.3, 1+0.3) -- (6-2.5-0.3, 1+0.3);
            \draw[dashed] (6-2.5-0.3, 1+0.3) -- (6-2.5-0.3, 2+0.3);
            \draw[dashed] (6-2.5-0.3, 1+0.3) -- (7-2.5-0.3, 1+0.3);
            \draw[dashed] (6-2.5-0.3, 1+0.3) -- (6-2.5-0.3, 0+0.3);
            \draw[dashed] (7-2.5-0.3, 0+0.3) -- (7-2.5-0.3, 2+0.3);
            \draw[dashed] (7-2.5-0.3, 1+0.3) -- (8-2.5-0.3, 1+0.3);
            \draw[dashed] (8-2.5-0.3, 1+0.3) -- (10-2.5-0.3, 1+0.3);
            \draw[dashed] (8-2.5-0.3, 0+0.3) -- (8-2.5-0.3, 2+0.3);
            \draw[dashed] (9-2.5-0.3, 0+0.3) -- (9-2.5-0.3, 2+0.3);

            \node[scale=0.65] at (6-2.5-0.4, 1+0.4) {$\bs{1}$};
            \node[scale=0.65] at (7-2.5-0.53, 1+0.4) {$\bs{1}$};
            \node[scale=0.65] at (8-2.5-0.53, 1+0.4) {$\bs{\omega^{a_3}}$};
            \node[scale=0.65] at (9-2.5-0.48, 1+0.4) {$\bs{\omega^{-a_3z_{e_3}}}$};

            \node[scale=0.75] at (6-2.5-0.4+0.3, 1+0.1) {$v_1$};
            \node[scale=0.75] at (7-2.5-0.4+0.3, 1+0.1) {$v_2$};
            \node[scale=0.75] at (8-2.5-0.4+0.3, 1+0.1) {$v_3$};
            \node[scale=0.55] at (9-2.5-0.4+0.3, 1+0.1) {$v_4$};

            \node[scale=0.55] at (6.6-2.5-0.4+0.3, 1-0.1) {$e_1$};
            \node[scale=0.55] at (7.6-2.5-0.4+0.3, 1-0.1) {$e_2$};
            \node[scale=0.55] at (8.6-2.5-0.4+0.3, 1-0.1) {$e_3$};

        \end{tikzpicture}
    }
\end{equation}
On the last step, we measure $\tilde{A}_{v_3}$: this allows us access to $a_3$ and applying a $\mcZ^{-a_3}_{e_3}$ correction operation eliminates all syndromes.

As discussed, the measurement of $\tilde{A}_{v_i}$ does not necessarily commute with $A_{v_i}$ operators. Thus, in the case that we get a non-trivial syndrome of $\omega, \omega^2$ from the measurement, we create a superposition of having a $B$ anyon or no anyon on the measured site, as discussed in Appendix \ref{S3Syndromes}.

\section{Magic state generated from the protocol}
\label{magicstate}

In this section, we explicitly show how our protocol can probabilistically generate magic states for qubit surface code. To set conventions for the logical states of both codes, we use the computational basis $\ket{b^{i}a^{j}}$ to describe the basis states of $\mathbb{C}^2 \otimes \mathbb{C}^3,$ where $i \in \{0, 1\}$ and $j \in \{0, 1, 2\}.$ In describing the matrices below, we order this basis by $\{1, ba, a^2, b, a, ba^2\}.$

In addition to a logical $C\mathcal{C}$ gate, we will further utilize the logical generalized Hadamard gate on the $\mathbb{Z}_2 \times \mathbb{Z}_3$ surface code that is given by $\overline{H} = \bigotimes_{e} H_e$ followed by code rotation. Given the fact that $\mathbb{Z}_2 \times \mathbb{Z}_3$ is isomorphic to $\mathbb{Z}_6$, we define $H$ for the $\mathbb{Z}_6$ qudit as
\begin{equation}
    H = \frac{1}{\sqrt{6}}\begin{pmatrix}
        1 & 1 & 1 & 1 & 1 & 1 \\
        1 & \omega & \omega^2 & \omega^3 & \omega^4 & \omega^5 \\
        1 & \omega^2 & \omega^4 & 1 & \omega^2 & \omega^4 \\
        1 & \omega^3 & 1 & \omega^3 & 1 & \omega^3 \\
        1 & \omega^4 & \omega^2 & 1 & \omega^4 & \omega^2 \\
        1 & \omega^5 & \omega^4 & \omega^3 & \omega^2 & \omega
    \end{pmatrix},
\end{equation}
where $\omega = e^{2 \pi i /6}.$ One can check this matrix is equivalent to $H_2 \otimes H_3^{\dagger}$ on the $\mathbb{C}^2 \otimes \mathbb{C}^3$ Hilbert space. Note that this operator is composed of operators in the second Clifford hierarchy of qubit and qutrit operators.

The controlled-charge conjugation gate is given by
\begin{equation}
    C\mathcal{C} = \begin{pmatrix}
        1 & 0 & 0 & 0 & 0 & 0 \\
        0 & 0 & 0 & 0 & 1 & 0 \\
        0 & 0 & 1 & 0 & 0 & 0 \\
        0 & 0 & 0 & 1 & 0 & 0 \\
        0 & 1 & 0 & 0 & 0 & 0 \\
        0 & 0 & 0 & 0 & 0 & 1
    \end{pmatrix}.
\end{equation}

Consider the following operation obtained by composing the above gates:
\begin{equation}
    H^{\dagger} C\mathcal{C} H = \frac{1}{3}\begin{pmatrix}
        3 & 0 & 0 & 0 & 0 & 0 \\
        0 & 1 & 0 & 1 -\sqrt{3}i & 0 & 1 + \sqrt{3}i \\
        0 & 0 & 1 & 0 & 0 & 0 \\
        0 & 1 + \sqrt{3}i & 0  & 1 & 0 & 1 - \sqrt{3}i \\
        0 & 0 & 0 & 0 & 1 & 0 \\
        0 & 1 - \sqrt{3}i & 0 & 1 + \sqrt{3}i & 0 & 1
    \end{pmatrix}.
\end{equation}

To obtain a qubit magic state, we start with the initial state $|\psi\rangle = \frac{1}{\sqrt{2}} \left(|0\rangle + |b\rangle\right)$. After applying the $H^{\dagger} C\mathcal{C} H$ gate, we get
\begin{equation}
    H^{\dagger} C\mathcal{C} H |\psi\rangle = \frac{1}{3\sqrt{2}} \left(3|0\rangle + \left(1 - \sqrt{3}i\right)|ba\rangle + |b\rangle + \left(1 + \sqrt{3}i\right)|ba^2\rangle \right). \label{eq:HCCH_1}
\end{equation}

Now we measure the $\mathbb{Z}_3$ layer in the onsite $\mathcal{Z}_{e}$ basis. This has the effect of measuring the logical state of the $\Z_3$ code in the $\overline{\mcZ}$ basis, or equivalently the $\{\ket{a^j}\}$ basis. Post-measurement, we obtain three possible states on the remaining $\Z_2$ code depending on the measured eigenvalue of $\overline{\mcZ}$. If the measured $\mathbb{Z}_3$ state has $\overline{\mcZ} = 1,$ which occurs with probability $\frac{5}{9},$ the $\mathbb{Z}_2$ state becomes 
    \begin{align}
        |\psi_1\rangle = \frac{1}{\sqrt{10}} \left(3 |0\rangle + |1\rangle \right).
    \end{align}
This state can be utilized as a magic state for the $\mathbb{Z}_2$ surface code since it can only be obtained by a Bloch sphere rotation by an angle that is not a rational multiple of $2\pi$~\cite{adleman1997quantum,barenco1995universal,barenco1995elementary,bernstein1993quantum,deutsch1989quantum,boykin1999universal}. In other two cases, the measured states are not magic states. However, by repeating this procedure many times, we can obtain at least one qubit magic state with probability arbitrarily close to $1.$

We can instead measure out the $\mathbb{Z}_2$ layer of the state in Eq.~\eqref{eq:HCCH_1} in the onsite $Z_{e}$ basis. When the measured $\mathbb{Z}_2$ state satisfies $\overline{Z} = -1$, the measured $\mathbb{Z}_3$ state we get is
\begin{align}
    |\psi_2\rangle = \frac{1}{3} |0\rangle + \left(\frac{1}{3} - \frac{i}{\sqrt{3}}\right) |1\rangle + \left(\frac{1}{3} + \frac{i}{\sqrt{3}}\right) |2\rangle.
\end{align}

This state is not a qutrit stabilizer state, but we leave it to future work to prove that this state can be utilized for universal qutrit computation.

% \ZJ{I suspect it's a magic state for qutrit surface code. I haven't come up with a proof yet. If you have any idea feel free to add.}

% \bibliographystyle{apsrev4-2}
% \bibliography{refs}

\end{document}